\documentclass[traditabstract,letter]{aa}  
\usepackage[pdf]{pstricks}
\usepackage{graphicx}
\usepackage[fleqn]{amsmath}
\usepackage{txfonts}
\usepackage{xcolor}
\usepackage{natbib}
\usepackage{soul}

\bibpunct{(}{)}{;}{a}{}{,}	

\begin{document}
   \title{The effect of viewing angle on the Kennicutt-Schmidt relation of the local molecular clouds}
      \author{J. Kainulainen\inst{1},	
              S. Rezaei Kh.\inst{2,1},
              A. Spilker\inst{1}, \and
              J. Orkisz\inst{1}
          }


	\institute{Chalmers University of Technology, Department of Space, Earth and Environment, SE-412 93 Gothenburg, Sweden \\
              	\email{jouni.kainulainen@chalmers.se}     
				\and 
				Max-Planck-Institute for Astronomy, K\"onigstuhl 17, 69117 Heidelberg, Germany \\
		        }
   \date{Received ; accepted }
  \abstract
{The Gaia data give us an unprecedented view to the 3-dimensional (3D) structure of molecular clouds in the Solar neighbourhood. We study how the projected areas and masses of clouds, and consequently the Kennicutt-Schmidt relation (KS-relation), depend on the viewing angle. We derive the probability distributions of the projected areas and masses for nine clouds within 400 pc from the Sun using 3D dust distribution data from the literature. We find that the viewing angle can have a dramatic effect on the observed areas and masses of individual clouds. The joint probability distributions of the areas and masses are strongly correlated, relatively flat, and can show multiple peaks. The typical ranges and 50\% quartiles of the distributions are roughly 100-200\% and 20-80\% of the median value, respectively, making viewing angle effects important for all individual clouds. The threshold value used to define the cloud areas is also important; our analysis suggests that the clouds become more anisotropic for smaller thresholds (larger scales). On average, the areas and masses of the plane-of-the-sky and face-on projections agree, albeit with a large scatter. This suggests that sample averages of areas and masses are relatively free of viewing angle effects, which is important to facilitate comparisons of extragalactic and galactic data. Ultimately, our results demonstrate that a cloud's location in the KS-relation is affected by viewing angle in a non-trivial manner. However, the KS-relation of our sample as a whole is not strongly affected by these effects, because the co-variance of the areas and masses causes the observed mean column density to remain relatively constant.
}
   \keywords{ISM: clouds - ISM: structure - stars: formation} 
  \titlerunning{The KS-relation of local molecular clouds}
  \maketitle


\section{Introduction}    


One central topic that permeates star formation studies across the scales in galaxies is decoding the information encapsulated in various scaling relations. In particular, the relationship between the gas and star formation rate surface densities, i.e., the Kennicutt-Schmidt relation (KS-relation; \citealt{Schmidt1959, Kennicutt1998}), has been a subject of intense study for decades \citep[for a review, see][]{kennicuttEvans2012review}. Right now, the accurate astrometric data from the Gaia satellite are opening a new door to gain insight into the key relations like the KS-relation. The Gaia data enable 3-dimensional (3D) mapping of the dust distribution in the interstellar medium \citep[e.g.,][]{Rezei2017method, Rezaei2018spiralarms, Green2019dustmap, Lallement2019dustmap, Leike2019Enslin, Leike2020}, which in turn enables determining the 3D shapes of star-forming molecular clouds \citep[e.g.,][]{Grosschedl2018Orion3D, Rezaei2020Orion3D, Roccatagliata2020Taurus3D, Zucker2021}. This provides a unique possibility to improve our interpretation of the observed KS-relations.

Revealing the 3D nature of molecular cloud properties and its impact on the KS-relation is important from various perspectives. First, our vantage point in the Milky Way dictates that we only see molecular clouds from one specific angle, parallel to the plane of the Galactic disk. This may bias the observed cloud properties as the clouds may be anisotropic, i.e., viewing them from different angles can lead to different properties. For example, galaxy-scale phenomena or processes linked to molecular cloud formation, such as shear, Galactic gravitational potential, and magnetic fields, can be hypothesised to influence cloud morphology in an anisotropic manner \citep[e.g.,][]{Dobbs2014}. Depending on the process in question, clouds could for example have preferential orientations in the Galactic disk. It is therefore important to consider the cloud properties and the emergence of the KS-relation from different viewing angles. Particularly interestingly, it has been pointed out by \citet{Lada2013schmidt} that there appears to be no KS-relation among a sample of nearby molecular clouds. Whether or not this result depends on the viewing angle is fundamental for understanding its relevance. 

Second, describing the possible effects of the viewing angle is of timely interest from the perspective of extragalactic studies. One focus point of the modern studies, spearheaded by the Physics at High Angular Resolution in Nearby GalaxieS (PHANGS) survey \citep{Leroy2021arXiv}, is to connect information from scales of entire galaxies to the information from the Milky Way clouds \citep[e.g.,][]{Leroy2016portrait, Schinnerer2019, Sun2020phangs, Spilker2021}. This inevitably means comparisons of extragalactic data, obtained in varying (often close-to face-on) viewing angles, with data from the Milky Way that are obtained from within the disk. Such comparisons require knowledge of the role played by the viewing angle.  


\begin{figure*}
\centering
\includegraphics[width=\textwidth]{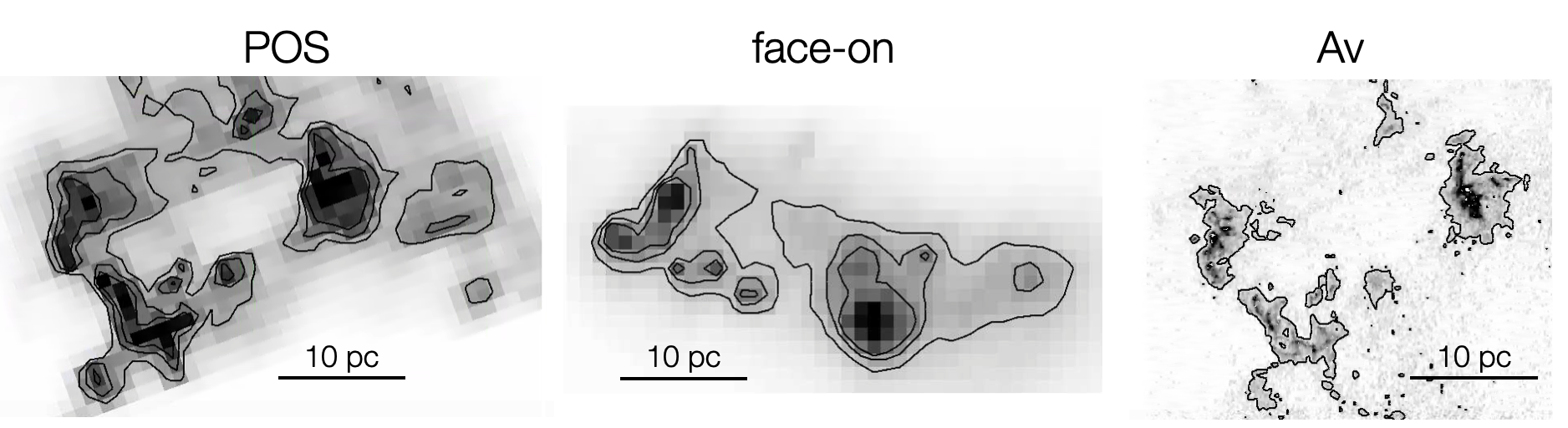}
\caption{Chamaeleon viewed from the POS (left) and face-on angles (centre); the data are from \citet{Leike2020} and the cloud definition from \citet{Zucker2021}. The contours are at $A_\mathrm{G} = {0.5, 0.75, 1}$ mag. The pixel size is 1 pc. The right frame shows a higher-resolution (2\arcmin) near-infrared  extinction map \citep{kainulainen2009probing}. For visibility, only one contour at $A_\mathrm{G} = 0.75$ mag ($A_\mathrm{V} \approx 1$ mag) is shown.}
\label{fig:viewingangles}
\end{figure*}


In this Letter, we address how the two key ingredients of the KS-relation, namely the projected area and mass of clouds, depend on the viewing angle. Our work is enabled by recent progress in mapping the 3D structures of nearby clouds. In particular, \citet{Leike2020} has derived relatively high-resolution (2 pc; 1 pc cell-size) 3D distribution of dust in the volume closer than about 400 pc of the Sun. These data enable us to consider clouds from all possible angles and describe the projection effects on the areas, masses, and ultimately on the KS-relation.

\section{Data}            


We use the 3D dust distribution of the volume closer than 400 pc of the Sun, derived and published by \citet{Leike2020}. For the description of the method, we refer to \citet[][see also \citealt{Leike2019Enslin}]{Leike2020}; we use the data as provided. 

We analyse a sample of nine nearby molecular clouds, listed in Table \ref{tab:parameters}. The sample choice is driven by selecting practically all major clouds within 400 pc that can be identified from the data reliably. We recall here the note from \citet{Zucker2021} that even though \citet{Leike2020} data cover the Orion molecular cloud, there are artefacts in the data due to it being close to the edge of the survey volume. Therefore, we do not include Orion in our sample. We define the volumes of the clouds with boxes as described in \citet[][see their Table 1]{Zucker2021}. 

The \cite{Leike2020} data are given in units of optical depth in the Gaia G band, $\tau_\mathrm{G}$ (673 nm; \citealt{Jordi2010gaia}), per 1 pc$^{3}$ sized cell. We convert this to column density using the same conversion factors as \citet{Zucker2021}. The $\tau_\mathrm{G}$ is first converted to G band extinction per 1 pc with a factor of 1.086. Then, the G band extinction is converted to column density using $A_\mathrm{G}/N_\mathrm{H} = 4 \times 10^{-22}$ cm$^{-2}$ \citep{Draine2009ASPC}, which stands for column density of hydrogen nuclei. At this point, the value is still per 1 pc cell of the \citet{Leike2020} data. The column densities per cell are later used to calculate the total column densities (after projection, see below), and from therein, the masses of the clouds. We note that the column density thresholds used to define the cloud areas are implemented on the projected data, as an observer would do.  


We derive the probability distributions of projected areas and masses for each of the nine molecular clouds. This is achieved by viewing a cloud from an array of viewing angles distributed on a hemisphere. Technically, we rotate the cubes of \citet{Leike2020} according to the viewing angle in question using successive repetitions of the IDL function \textrm{ROT} and then collapse the rotated data onto a 2D plane. This retains the centre of the data cube as defined by \citet{Zucker2021} as the centre of the projection and the original 1 pc cell size. 
The rotation of the data with \textrm{ROT} involves interpolation onto a new grid; we check the total mass of the cube before and after the rotation to ascertain that the mass is conserved. Once projected, we determine the cloud area and mass above an extinction threshold, $A_\mathrm{G,th}$. As an example, Fig. \ref{fig:viewingangles} illustrates two viewing angles for one cloud (Chamaeleon). The array of viewing angles are chosen so that they sample equal areas on the hemisphere, i.e., we create a probability distribution per unit solid angle. We sample the hemisphere using about 21\,000 points. 

We use three extinction thresholds, $A_\mathrm{G,th} = \{0.5, 0.75, 1\}$ mag, to study the effect of the used threshold. These thresholds outline structures in the regions in somewhat different ways (see Figs. \ref{fig:viewingangles}, \ref{fig:different_angles_dir1}, and \ref{fig:different_angles_dir2} for examples), and they enclose discrete regions clearly smaller than the total area of the projected volume. The volumes defined by \citet{Zucker2021} contain virtually all high density gas in the studied complexes; our approach is to study how that gas reservoir appears from various viewing angles. We note that it is not possible to use larger thresholds. The resolution of the \citet{Leike2020} data is still coarse in the context of the cloud substructure; averaging over the cell size of 1 pc$^3$ reduces the dynamical range of the projected density data. Further, the \citet{Leike2020} data is likely to miss the densest parts of the clouds, owing to the direct parallax uncertainty cut (30\%) of the Gaia DR2 data. As a result, some clouds have only small areas above the threshold of $A_\mathrm{G,th} = 1$ mag (e.g., Corona Australis) and some mass at high column densities can be missed. However, as noted by \citet{Zucker2021}, the \citet{Leike2020} data recover the total masses reasonably well; most of the mass is at low (column) densities. We choose data derived using the threshold of $A_\mathrm{G,th} = 0.75$ mag in the examples of this Letter. However, we note that the exact shapes of the probability distributions of individual clouds depend on the chosen threshold (often strongly). Our conclusions regarding KS-relation do not fundamentally change for different thresholds. We demonstrate the effect of the threshold choice further in Appendix \ref{sec:appendix_threshold}.

\section{Results and discussion} 

\subsection{Effect of the viewing angle on areas and masses}

\begin{figure*}
\centering
\includegraphics[width=0.33\textwidth]{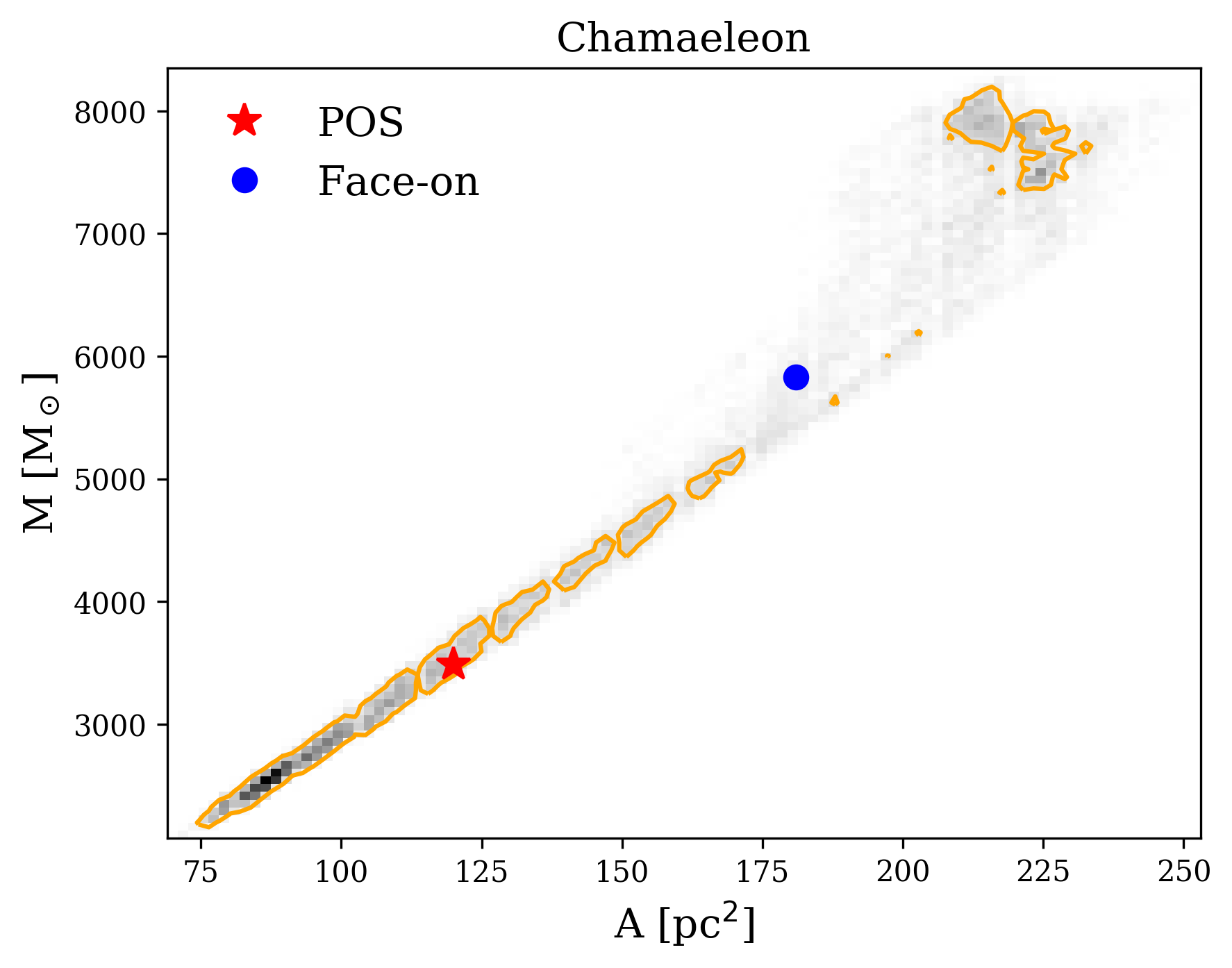}
\includegraphics[width=0.33\textwidth]{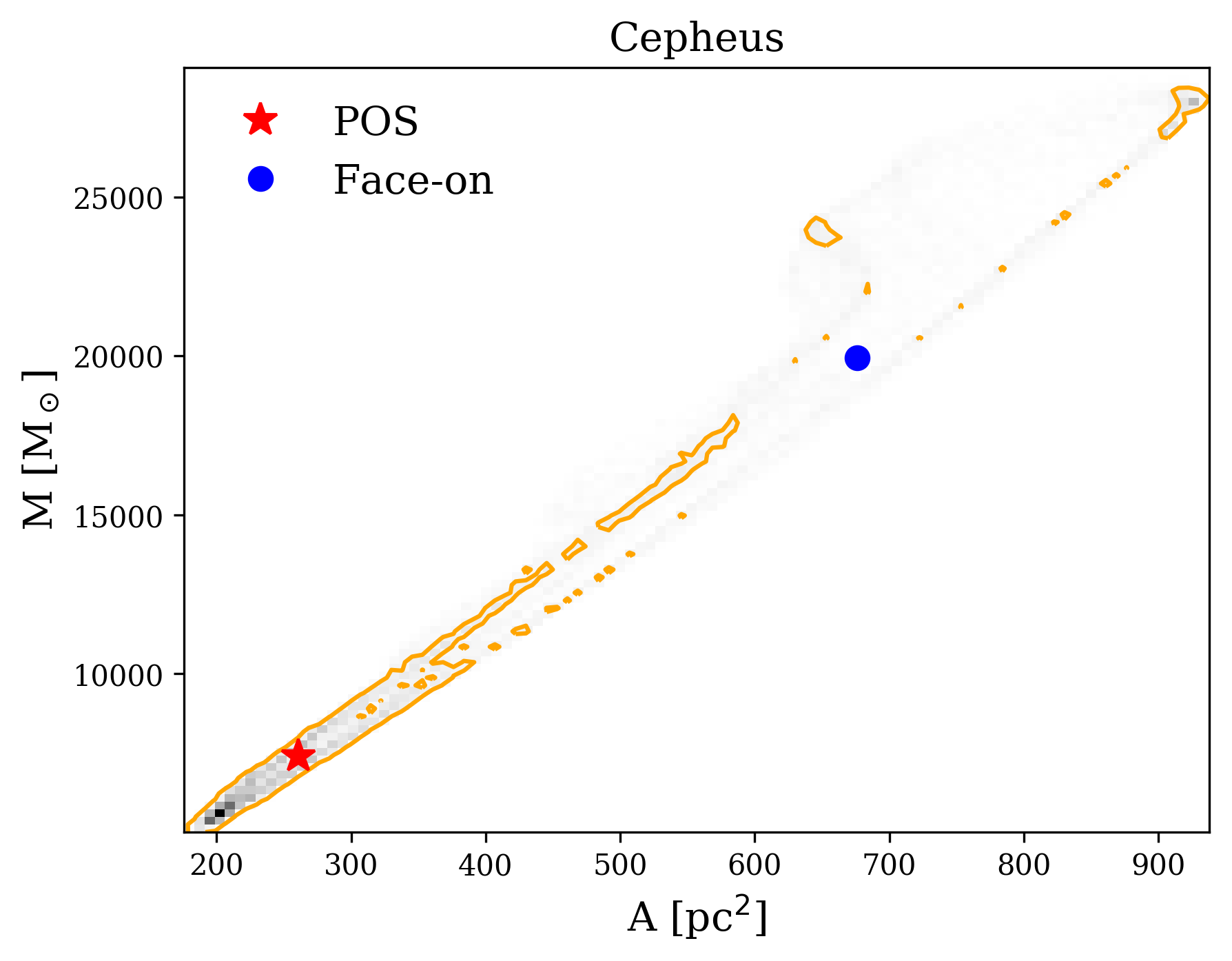}
\includegraphics[width=0.33\textwidth]{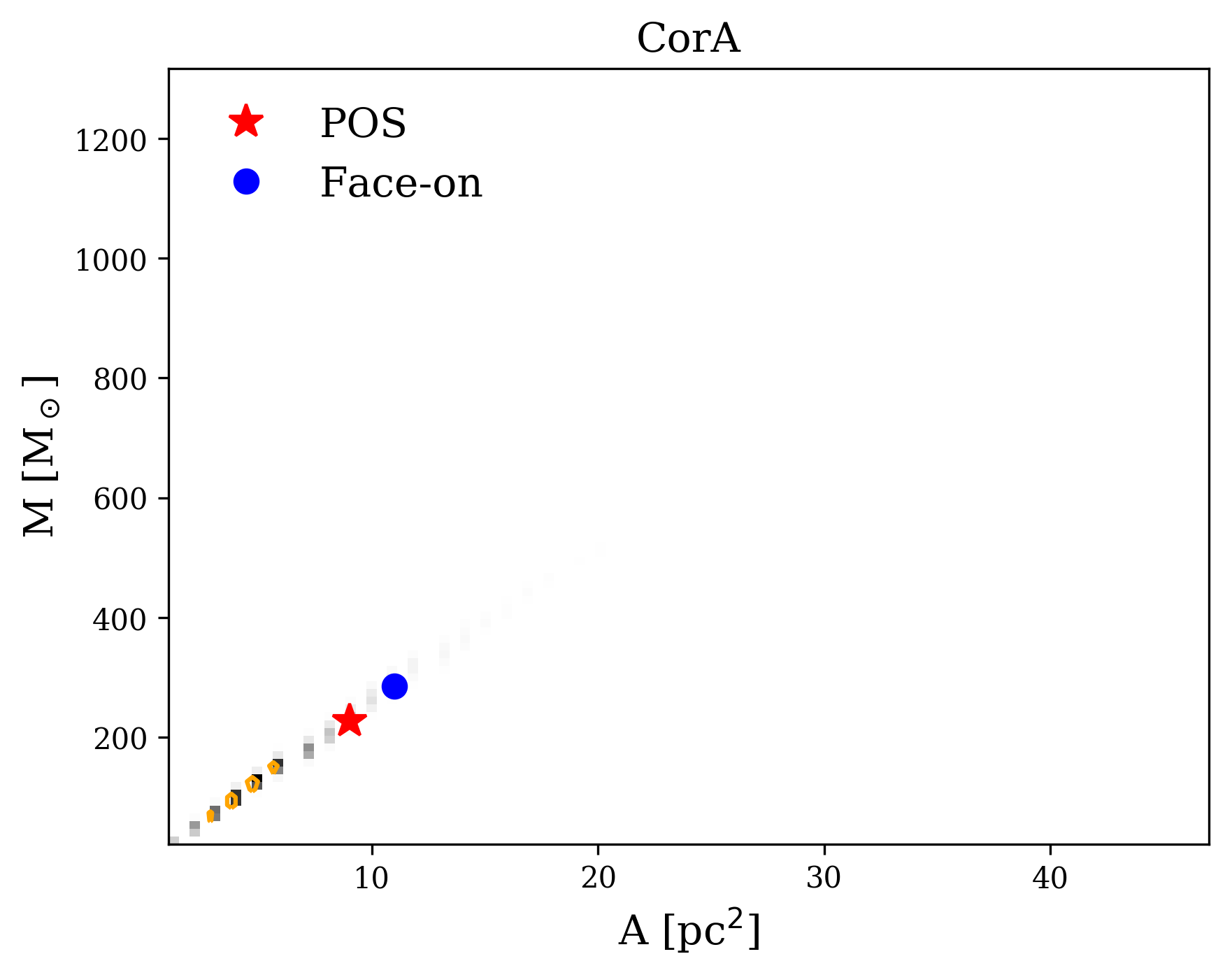}
\includegraphics[width=0.33\textwidth]{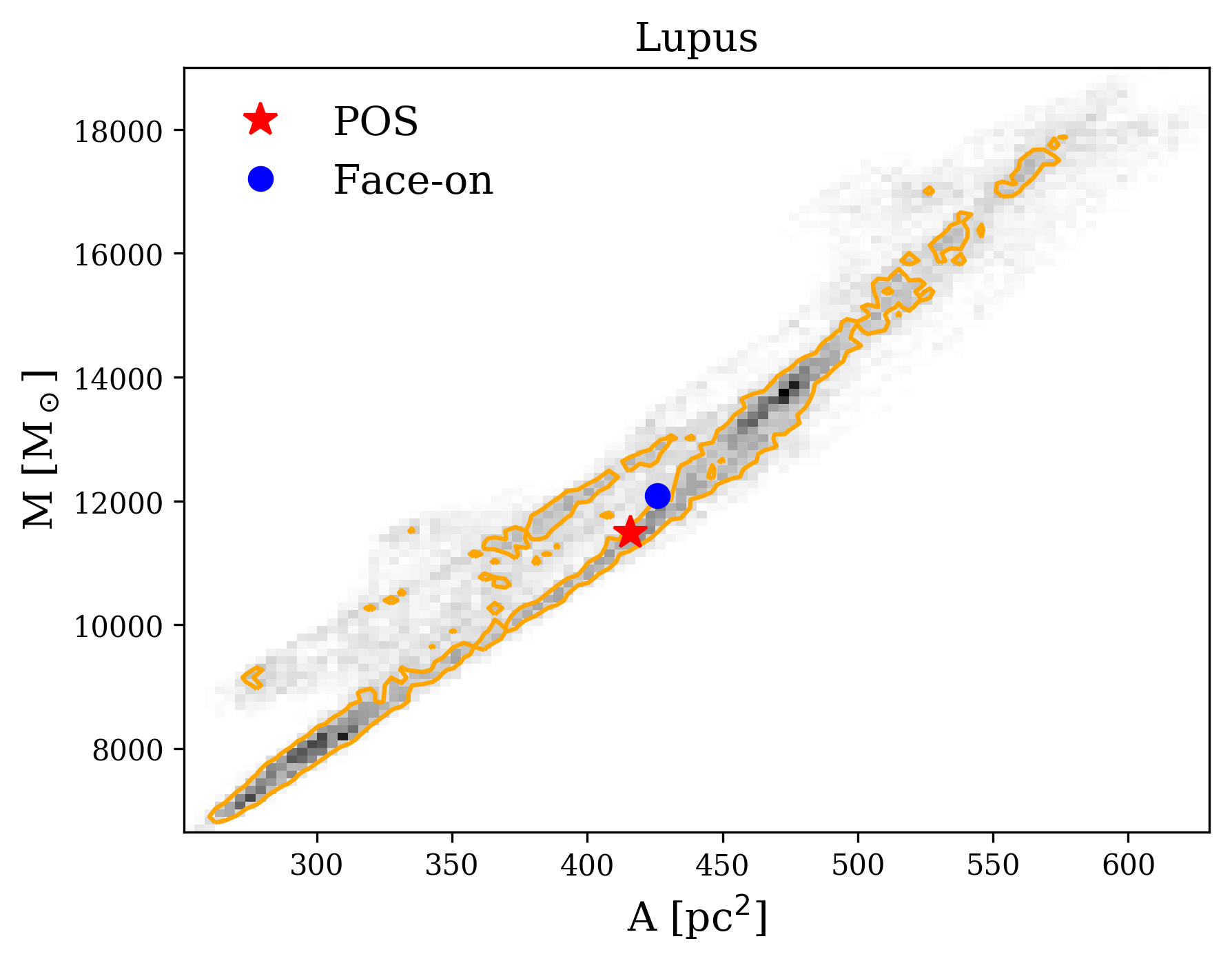}
\includegraphics[width=0.33\textwidth]{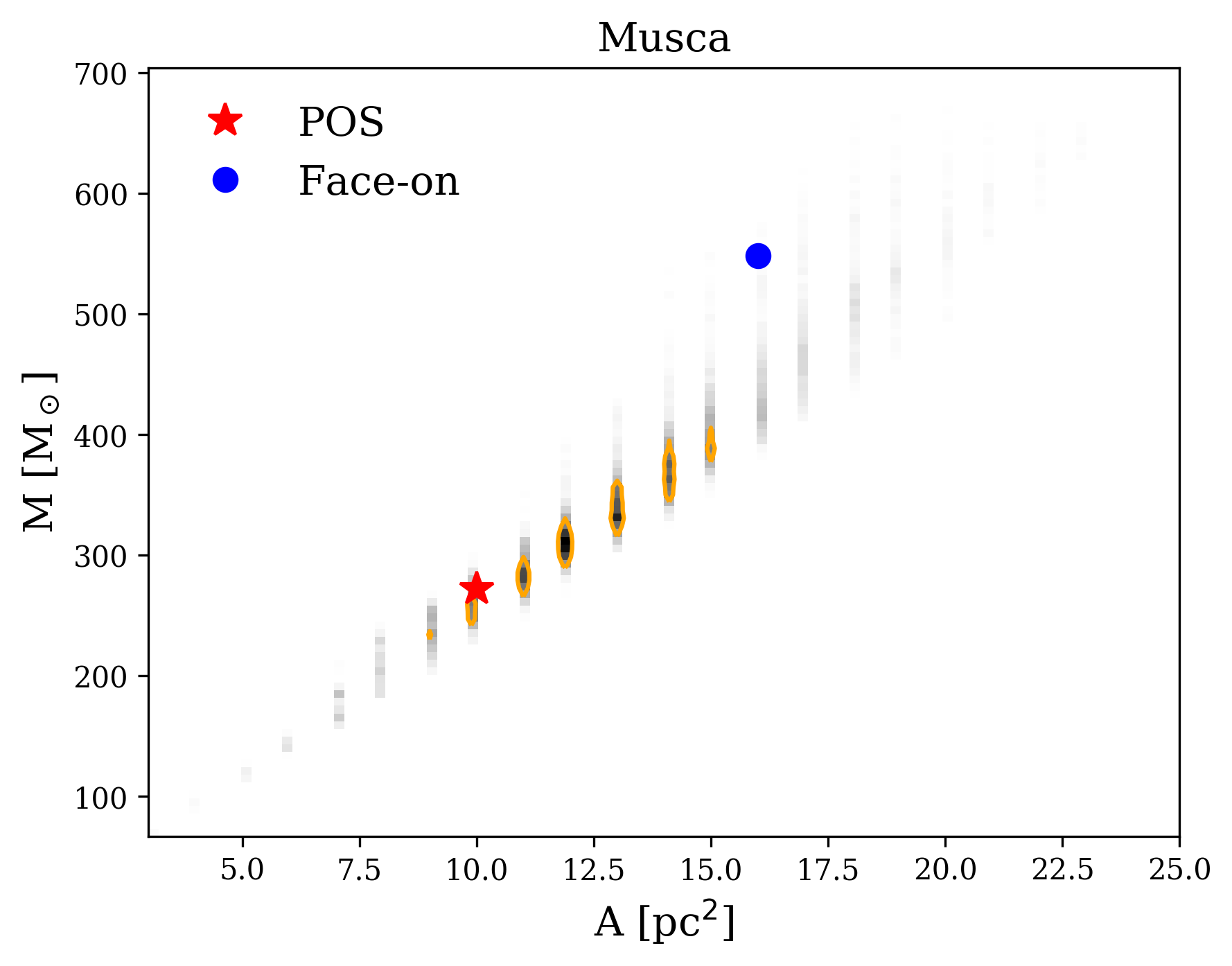}
\includegraphics[width=0.33\textwidth]{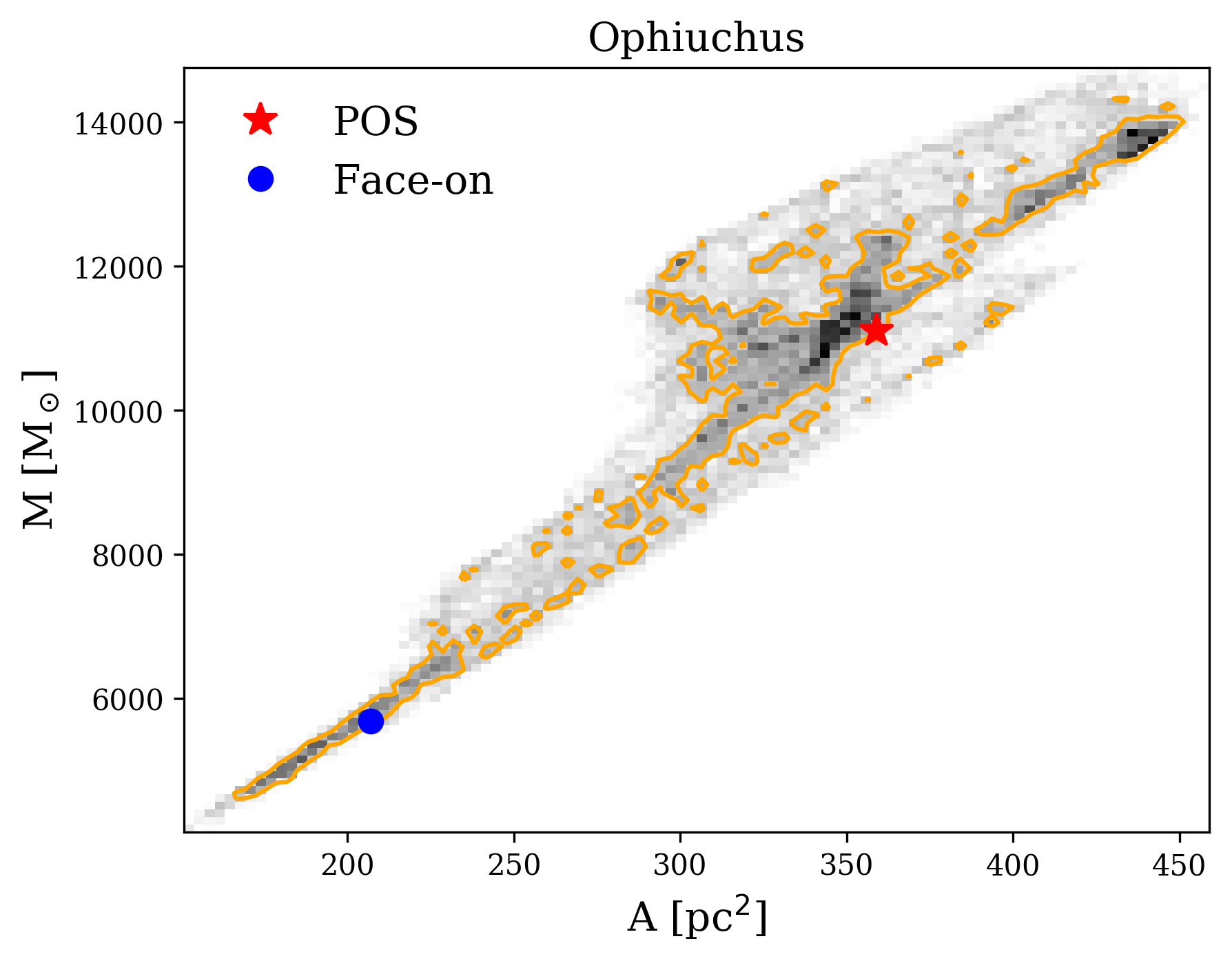}
\includegraphics[width=0.33\textwidth]{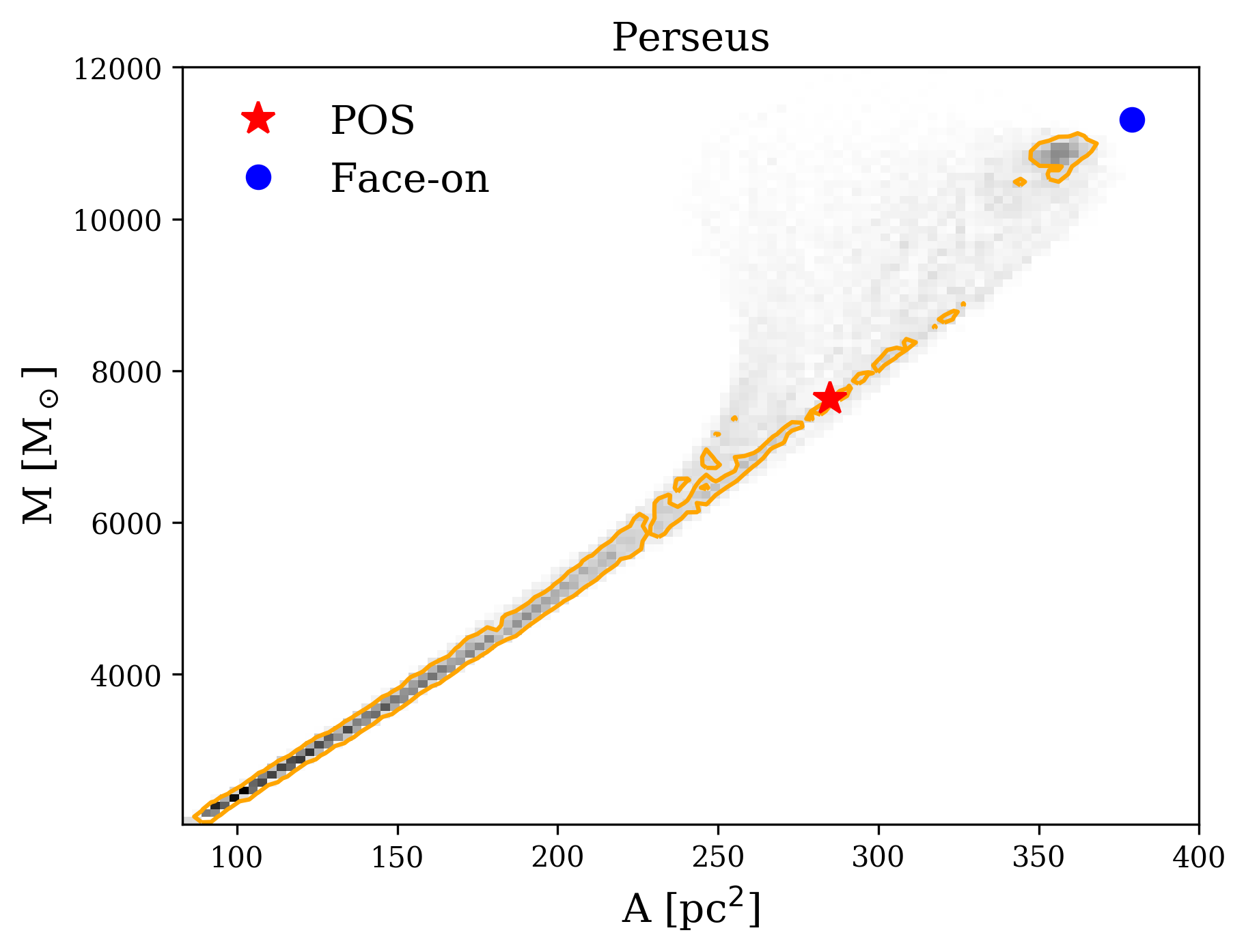}
\includegraphics[width=0.33\textwidth]{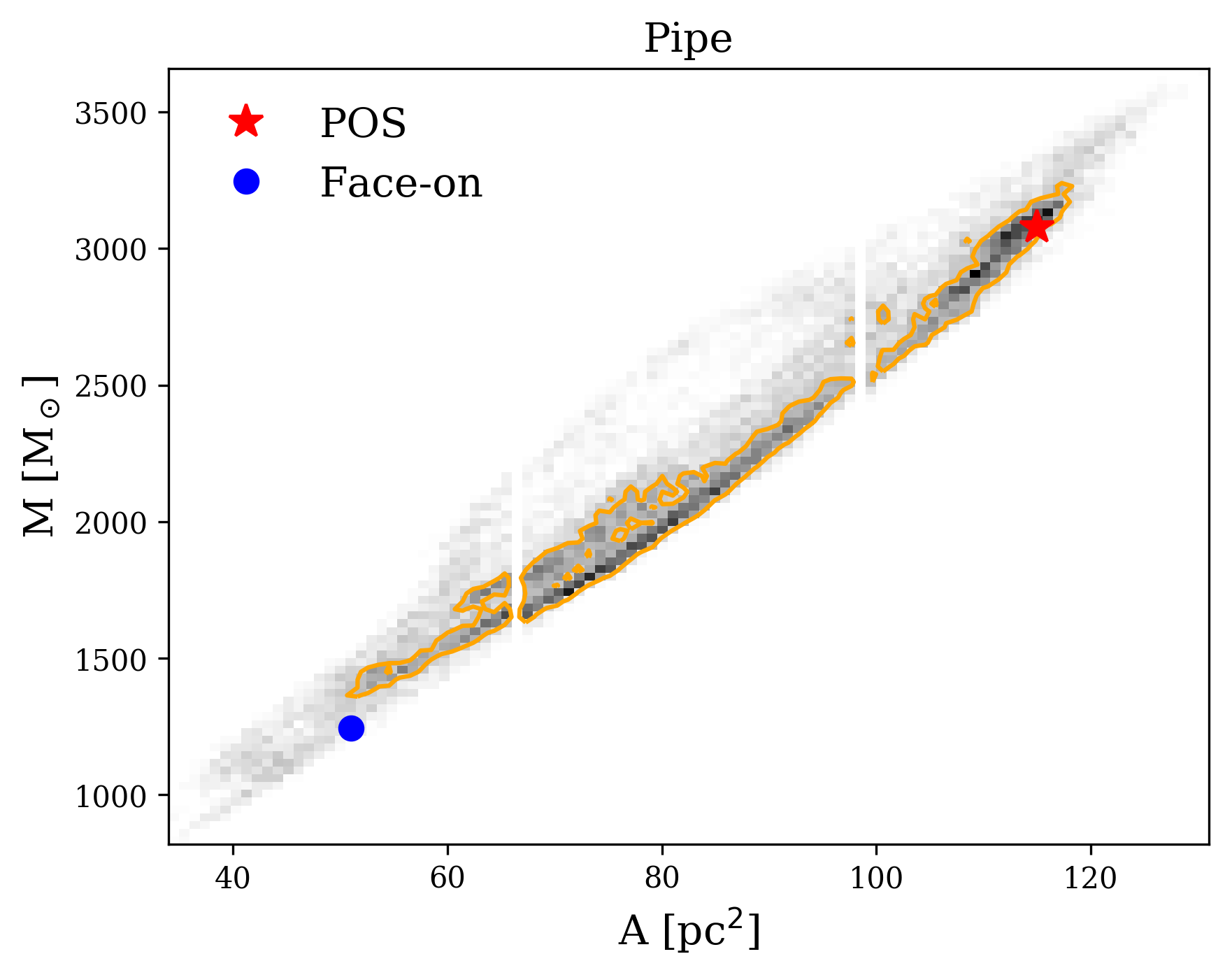}
\includegraphics[width=0.33\textwidth]{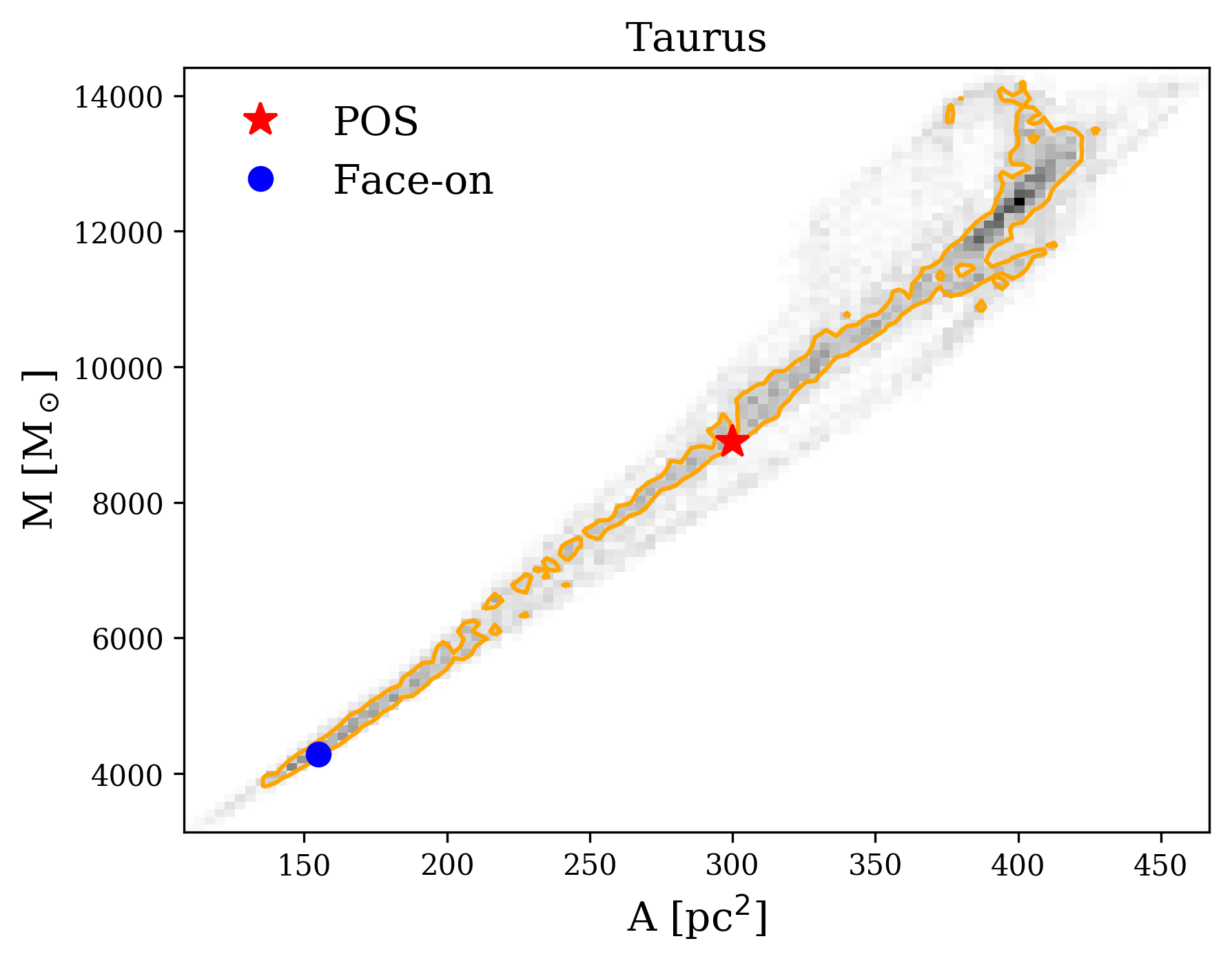}
\caption{Joint probability distributions of the projected mass and area for the nine clouds of our sample (grey-scale). The red star shows the value from the POS angle and the blue circle from the face-on angle. The orange contour delineates the area within which the cumulative probability is 50\% (computed in the order of descending probability from the peak).} 
\label{fig:2dhist}
\end{figure*}

We first examine the joint probability distributions of the projected areas and masses. Figure \ref{fig:2dhist} shows the distributions for our nine clouds. Note that for some clouds the area above the chosen threshold is only a few, or tens of, pixels; the 2D histograms are discretised due to this (Musca, Corona Australis). We also show in Appendix \ref{sec:appendix_histograms} the probability distributions of areas and masses separately. The joint distributions in Fig. \ref{fig:2dhist} give rise to several qualitative results. Both the areas and masses depend significantly on the viewing angle; for most clouds, the distributions span wide ranges in both variables. It is important to emphasise that not only the cloud area changes as a function of the viewing angle, but also the mass. This is a direct effect of the observational practice to define clouds using a threshold column density; the total cloud mass is never recovered, only the mass above the chosen threshold. Depending on the exact morphology of the cloud, this can have a strong effect on the recovered mass.


The joint distributions can have several maxima, or not clearly-defined maxima (e.g., Ophiuchus, Lupus, Pipe). This indicates that the characteristics of the probability distributions are not well captured by only their mean/median and an associated uncertainty. Figure \ref{fig:2dhist} shows contours that enclose 50\% of the cumulative probability, computed by integrating the probability (histogram counts in the 2D histograms of Fig. \ref{fig:2dhist}). The encircled areas are elongated and can have multiple islands. Finally, the distributions of masses and areas are tightly correlated, along an approximately linear relationship. This seems to arise from the relatively narrow dynamic range of the \citet{Leike2020} data. To the first degree, this indicates that the probability distributions of areas and masses are similar. However, some clear deviations can be found (e.g., Ophiuchus, Pipe), especially with the decreasing extinction threshold (see Fig. \ref{fig:2dhist_t0_50}). This means that the clouds become more isotropic when one focuses on higher column densities. Visual inspection of the data indicates that when the threshold increases the structures become more isolated and, qualitatively, spherical. With low thresholds, structures can be elongated, complex, or sheet-like, in which case the viewing angle clearly affects the results more.

Interestingly, the constant $M-A$ relation implies a relatively constant column density above the chosen threshold independently of the viewing angle. This is reminiscent of the Larson's third relation \citep{larson1981}, i.e., the relation between the clouds' mean density and size (which translates to a $M \propto R^2 \propto A$ relationship). The relation has been studied in detail in the past, both for a sample of clouds and individual clouds \citep[e.g.,][]{kegel1989interpretation, Lombardi2010larson, Beaumont2012}. In particular, \citet{Lombardi2010larson} found increasing scatter and flattening in the $M \propto R$ relation for the clouds individually. Our viewing angle analysis points out a new insight into this: at large scales (i.e., lower thresholds) the $M \propto A$ relation seems to break down (cf., Fig. \ref{fig:2dhist_t0_50}), as it becomes increasingly dependent on the viewing angle. The increasing anisotropy at larger scales seems then the likely reason for the break-up of the Larson's third law when considering individual clouds.

We next quantify the ranges of the mass and area probability distributions. The ranges are useful as maximum uncertainties for any applications that consider the areas and masses of molecular clouds. We give in Table \ref{tab:parameters} the full ranges of the distributions and also the ranges relative to the median values. To describe where most of the probability is, we give the 50\% quartiles (Q50) of the distributions. In most cases, the ranges of the distributions vary strongly but are typically 100-200\% of the median value for both areas and masses; the Q50 ranges are about 20-80\% of the median value. These ranges are much larger than the variance due to the inherent uncertainty of the \citet{Leike2020} data, which is about 10\% (Appendix \ref{sec:appendix_allsamples}). Importantly, the large ranges of the parameters show that in case of individual clouds, the viewing angle can have a dramatic effect on both the area and mass of the cloud. 


Finally, we consider two special viewing angles: the plane-of-the-sky (POS) and face-on angles. The former refers to the angle as observed on the sky and the latter as the cloud would be viewed from the direction perpendicular to the Galactic disk. The areas and masses viewed from these two angles are commonly offset from the maxima of the probability distributions (e.g., Chamaeleon, Lupus, Taurus, see Fig. \ref{fig:2dhist}). This is expected, as the distributions are not strongly peaked (see Figs. \ref{fig:areahistograms}); the probability of observing an area and mass significantly offset from the peak or median value is large. We cannot find any significant systematic differences between the areas and masses when viewed from the POS and face-on angles (Appendix \ref{sec:appendix_POS-vs-faceon}). The ratios of the areas and masses from the two angles are on average consistent with unity, albeit with a large scatter for individual clouds. The scatter depends on the chosen extinction threshold, becoming smaller with increasing threshold. This is a consequence of the limited dynamic range of the data: when the threshold is low, the mass becomes more and more tightly correlated with the total volume (a result analogous to that derived for POS-angle column densities by \citealt{Lombardi2010larson}). While our small sample does not enable a robust description of systematic effects, it suggests that the effects are minor for clouds in the Solar neighbourhood. This is encouraging for works that compare Galactic and extra-galactic cloud properties and require assumptions about the effects of the viewing angle.


\subsection{Effect of the viewing angle on the KS-relation}

We then consider the effect of viewing angle on the KS-relation, i.e., the relation between the gas surface density ($\Sigma_\mathrm{gas} = M/A$) and star formation rate (SFR) surface density ($\Sigma_\mathrm{SFR} =$SFR/$A$). To construct the KS-relation, we use our mass and area data from different viewing angles and the SFRs derived for the clouds by \citet{lada2010}. The SFRs were derived from the number of young stellar objects (YSOs) in the clouds, multiplied by the mean mass of a star (0.5 M$_\odot$) and divided by the lifetime of the Class II YSOs (2 Myr). For Musca that was not included in the \citet{lada2010} sample, we adopt a reference value of 0.25 M$_\odot$ Myr$^{-1}$ (1 YSO), reflecting its known low star formation activity \citep{Vilas-Boas1994musca,Juvela2010}. We conjecture in this analysis that the observed SFR of the cloud does not depend on the viewing angle. The SFRs of \citet{lada2010} are based on the total number of YSOs in the vicinity of the cloud; there is no column density selection involved in the selection. The reservoir of YSOs is thus unaffected by the main selection effect related to areas and masses. 

\begin{figure}
\centering
\includegraphics[width=\columnwidth]{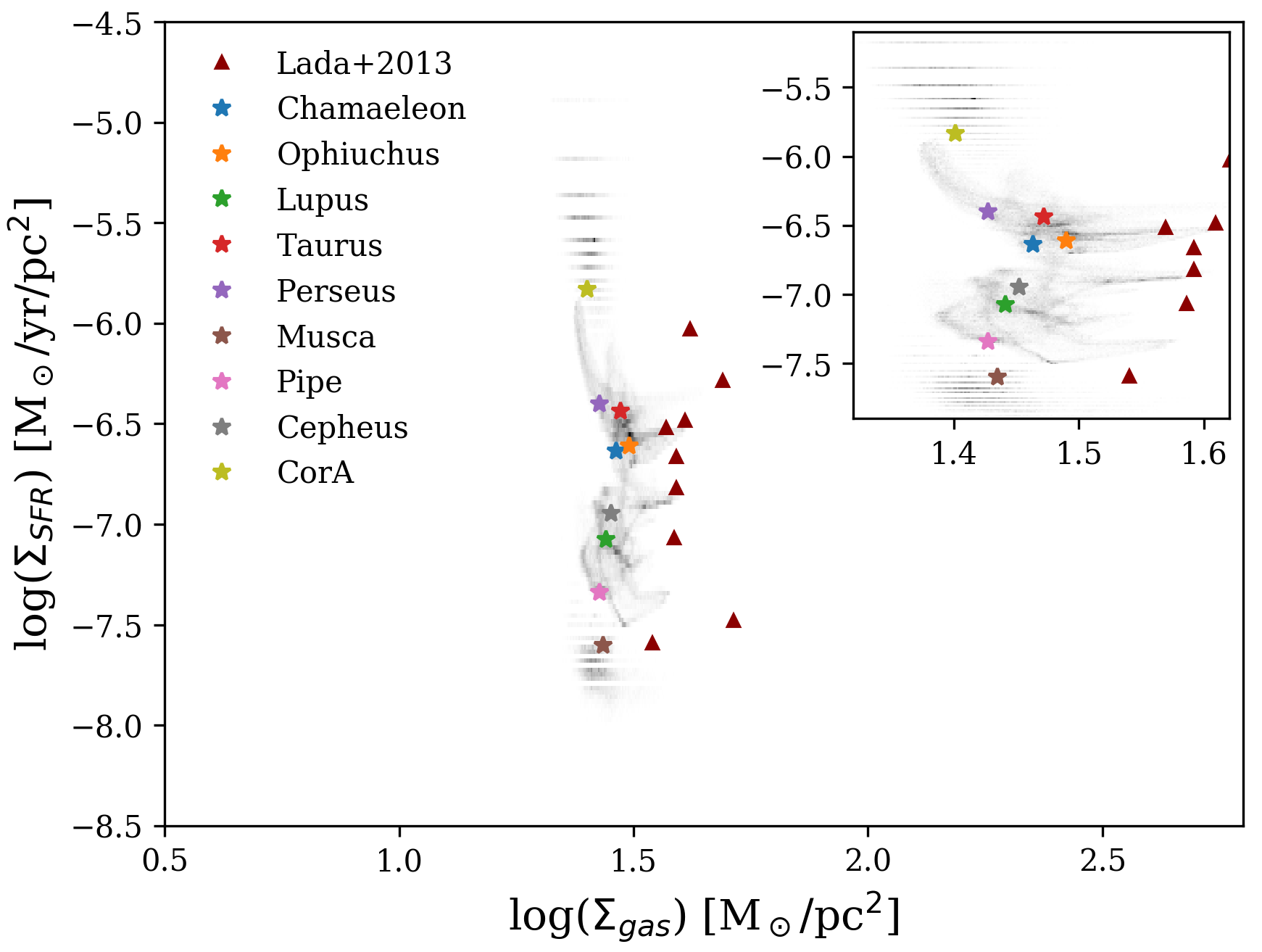}
\caption{Probability distributions of the sample clouds from various viewing angles in the KS-plane (grey-scale). The observed (POS) values of the clouds are shown with stars. The SFR data is from \citet{lada2010}. The dark red triangles show data from \citet{Lada2013schmidt}. The inset shows a zoom-in to highlight the shape of the distributions.}
\label{fig:KS}
\end{figure}

We present the resulting probability distribution of the clouds in the KS-plane in Fig. \ref{fig:KS}. The joint probability distributions of the cloud masses and areas translate into curved relationships in the KS-plane. These relationships, however, span only a relatively small range in surface density. This is expected from the behaviour of the joint distributions (Fig. \ref{fig:2dhist}); a close to linear $M$ vs. $A$ relationship implies close to constant column density. However, it is important to notice that the joint probability of $A^{-1}$ vs $M/A$ is not simple in the KS-plane (see the inset of Fig. \ref{fig:KS}), especially for lower thresholds. This highlights that the exact effect of the viewing angle on the KS-relation is not trivial to account for (the uncertainty is not gaussian). Regardless, the ranges provided in Table \ref{tab:parameters} give means to put limits on the maximum viewing angle effect. Finally, we find a slight change ($\sim$0.2 dex) in the mean surface density of the clouds as a function of the adopted threshold; this effect is similar to that found out by \citet{Lombardi2010larson} for POS-angle data. 

For a qualitative comparison, we also show in Fig. \ref{fig:KS} the data points for clouds from \citet{Lada2013schmidt}. The \citet{Lada2013schmidt} data are offset from our data by a factor of 1.3-1.4. Several reasons can contribute to this difference, e.g., the different column density ranges probed by the optical Gaia and near-infrared extinction data, different zero-point calibrations of the data, and uncertainty in the extinction law. While we consider that such differences primarily affect the scaling of the masses, and thus are not important for our conclusions, they should be kept in mind when comparing in detail data derived using different observational techniques.

It is necessary to note that the clouds in our sample are relatively small. Unfortunately, the \citet{Leike2020} data becomes unreliable above 400 pc distance, excluding the most nearby massive clouds like Orion and the California Cloud \citep[as established by][]{Zucker2021}. The recent work of \citet{RezaeiKh2021Kainulainen} shows that the projected area of California changes by more than an order of magnitude between the POS and face-on angles, owing to its extended sheet-like structure. It remains unclear how the joint distribution of mass and area of the more massive, potentially more complex clouds like California looks like. Similarly, it is necessary to consider if different 3D mapping techniques result in differing 3D structures of the clouds; for this reason, we decided not to directly compare the \citet{RezaeiKh2021Kainulainen} data to this work that uses the \citet{Leike2020} data; the data are based on different 3D mapping techniques. Clearly, further work on the 3D structures of clouds is needed and can be highly beneficial to further understand the observed KS-relations. 


\begin{table*}
\centering
\caption{The sample clouds and their properties.}
\begin{tabular}{lcccccccc}
\hline\hline
Cloud 		&   Range $A$   &   Range $M$   & $\Delta A / \tilde{A}$   &    $Q50(A)/ \tilde{A}$   &   $\Delta M / \tilde{M}$ & $Q50M / \tilde{M}$ &  $A_\mathrm{POS}$ &  $M_\mathrm{POS}$  	\\
			&	pc$^2$	    &  M$_\odot$    & 		\%	                    &            \%                   &          \%                  &               \%        &         pc$^2$          &         M$_\odot$              \\
\hline
Chamaeleon	&   [69, 253]   & [2\,100, 8\,300]  &	108	&   57		&	120		&	75  & 120   & 3\,500    \\ 
Cepheus		&   [176, 938]  & [5\,100, 29\,100] & 	164	&	83		&	177 	&	99  & 261   & 7\,400  	\\ 
CrA Cloud	&   [1, 47]     & [21, 1\,300]      &	920	&	60		&	994 	&	65  & 9     & 220  	    \\ 
Lupus		&   [251, 630]  & [6\,600, 19\,000] &	89	&	35		&	101		&	37  & 416   & 11\,400 	\\ 
Musca		&   [3, 25]     & [66, 700]         &	117	&	41		&	131	    &   46	& 10    & 270       \\ 
Ophiuchus	&   [151, 459]  & [4\,200, 14\,800] &	93	&	22		&	 96		&	29  & 359   & 11\,100     \\
Perseus		&   [83, 377]   & [2\,100, 12\,000]  &	115	&   53		&	141	    &	73  & 285   & 7\,800     \\
Pipe	    &   [34, 131]   & [820, 3\,600]     &	117	&	41		&	131	    &	45  & 115   & 3\,100  	\\
Taurus		&   [108, 467]  & [3\,100, 14\,300] &	104	&	34		&	106	    &	40  & 300   & 8\,900  	\\	
\hline
\end{tabular}
\tablefoot{The columns from left to right: (1) cloud name; (2, 3) the ranges of the areas and masses from all analysed viewing angles; (4, 5) the range and 50\% quartile, Q50, of the cloud areas divided by the median area, $\tilde{A}$; (6, 7) same as 4 and 5, but for cloud masses; (8, 9) the area and mass in the POS viewing angle.
}
\label{tab:parameters}
\end{table*}


 \section{Conclusions}   

In this paper, we analysed the masses and areas of a sample of nine nearby molecular clouds from different viewing angles exploiting Gaia-based, 3D dust distribution data from literature \citep{Leike2020}. Our conclusions are as follows.

\begin{enumerate}

\item The probability distributions of the projected areas and masses of the clouds are usually asymmetric, relatively flat, and they can be multi-peaked. The ranges and 50\% quartiles of the probability distributions are roughly 100-200\% and 20-80\%, respectively, with a large scatter for individual clouds. These properties imply that the clouds are not isotropic and that the viewing angle can have a strong effect on the areas and masses of individual clouds.  

\item The column density threshold used to define the clouds has a significant effect on the exact shape of the mass and area probability distributions. The distributions become more complex with our lowest threshold, $A_\mathrm{G} = 0.5$ mag. This suggests the clouds are more anisotropic at large scales (smaller thresholds) than at small scales (larger thresholds). The threshold effects caution against comparisons of cloud property catalogues derived using different parameters.  

\item The effect of viewing angle to the location of clouds in the KS-plane is asymmetric, but relatively minor due to the correlation of the mass and area probability distributions and the effect of defining the clouds using a threshold value. This means that the same clouds, viewed from any angle, would not show a KS-relation. This further suggests that the result of \citet{Lada2013schmidt} is not affected by the angle in which we view the clouds, despite some observational differences between this work and \citet{Lada2013schmidt}.

\item We find that the masses and areas measured from the POS and face-on angles are on average in agreement, even if the scatter is high. This suggests that the viewing angle of clouds is not a crucial factor determining average properties of cloud samples. This is relevant for, e.g., comparisons of extragalactic and galactic cloud samples that typically represent (close-to) face-on and POS angles, respectively.   

\end{enumerate}

Our results are based on a small sample of nine clouds and thus suggestive; while they provide important first insight, they also highlight the importance of broader studies on the impact of clouds' 3D morphology on the canonical scaling relations.

\begin{acknowledgements}
J.O. acknowledges funding from the Swedish Research Council, grant No. 2017-03864. The authors thank the referee, Catherine Zucker, for the constructive review that significantly improved the paper. 
\end{acknowledgements}

\bibliographystyle{aa} 
\bibliography{ref}

\pagebreak

\appendix

\section{Additional figures}
\label{sec:appendix_additionalfigures}

\subsection{Area, mass, and surface density histograms}
\label{sec:appendix_histograms}

\begin{figure*}
\centering
\includegraphics[width=0.32\textwidth]{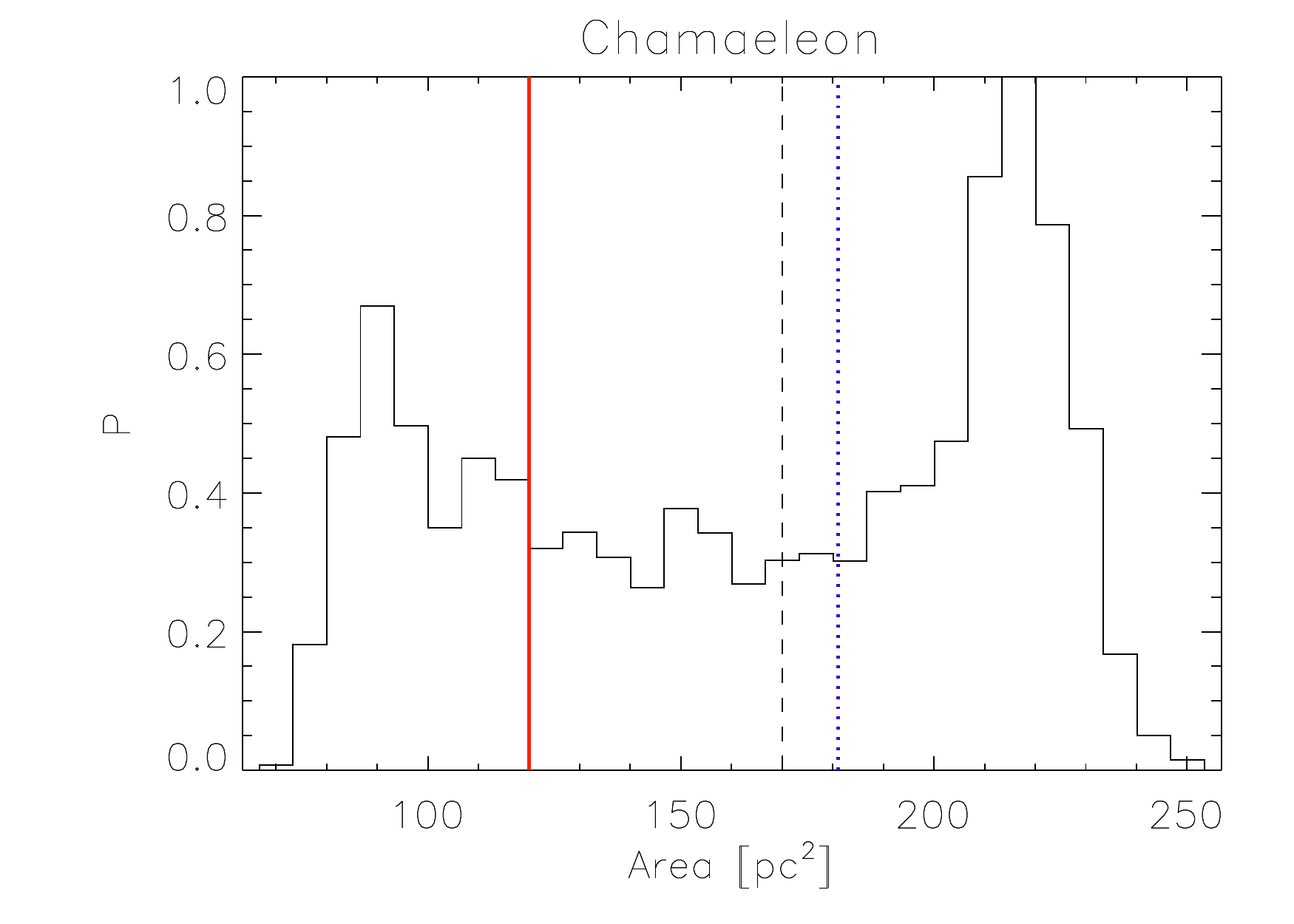}\includegraphics[width=0.32\textwidth]{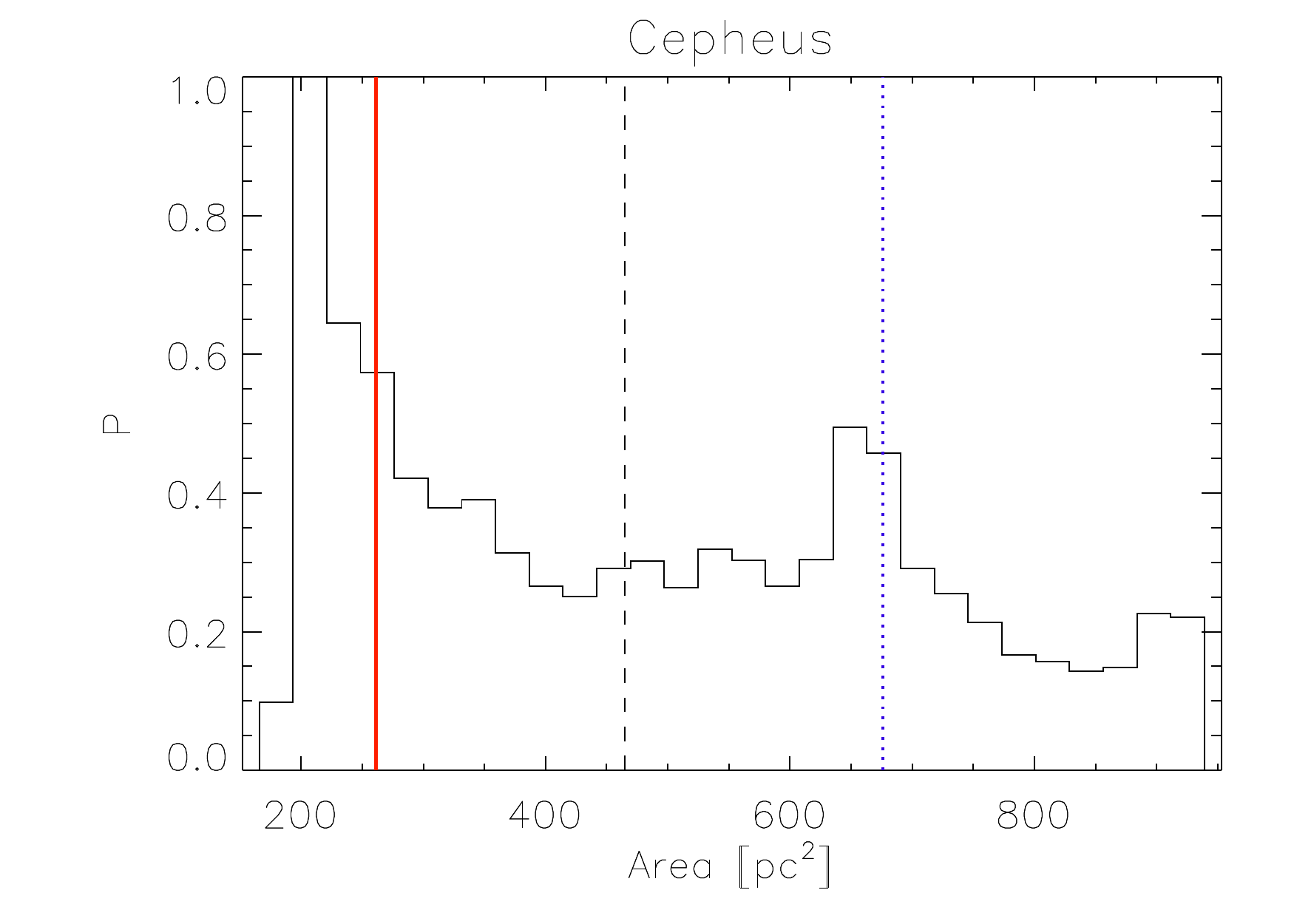}\includegraphics[width=0.32\textwidth]{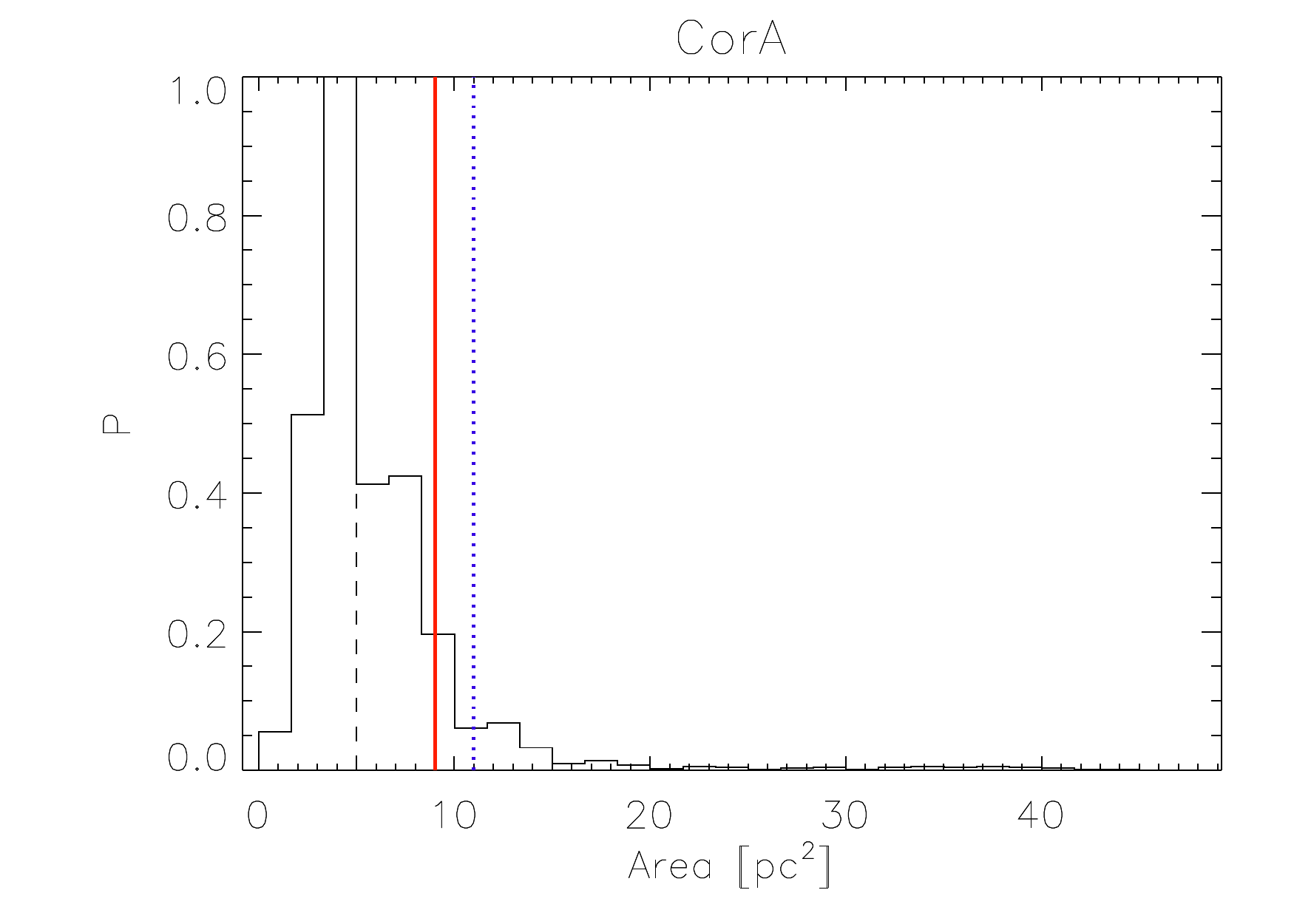}
\includegraphics[width=0.32\textwidth]{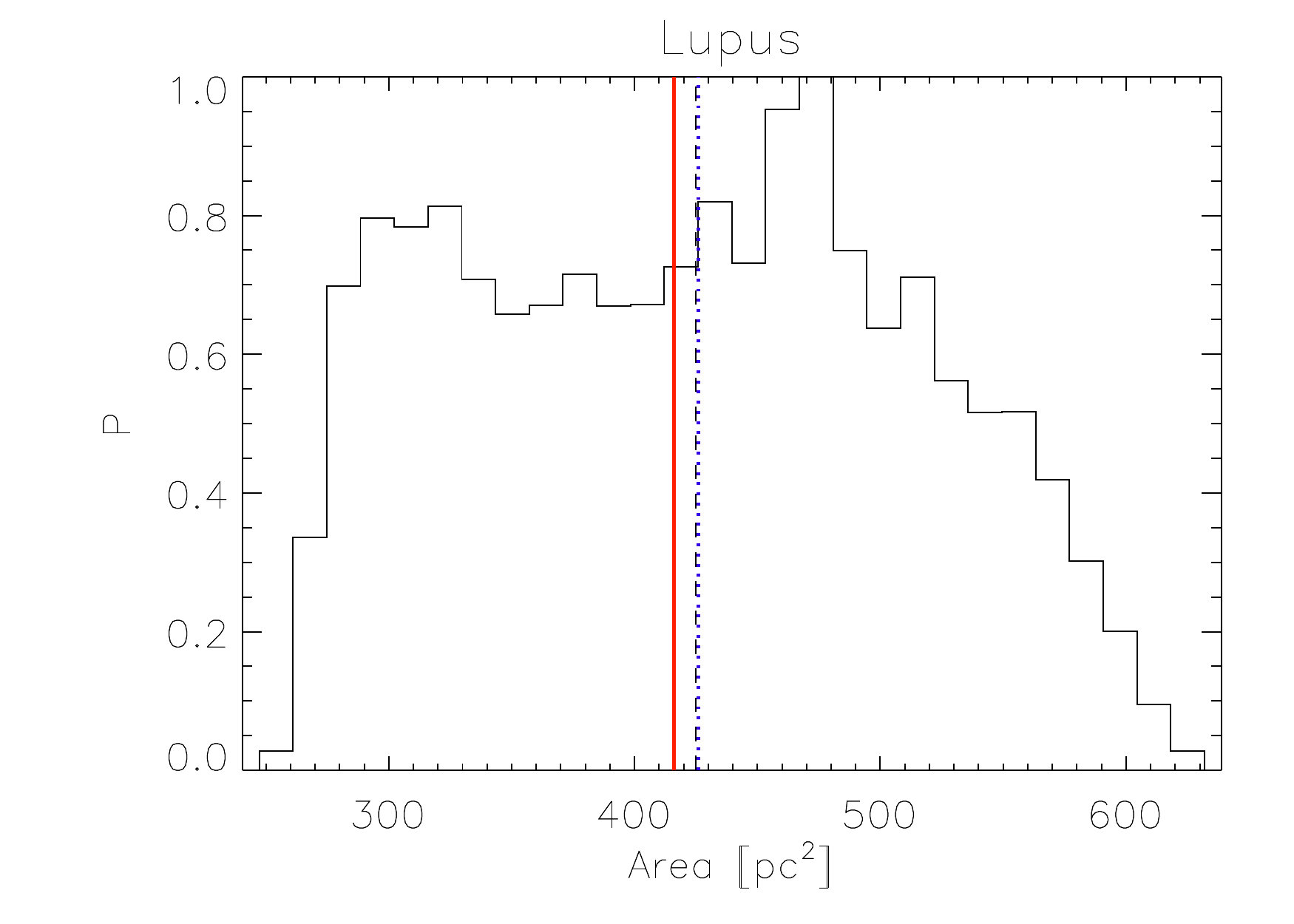}\includegraphics[width=0.32\textwidth]{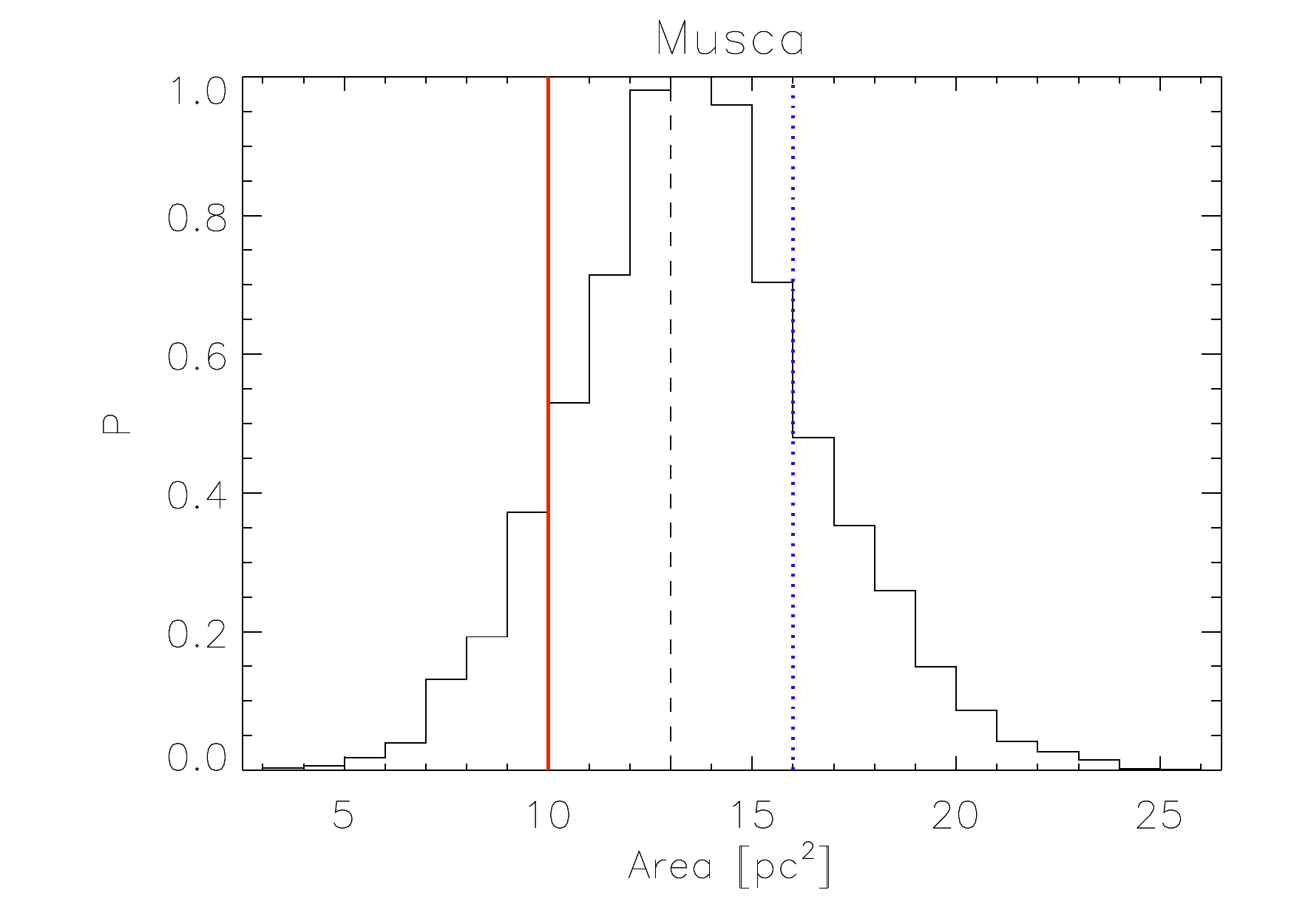}\includegraphics[width=0.32\textwidth]{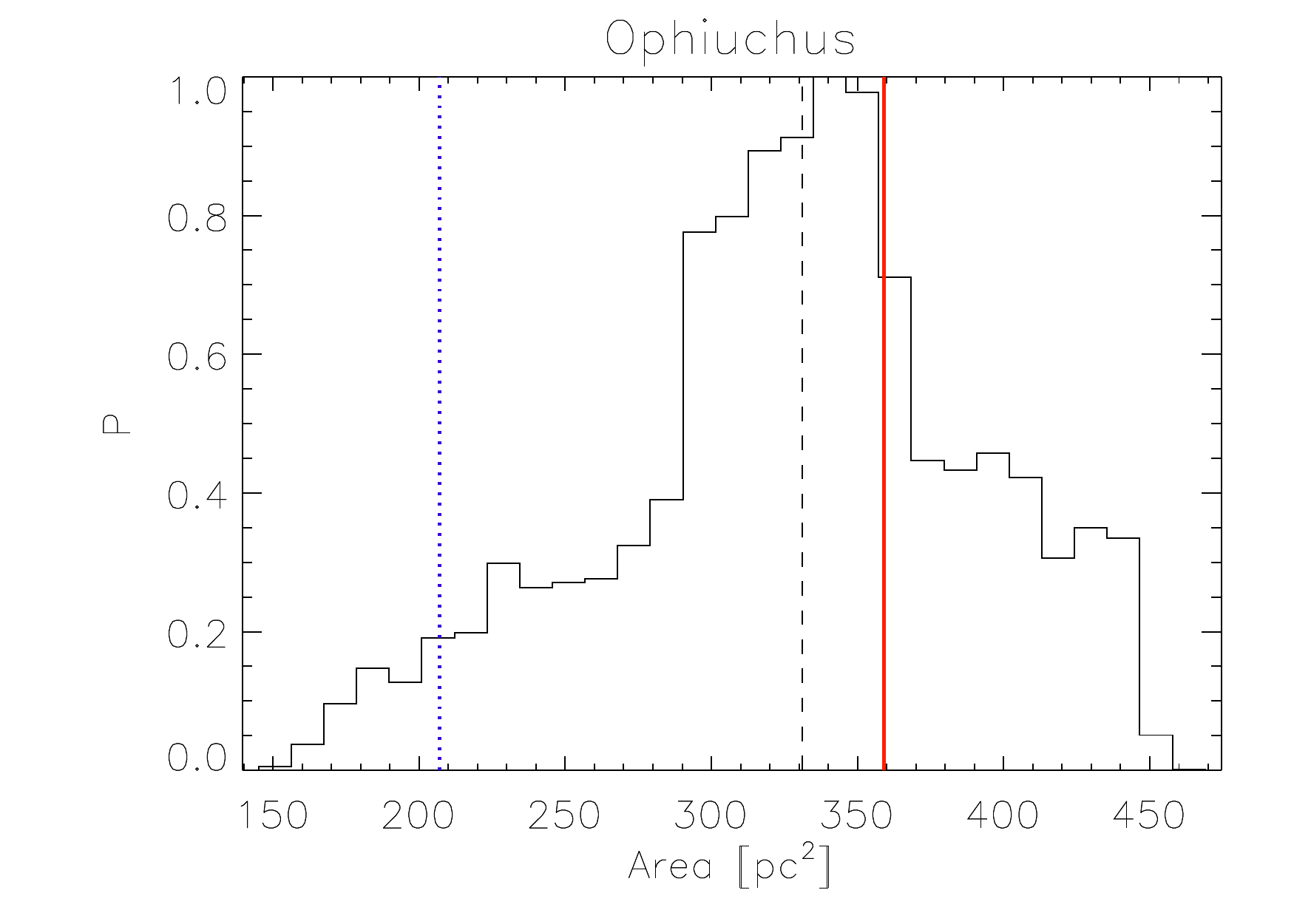}
\includegraphics[width=0.32\textwidth]{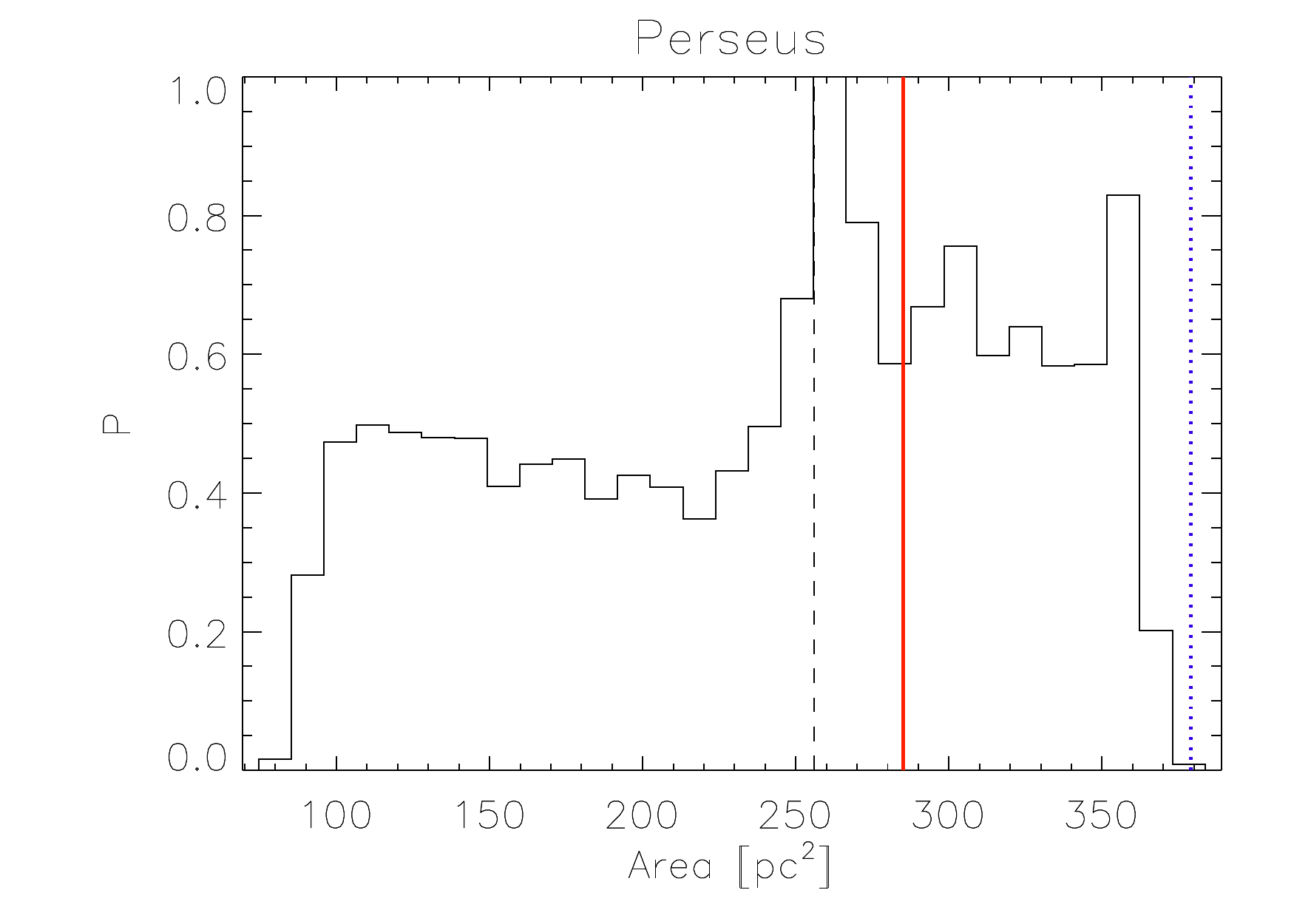}\includegraphics[width=0.32\textwidth]{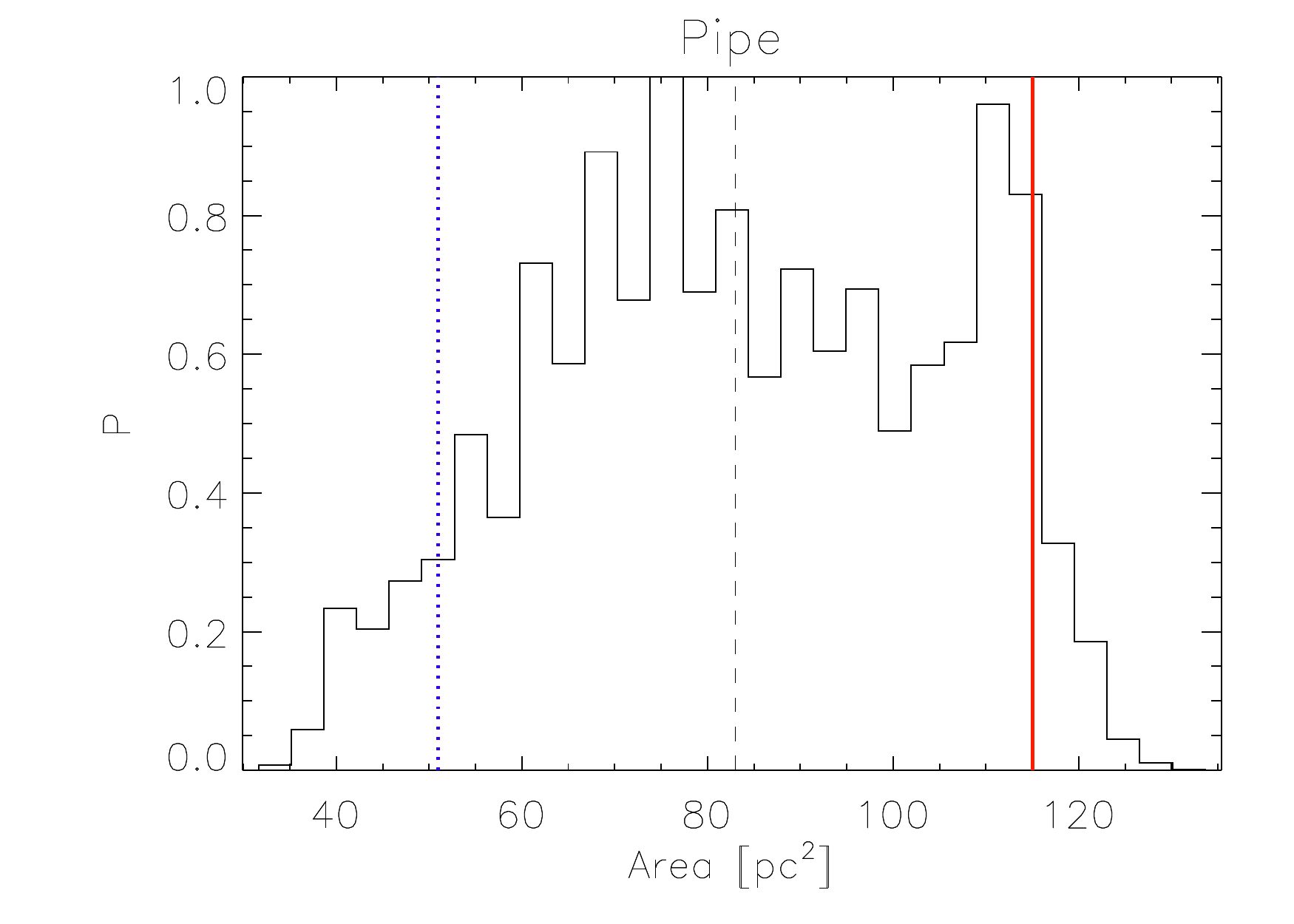}\includegraphics[width=0.32\textwidth]{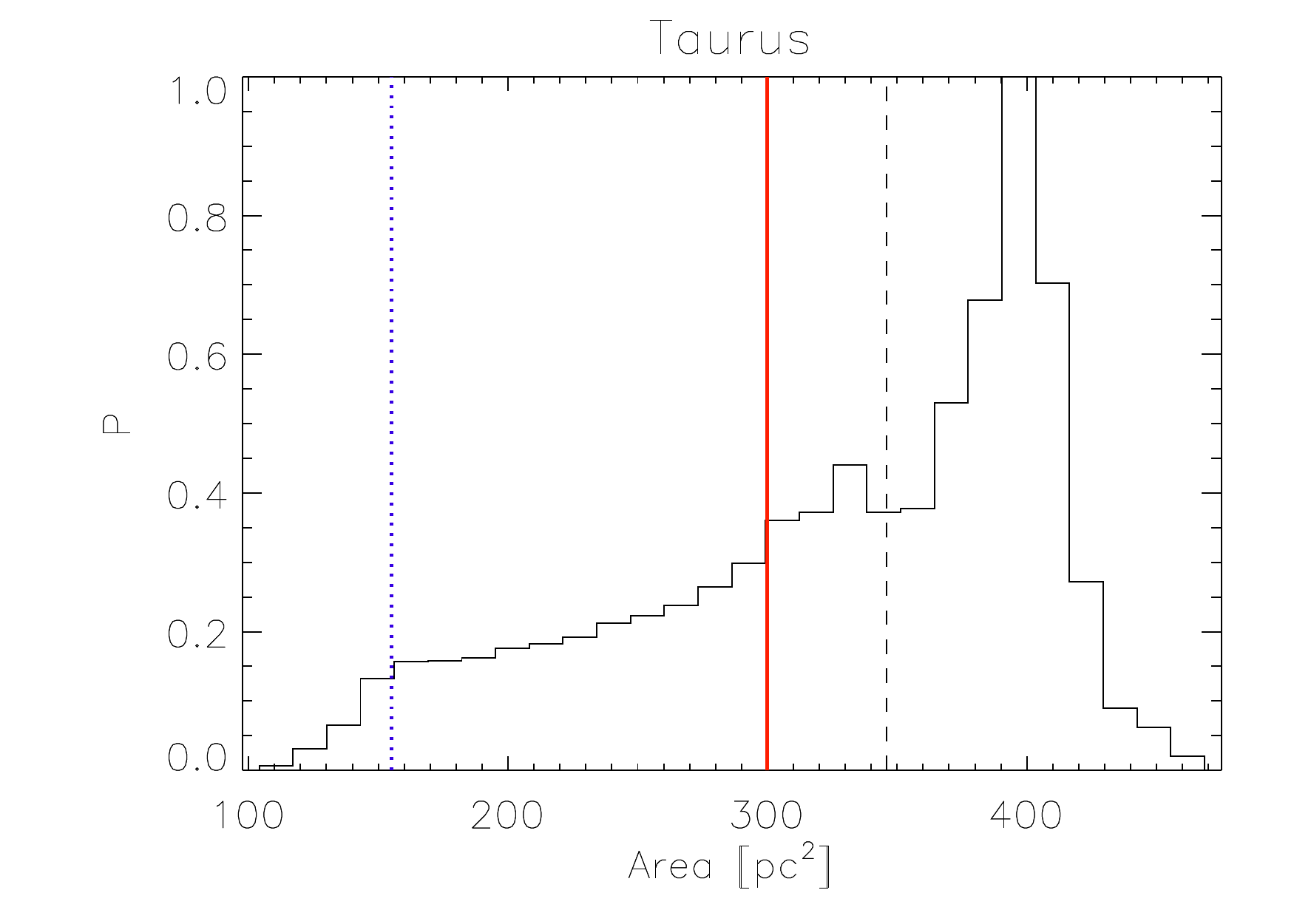}
\caption{Frequency distributions of cloud areas for the molecular clouds in our sample, normalized to the peak. The data are derived using the threshold of $A_\mathrm{G} = 0.75$ mag. The dashed vertical line indicates the median value. The red line shows the area from the plane-of-the-sky perspective (i.e., the observed area) and the blue dotted line from the face-on perspective (perpendicular to the Galactic disk). }
\label{fig:areahistograms}
\end{figure*}

\begin{figure*}
\centering
\includegraphics[width=0.32\textwidth]{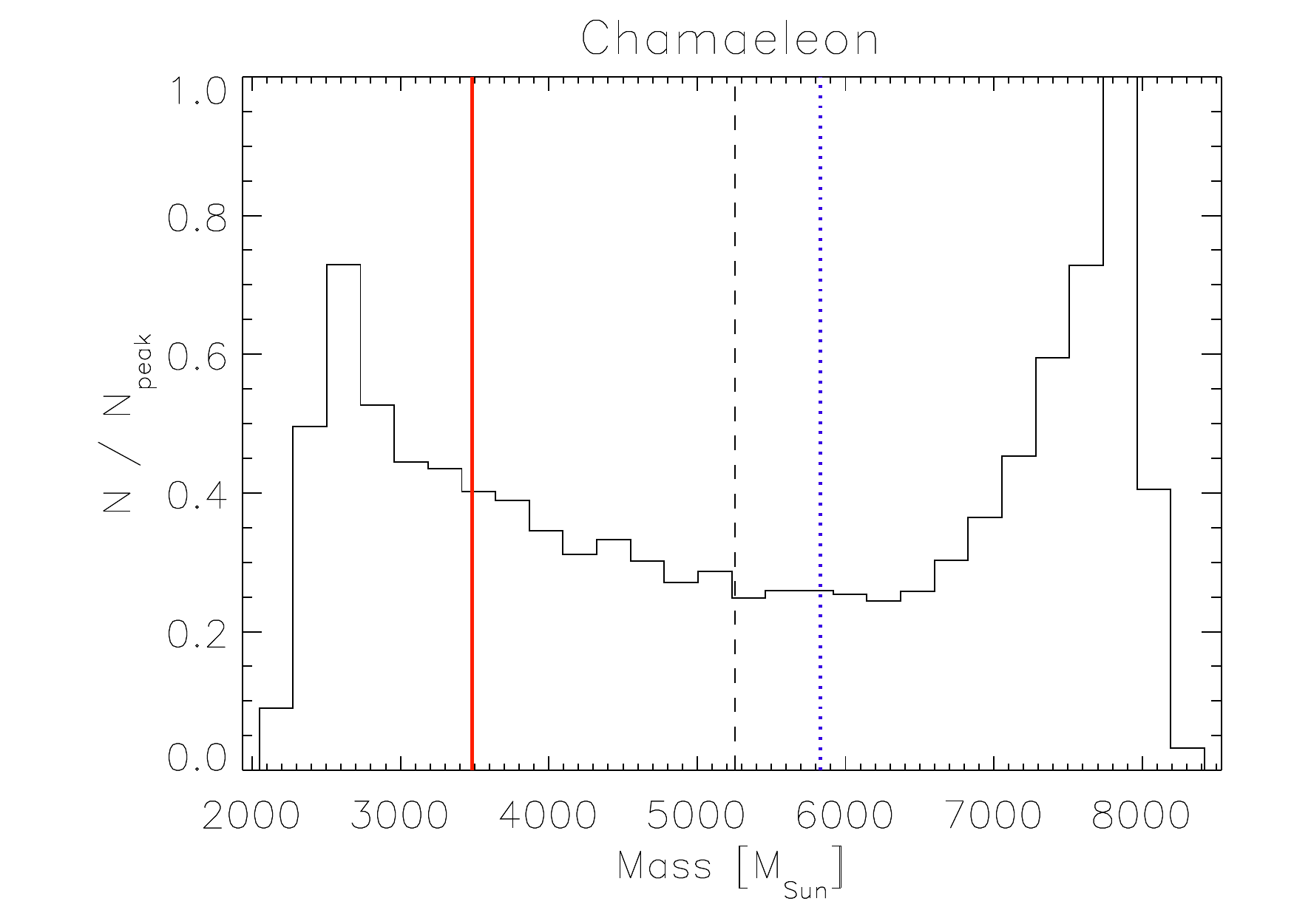}\includegraphics[width=0.32\textwidth]{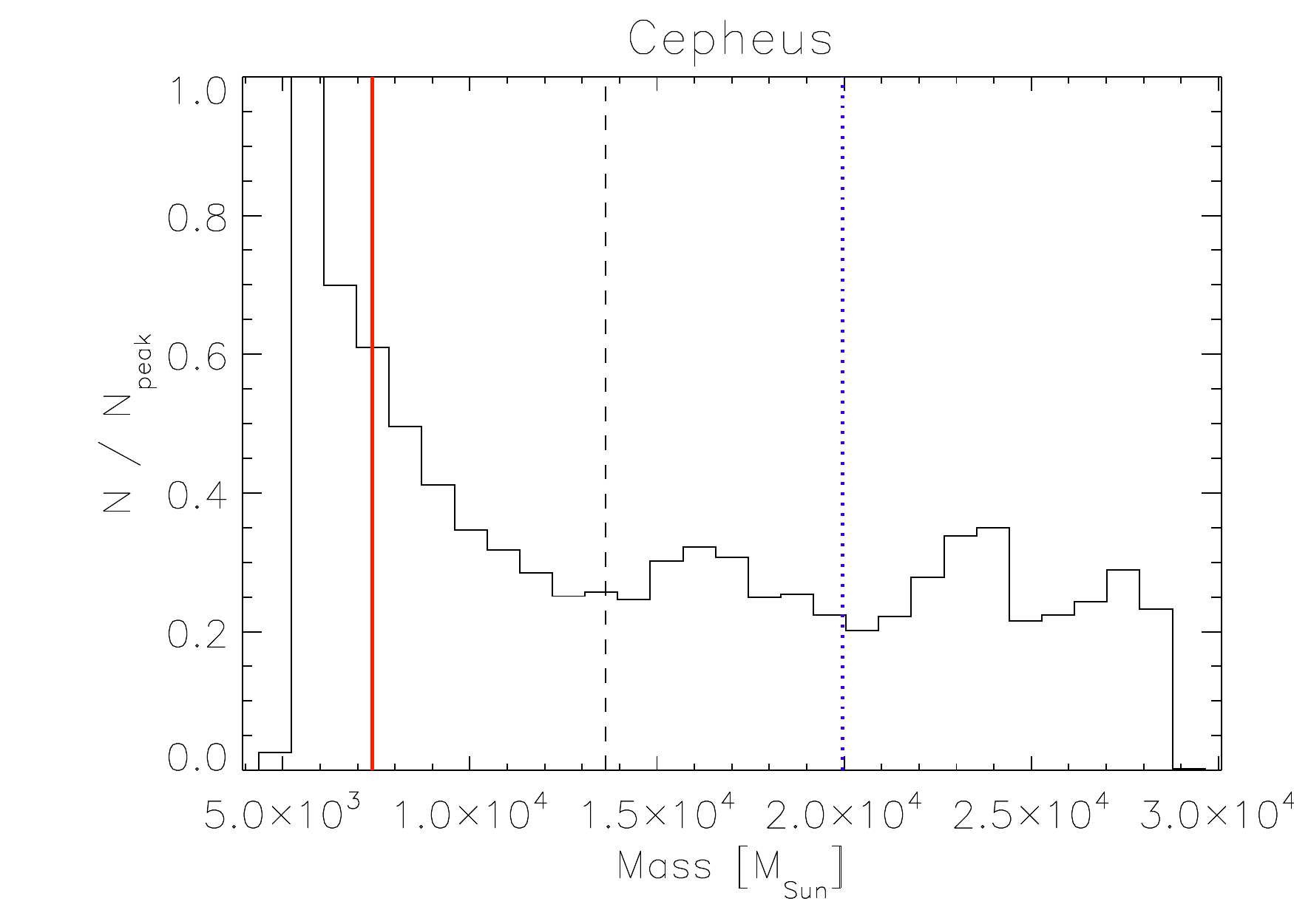}\includegraphics[width=0.32\textwidth]{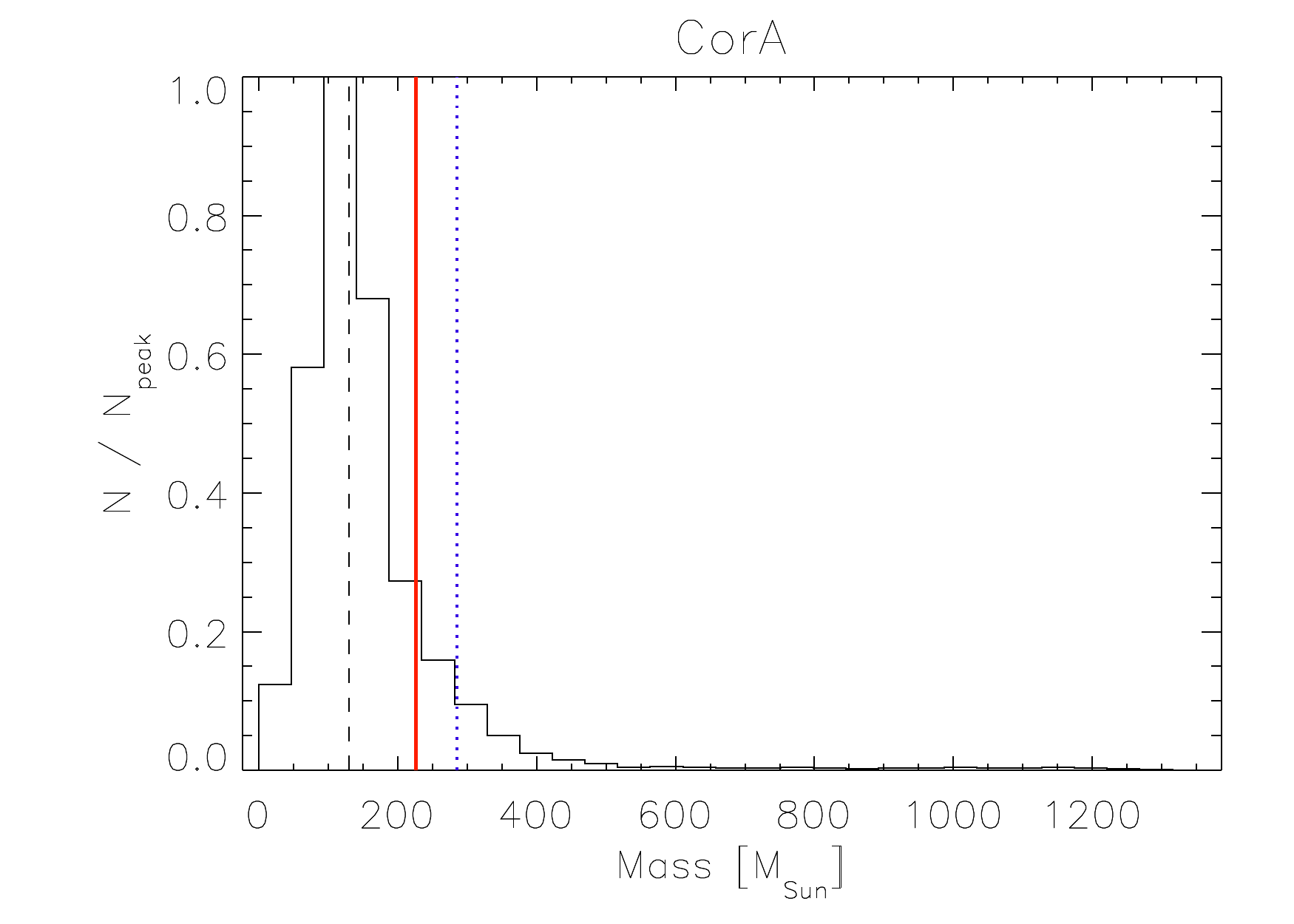}
\includegraphics[width=0.32\textwidth]{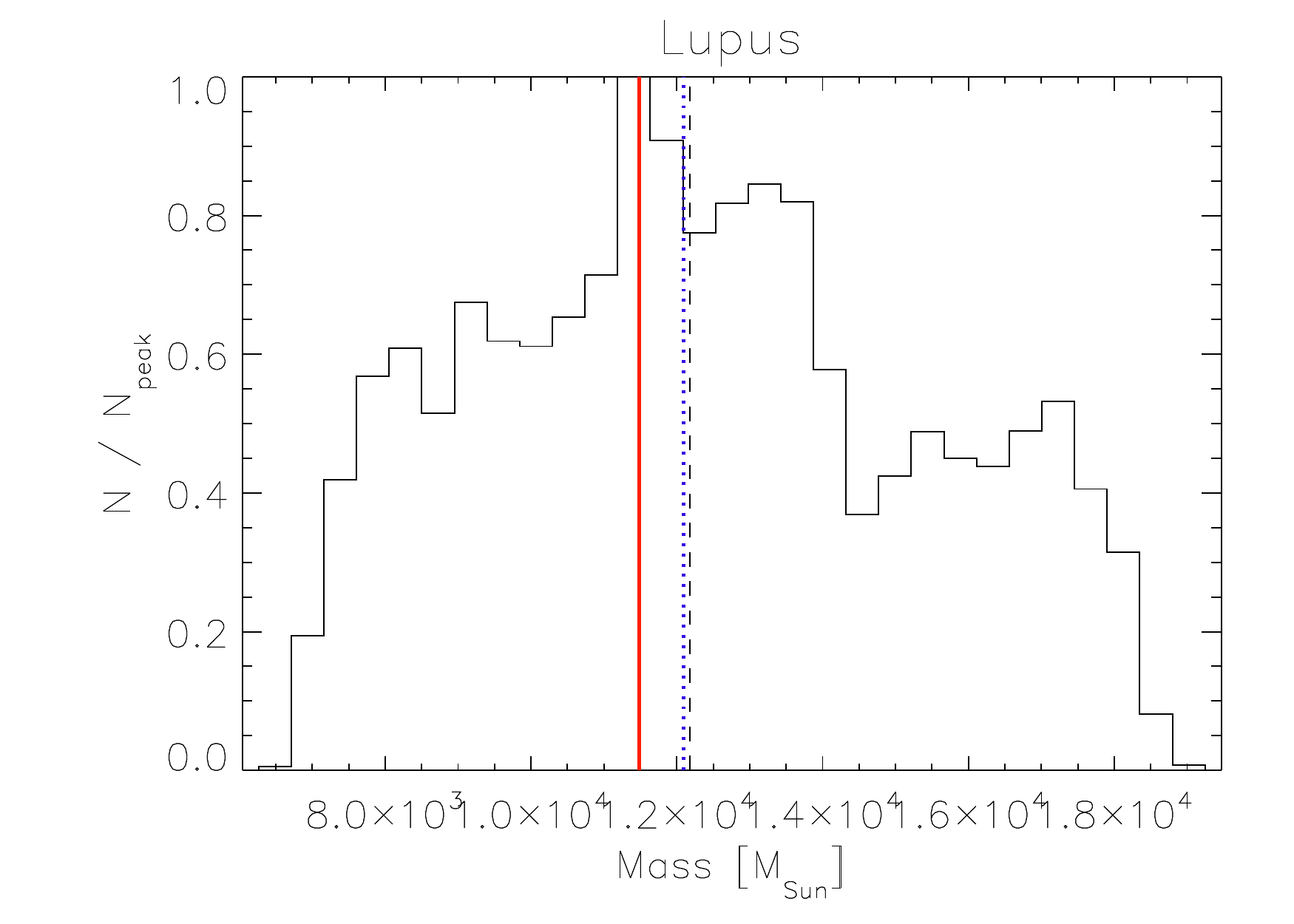}\includegraphics[width=0.32\textwidth]{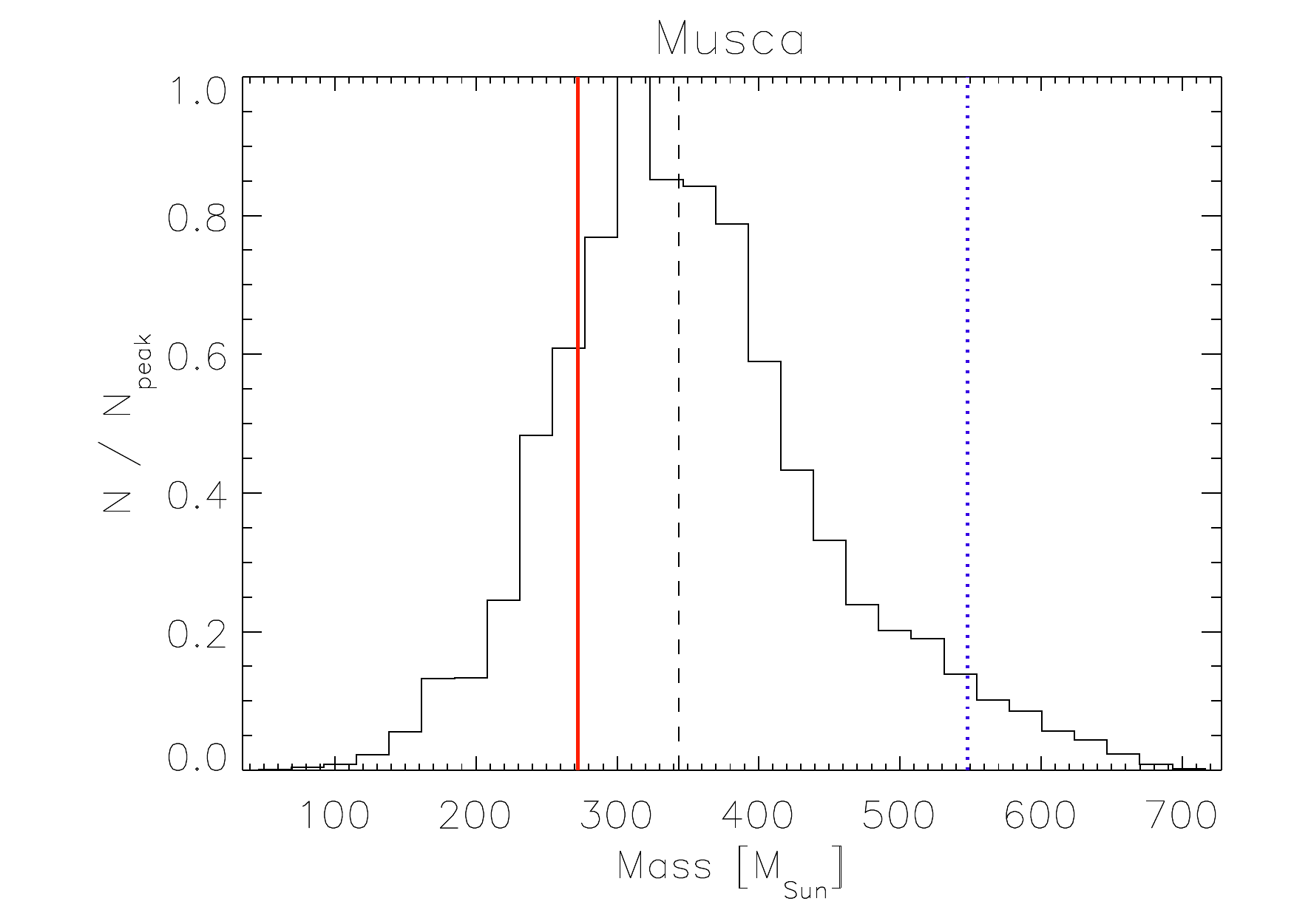}\includegraphics[width=0.32\textwidth]{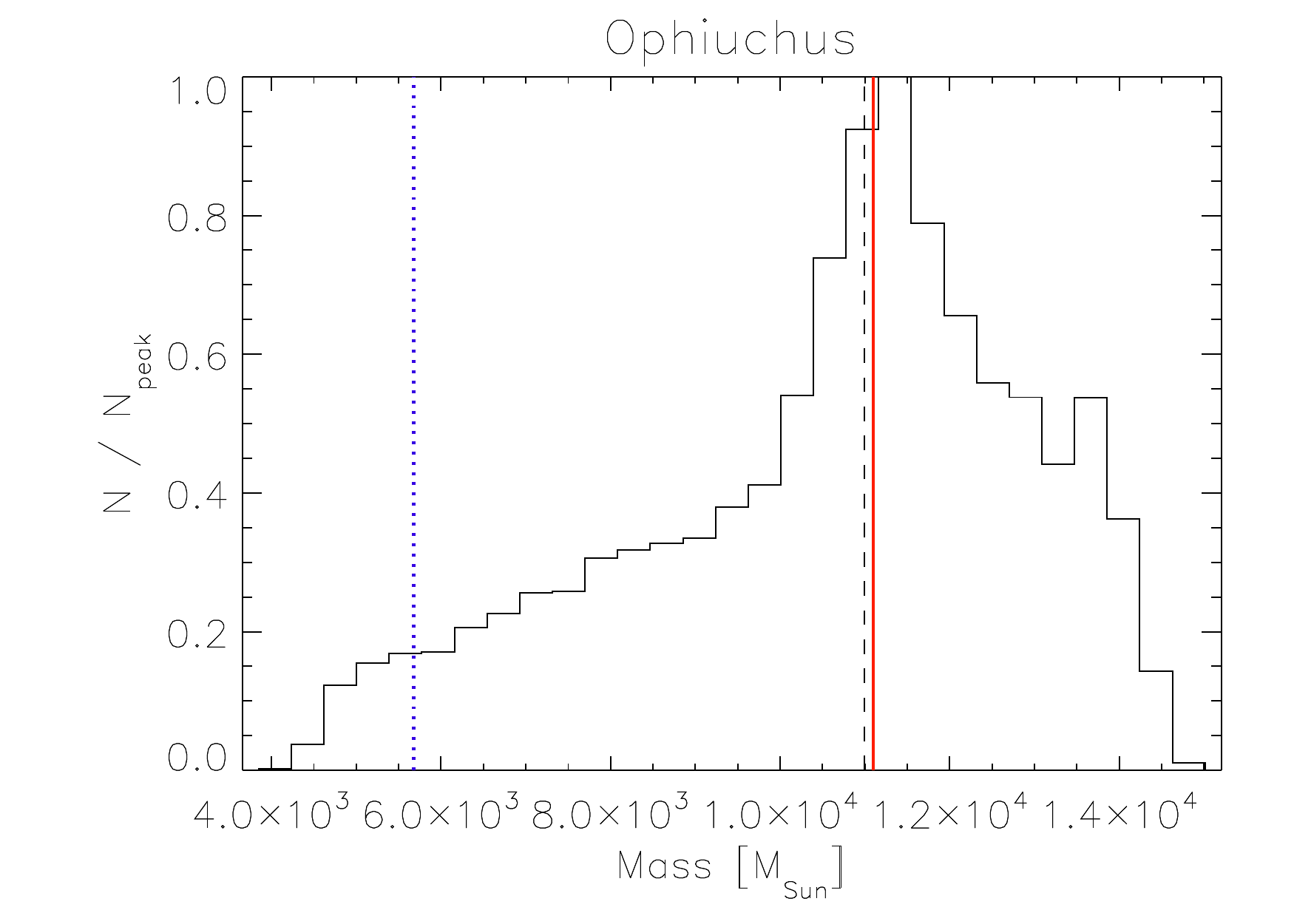}
\includegraphics[width=0.32\textwidth]{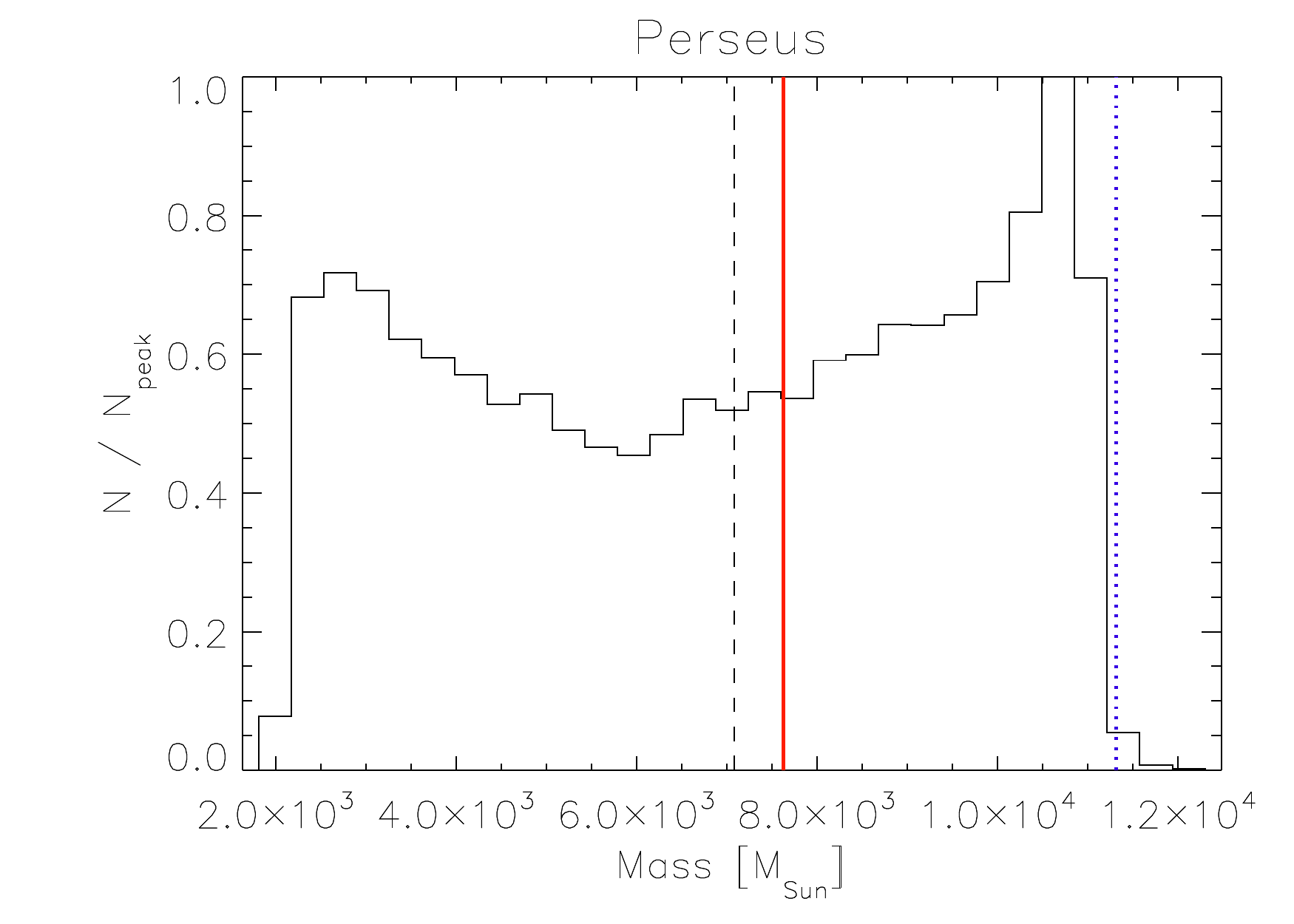}\includegraphics[width=0.32\textwidth]{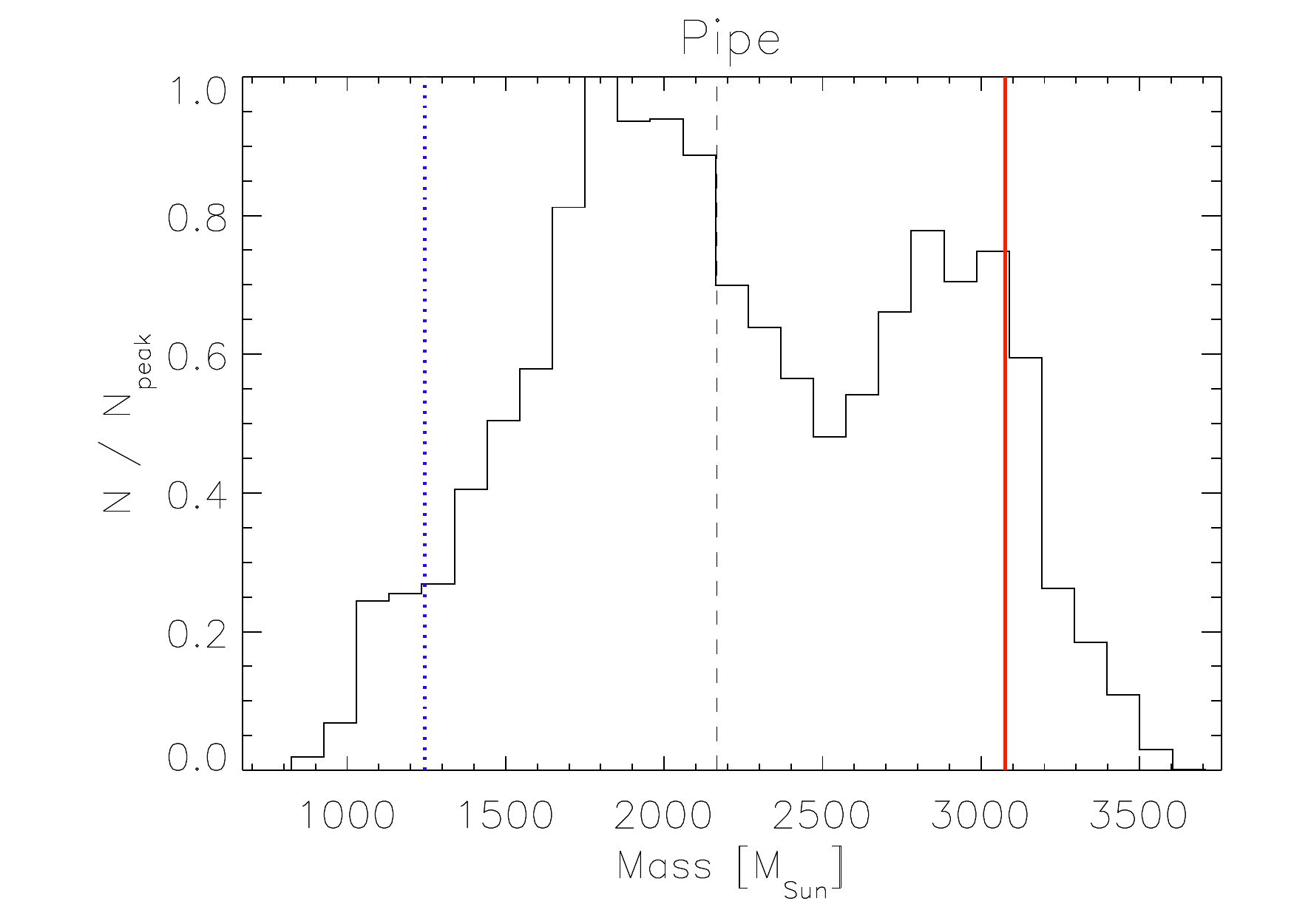}\includegraphics[width=0.32\textwidth]{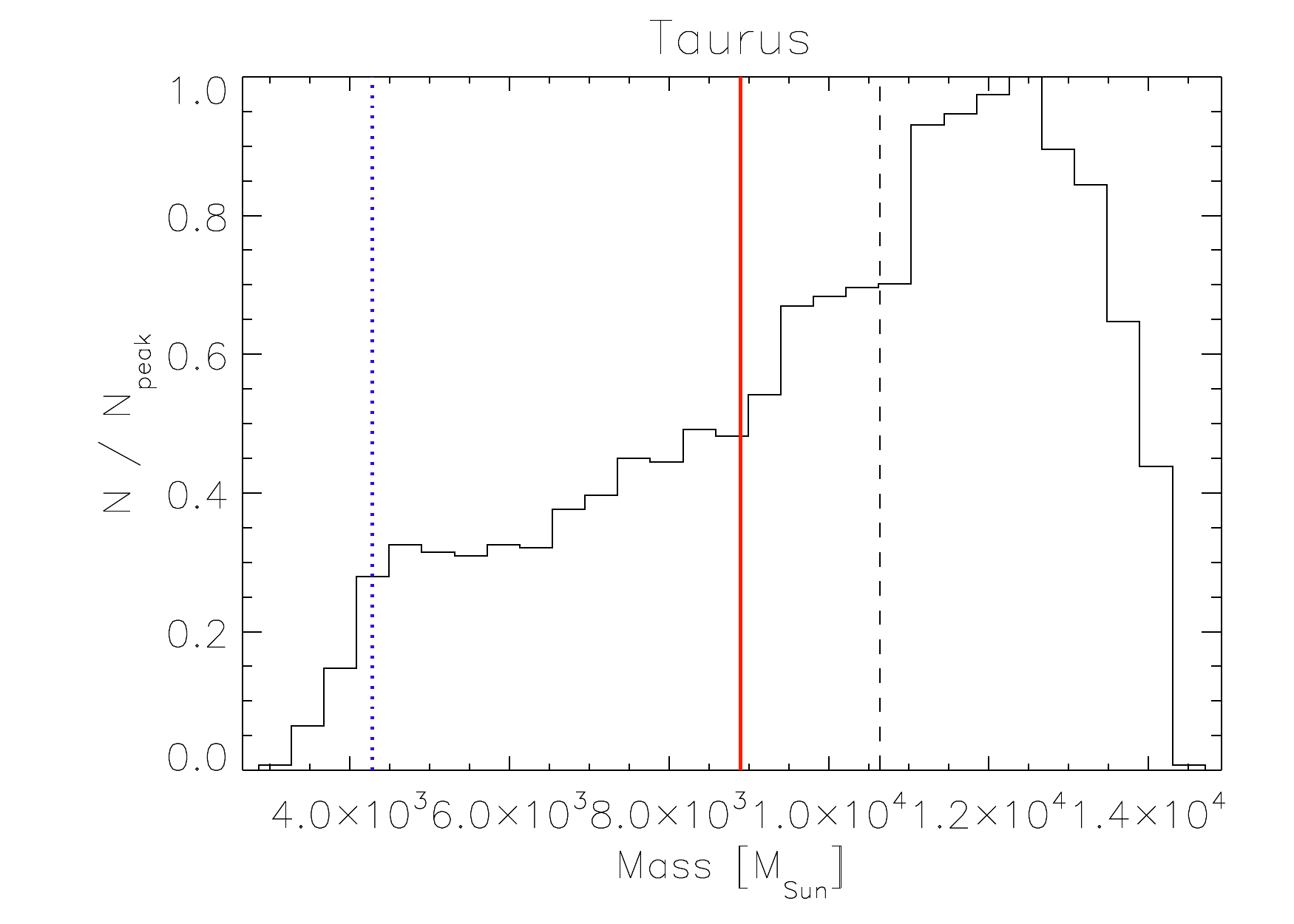}
\caption{Frequency distributions of cloud masses for the molecular clouds in our sample, normalized to the peak. The data are derived using the threshold of $A_\mathrm{G} = 0.75$ mag. The dashed vertical line indicates the median value. The red line shows the mass from the plane-of-the-sky perspective (i.e., the observed mass) and the blue dotted line from the face-on perspective (perpendicular to the Galactic disk). }
\label{fig:masshistograms}
\end{figure*}

\begin{figure*}
\centering
\includegraphics[width=0.32\textwidth]{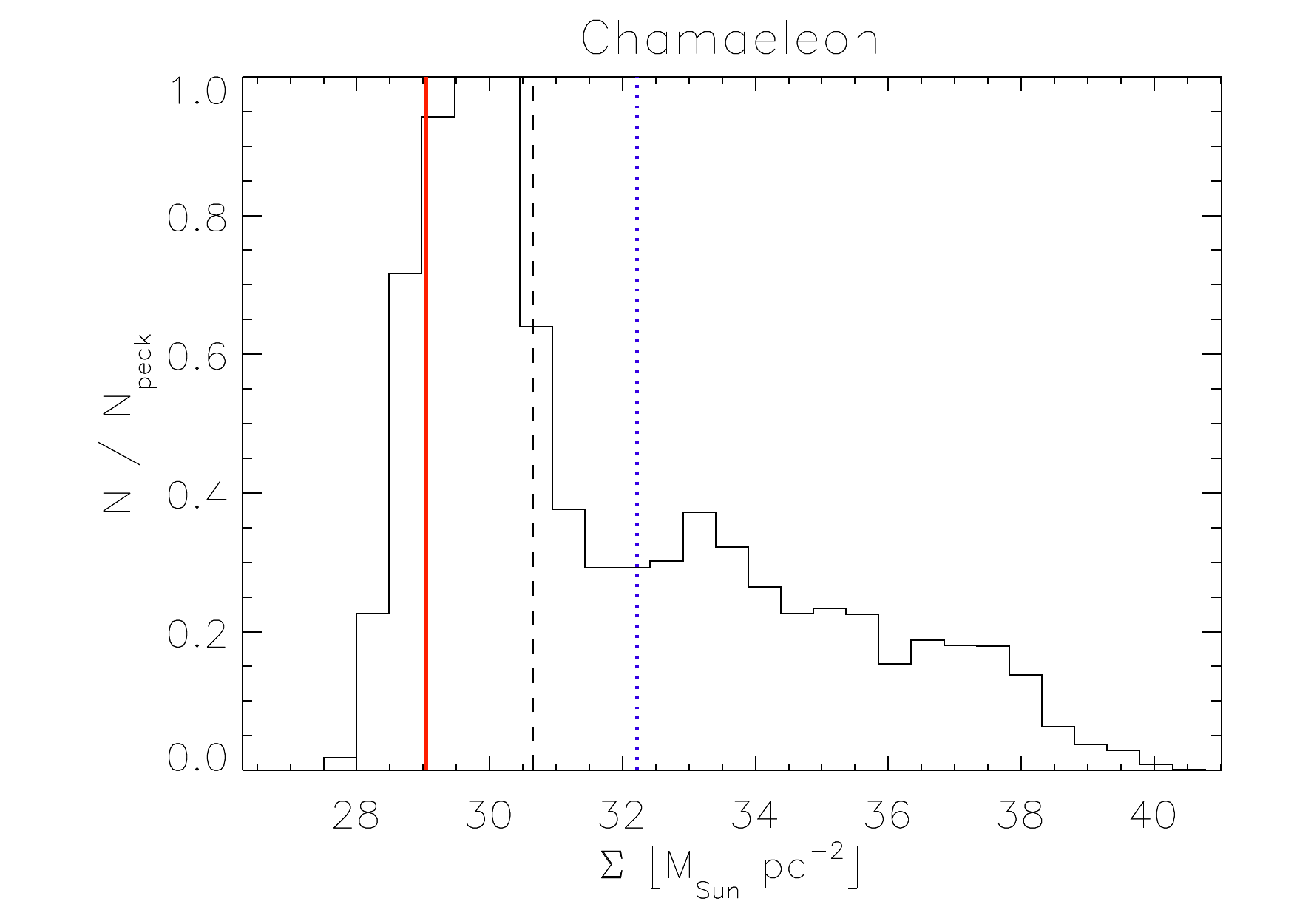}\includegraphics[width=0.32\textwidth]{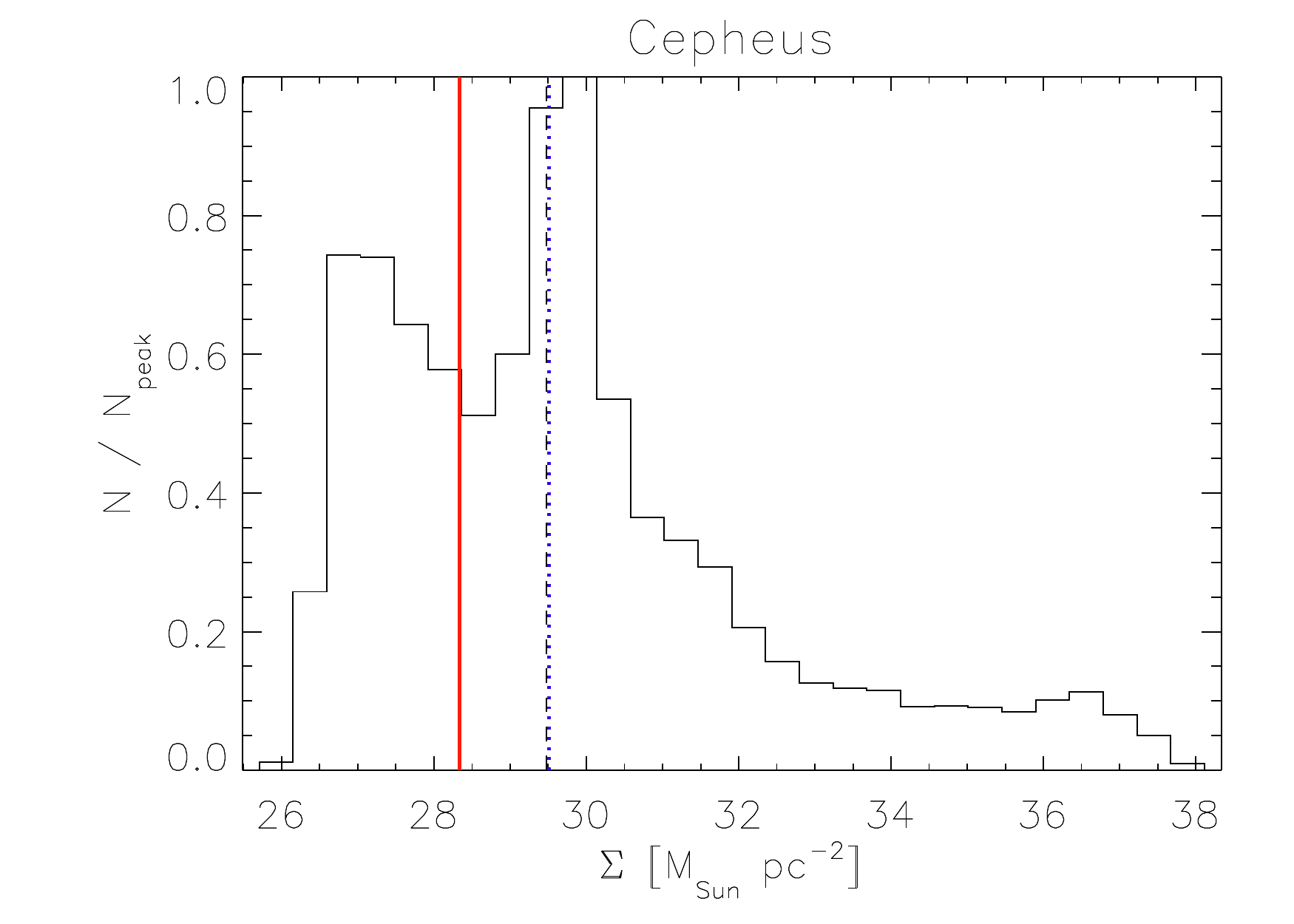}\includegraphics[width=0.32\textwidth]{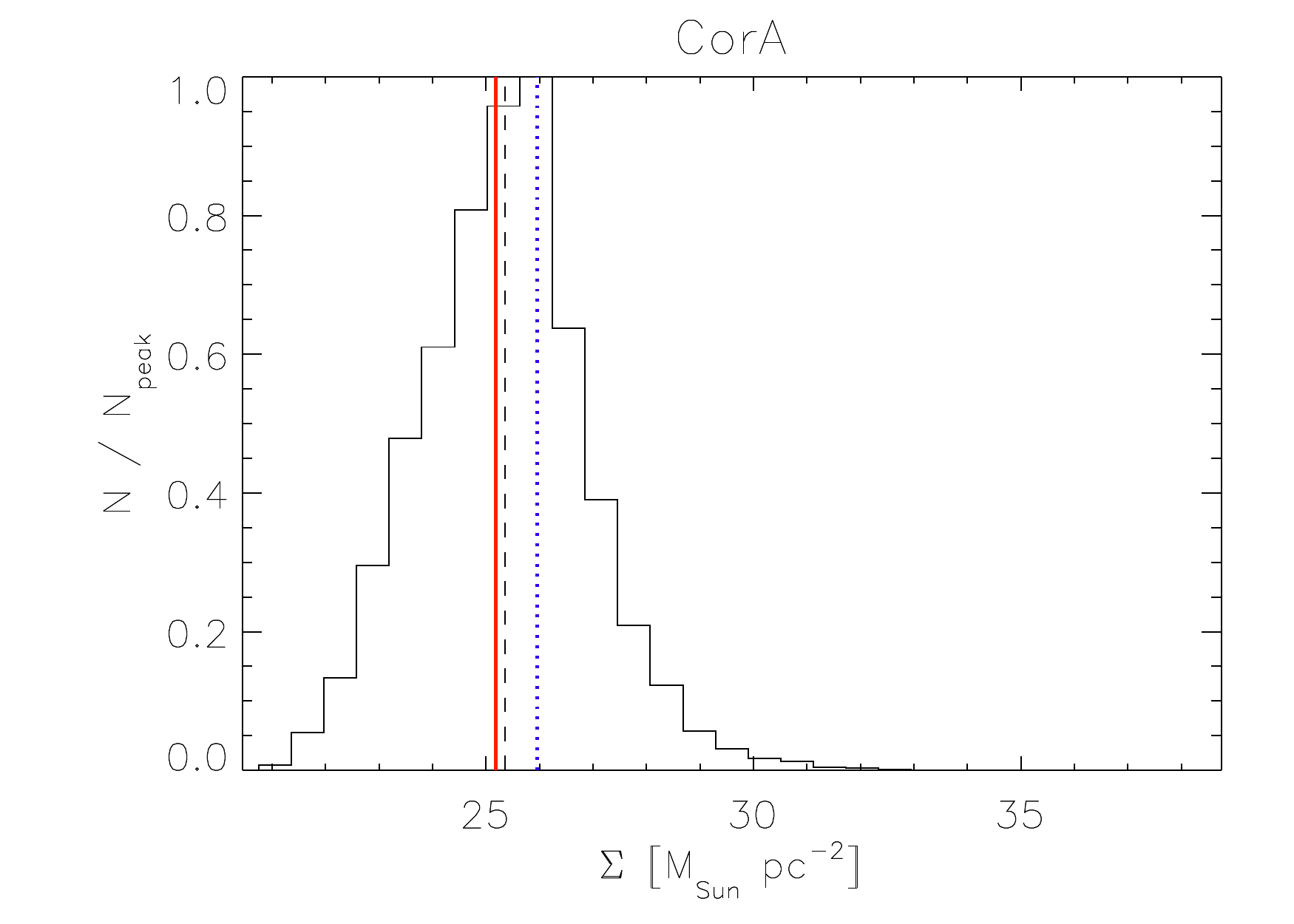}
\includegraphics[width=0.32\textwidth]{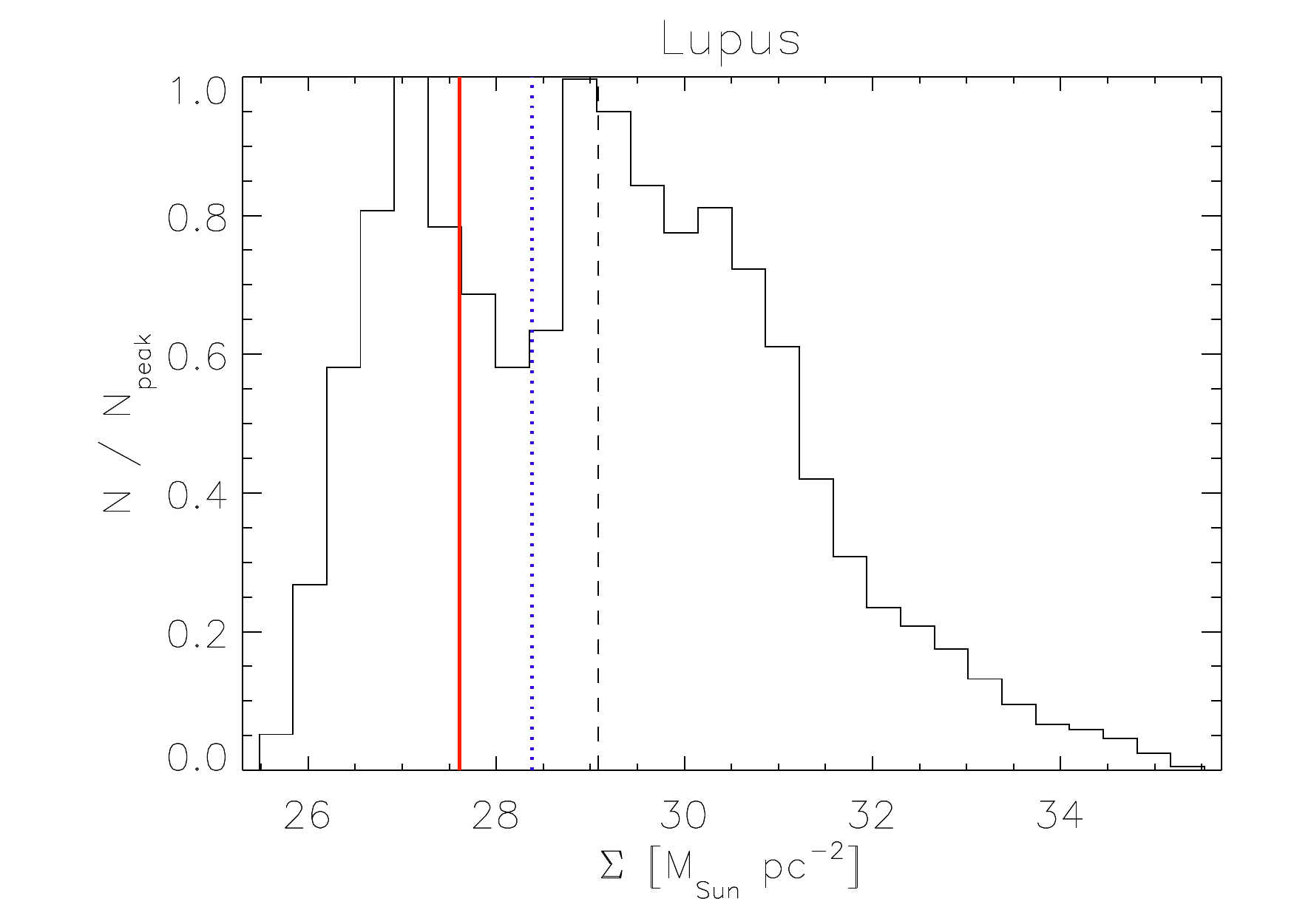}\includegraphics[width=0.32\textwidth]{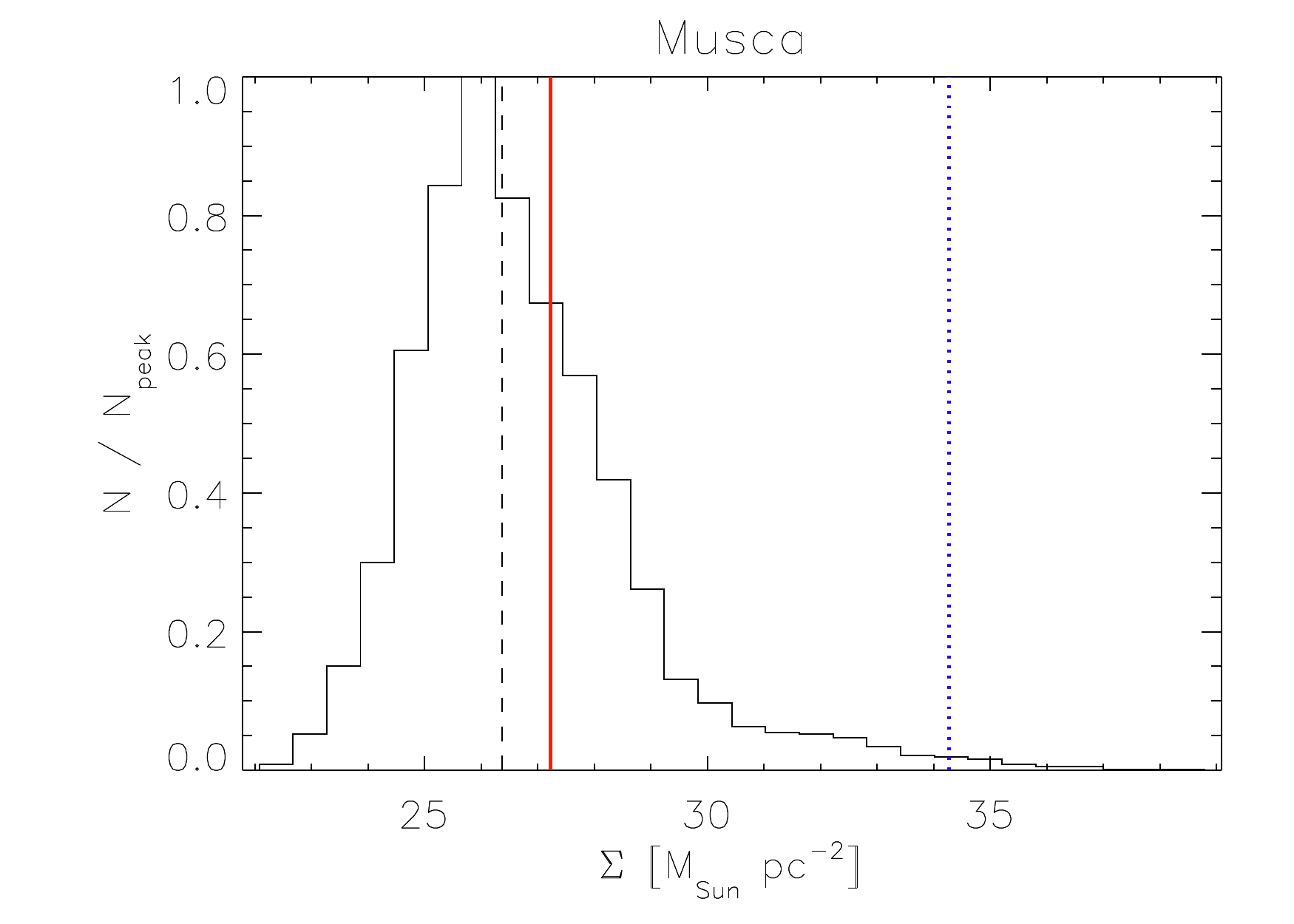}\includegraphics[width=0.32\textwidth]{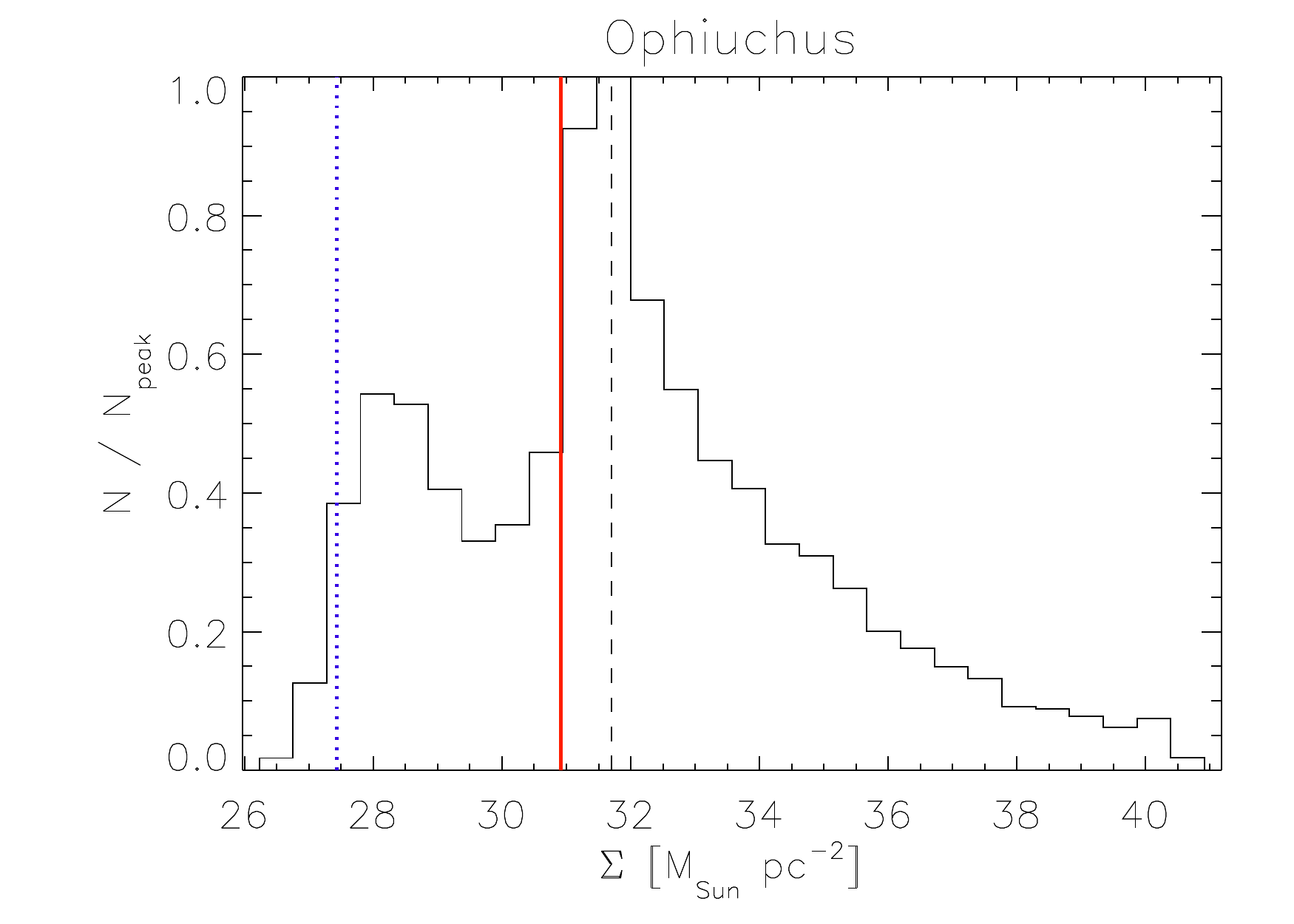}
\includegraphics[width=0.32\textwidth]{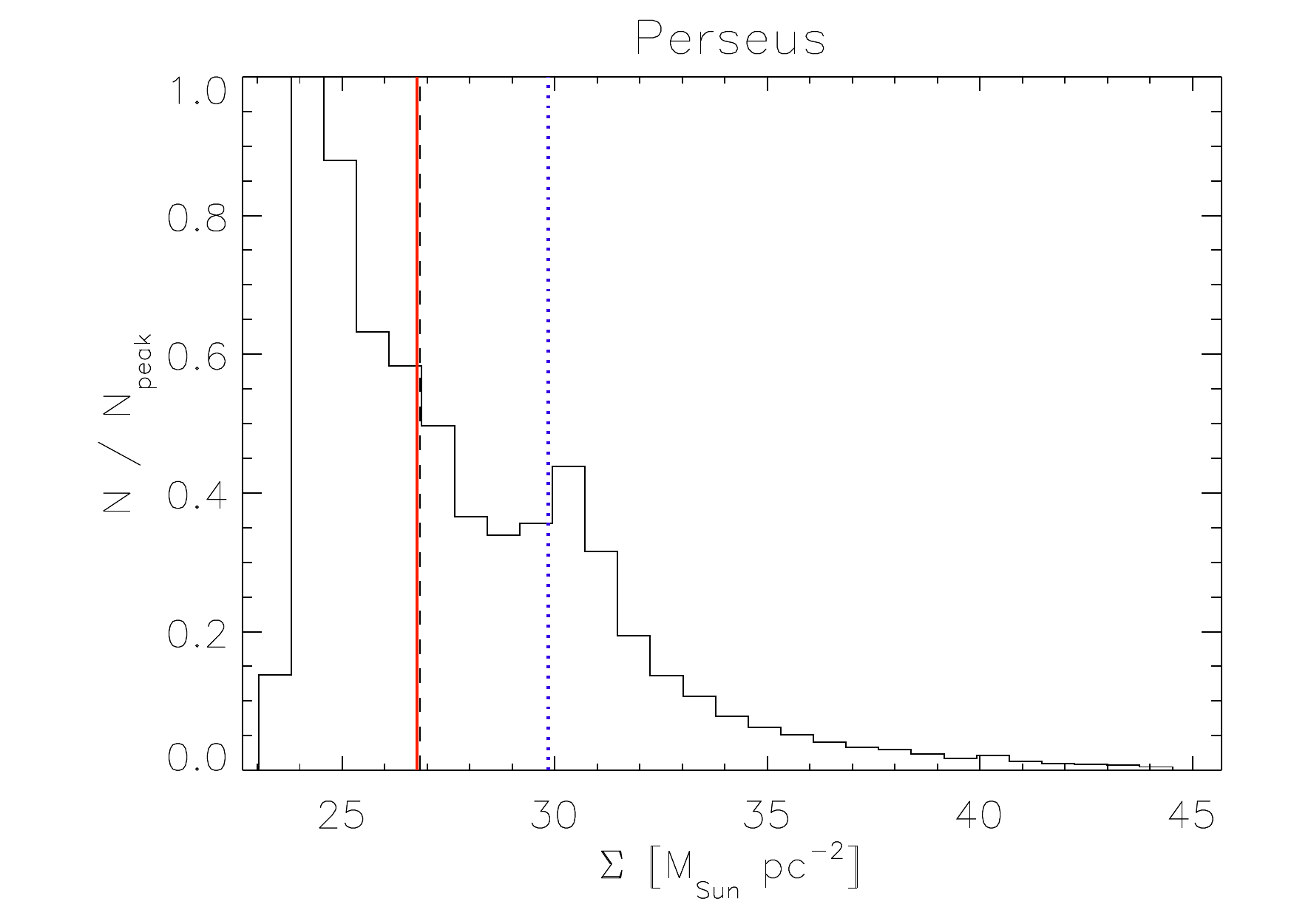}\includegraphics[width=0.32\textwidth]{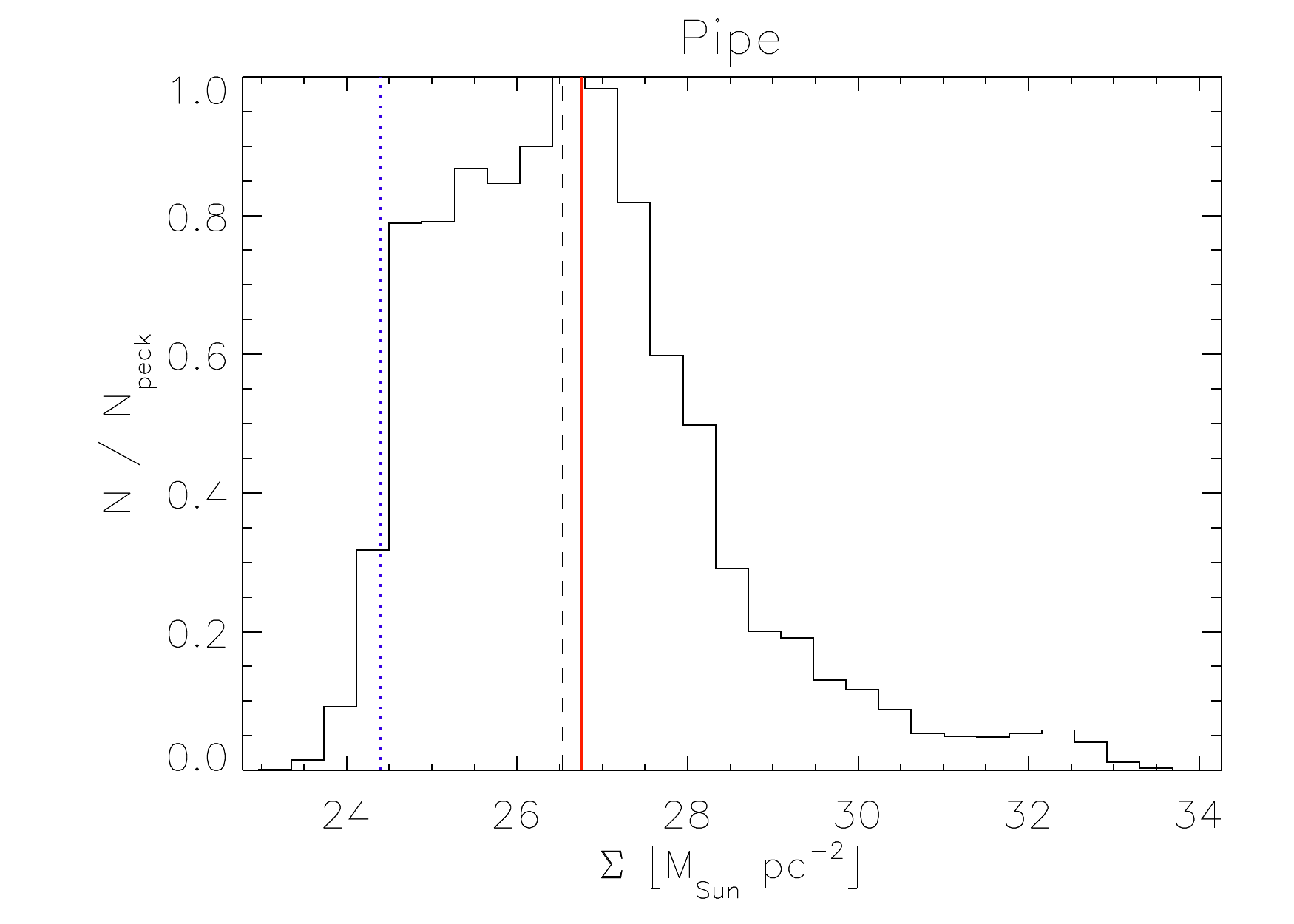}\includegraphics[width=0.32\textwidth]{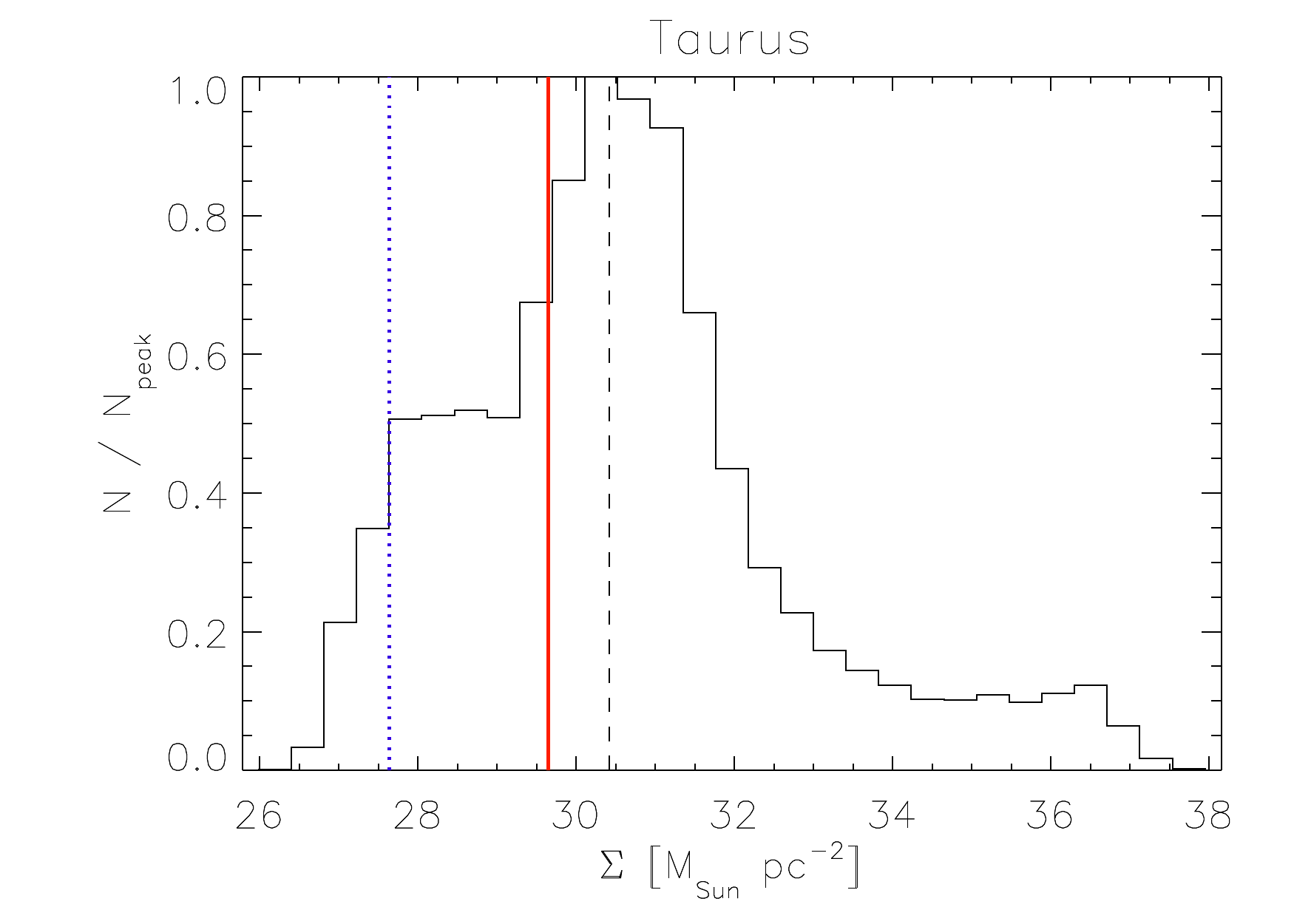}
\caption{Frequency distributions of cloud surface densities for the molecular clouds in our sample, normalized to the peak. The data are derived using the threshold of $A_\mathrm{G} = 0.75$ mag. The dashed vertical line indicates the median value. The red line shows the surface density from the plane-of-the-sky perspective (i.e., the observed value) and the blue dotted line from the face-on perspective (perpendicular to the Galactic disk). }
\label{fig:sigmahistograms}
\end{figure*}

We show here in Fig. \ref{fig:areahistograms} and \ref{fig:masshistograms} the probability distributions of areas and masses for the clouds in our sample (see Fig. \ref{fig:2dhist} for the joint distributions). We first note that the distributions of masses and areas are largely similar, differing only slightly in the detailed shapes. The distributions are in most cases clearly asymmetric, characterised by multiple peaks (e.g., Chamaeleon, Perseus, Pipe), or plateaus on either side of the maxima (e.g., Cepheus, Ophiuchus, Taurus). This asymmetry indicates that most clouds are anisotropic at the scales probed by the data. %
The distributions of surface densities (Fig. \ref{fig:sigmahistograms}) are relatively similar for different clouds, with maximum at the low-column density side and a tail towards higher surface densities. The full range of the distribution is typically about 40\% of the median value. The full-width-at-half-max of the distributions are roughly 20\% of the median value. Overall, an interesting follow-up of this work would be to explore the morphological information encoded in the probability distributions with the help of simulations. Constructing and analysing simulated probability distributions for simple model shapes like 3D Gaussians, sheets, and filaments could be compared with the observed distributions; this can provide further insight into the 3D shapes of the clouds. 


\subsection{Demonstration of various viewing angles}
\label{sec:appendix_multipleangles}

\begin{figure*}
\centering
\includegraphics[width=\textwidth]{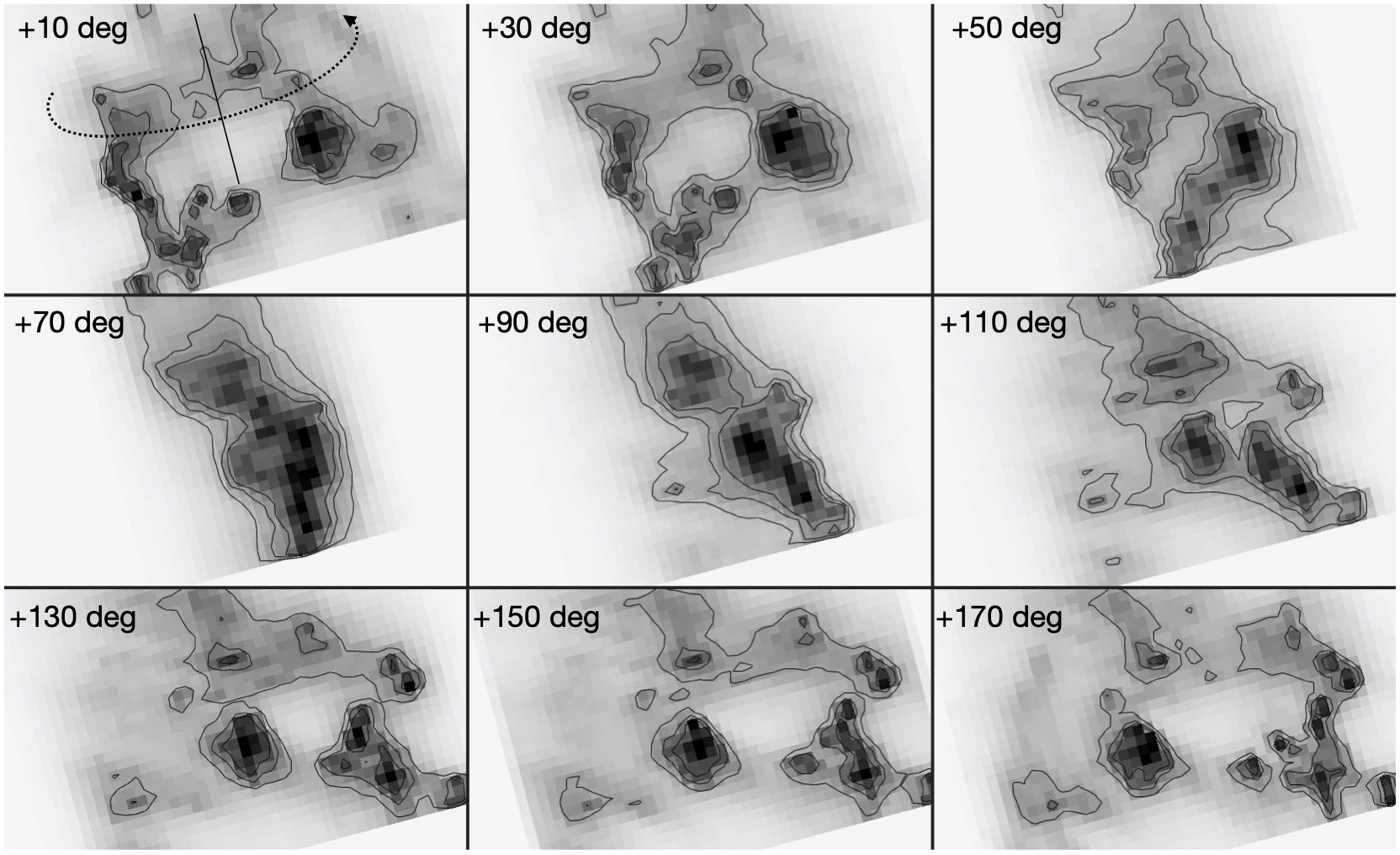}
\caption{Chamaeleon cloud complex viewed from various viewing angles rotating with respect to the z-axis. The number in the panels indicates the rotation from the plane-of-the-sky angle. The rotation direction is indicated in the top left panel. The contours show the levels of $A_\mathrm{G} = \{ 0.5, 0.75, 1\}$ mag. }
\label{fig:different_angles_dir1}
\end{figure*}

\begin{figure*}
\centering
\includegraphics[width=\textwidth]{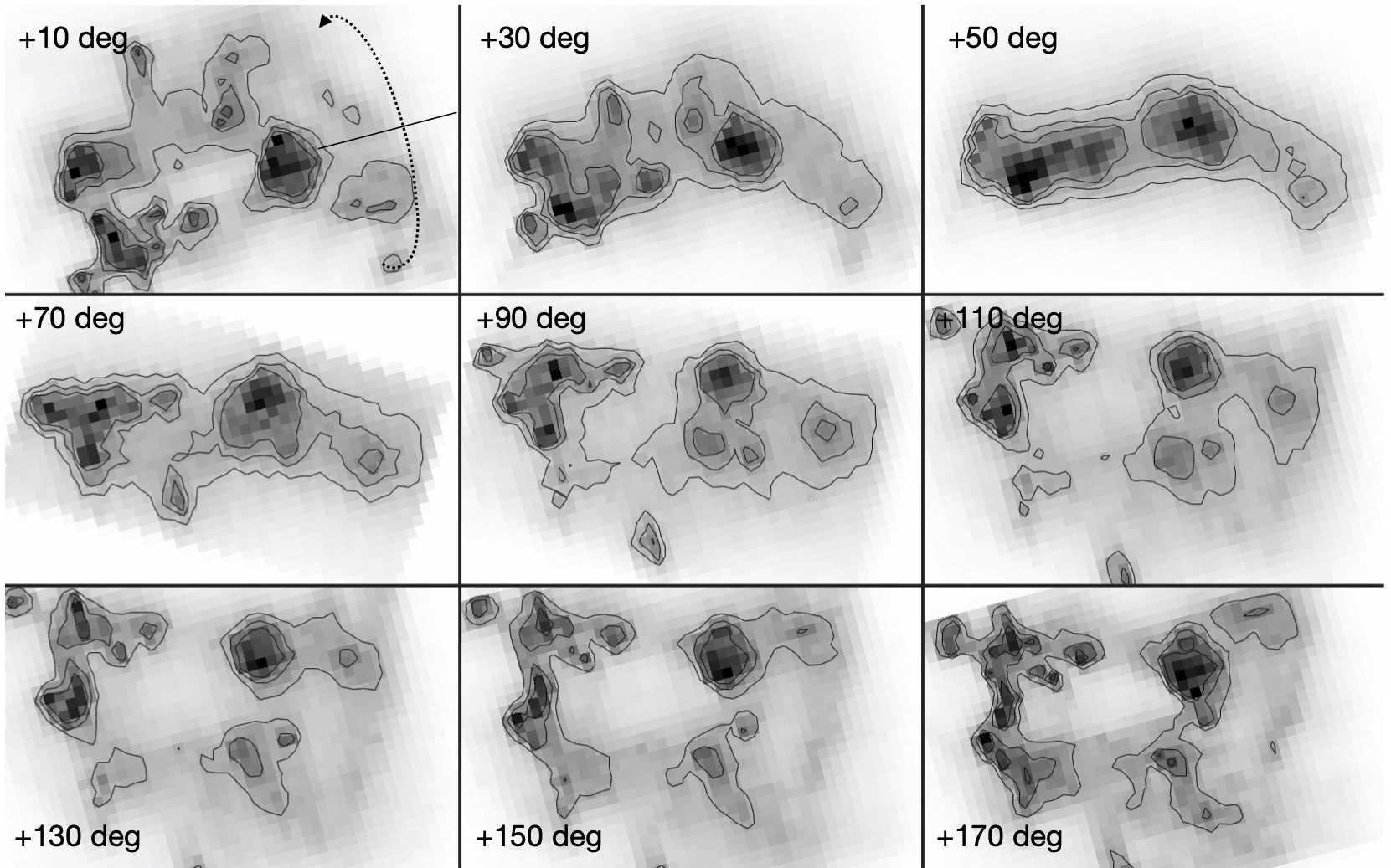}
\caption{Chamaeleon cloud complex viewed from various viewing angles rotating with respect to the y-axis. The number in the panels indicates the rotation from the plane-of-the-sky angle. The rotation direction is indicated in the top left panel. The contours show the levels of $A_\mathrm{G} = \{ 0.5, 0.75, 1\}$ mag.}
\label{fig:different_angles_dir2}
\end{figure*}

We illustrate in Figs. \ref{fig:different_angles_dir1} and \ref{fig:different_angles_dir2} what one cloud, Chamaeleon, looks like from an array of viewing angles. The figures show the cloud complex rotated in 20 degree increments with respect to the $z$ and $y$ axes, starting from the POS view. The top left panel in the figures is close to the POS view. The figures demonstrate how the cloud morphology above a given column density (extinction) threshold dramatically changes for different viewing angles.

\section{Inherent variance of the Leike et al. (2020) data} 
\label{sec:appendix_allsamples}

\citet{Leike2020} provides twelve posterior samples for their dust distribution model, enabling an assessment of the variance due to the uncertainty of the model. We re-calculate here the area, mass, and surface density probability distributions for each posterior sample of the Chamaeleon cloud to describe the inherent uncertainty of the model.

\begin{figure*}
\centering
\includegraphics[width=0.33\textwidth]{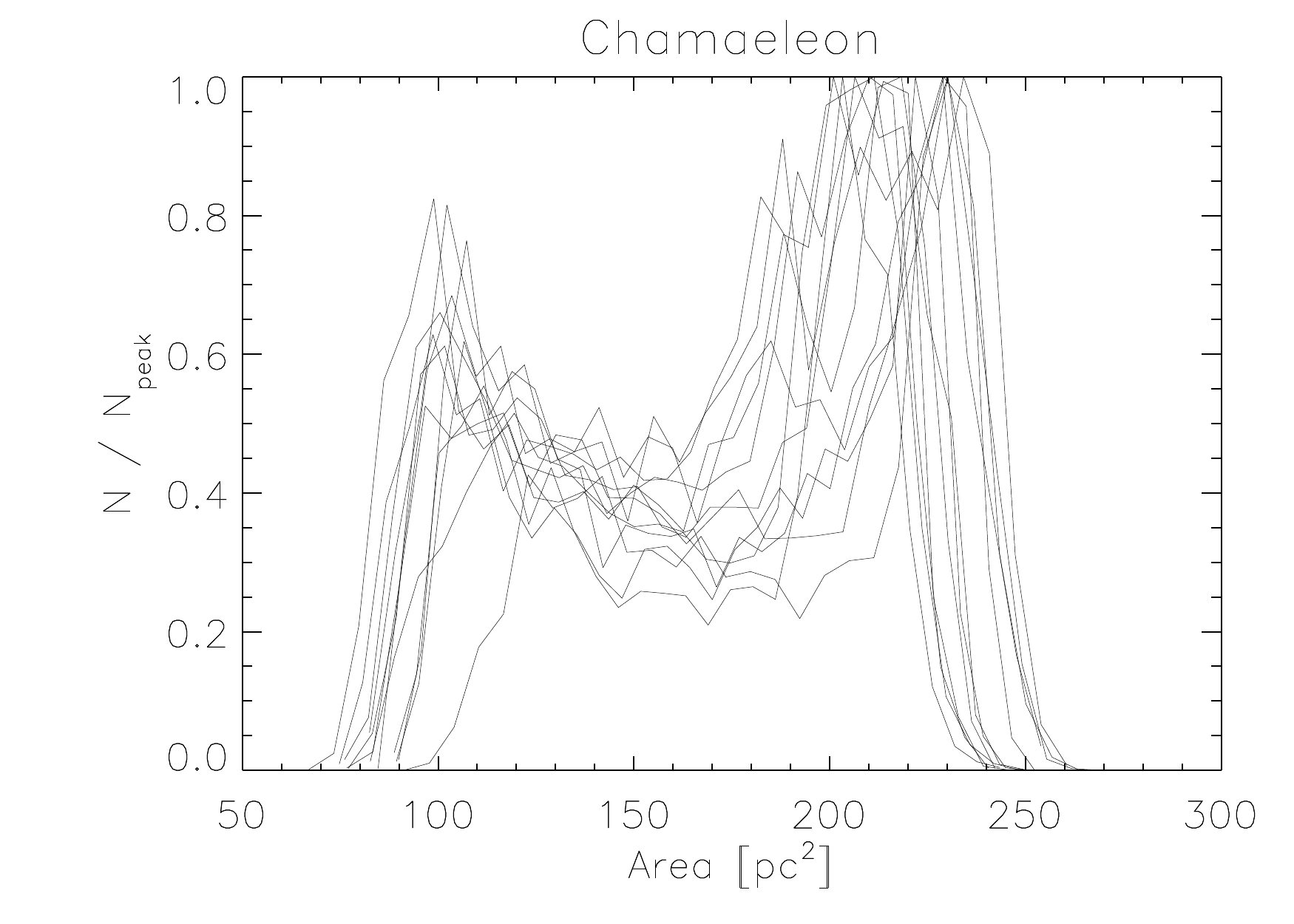}\includegraphics[width=0.33\textwidth]{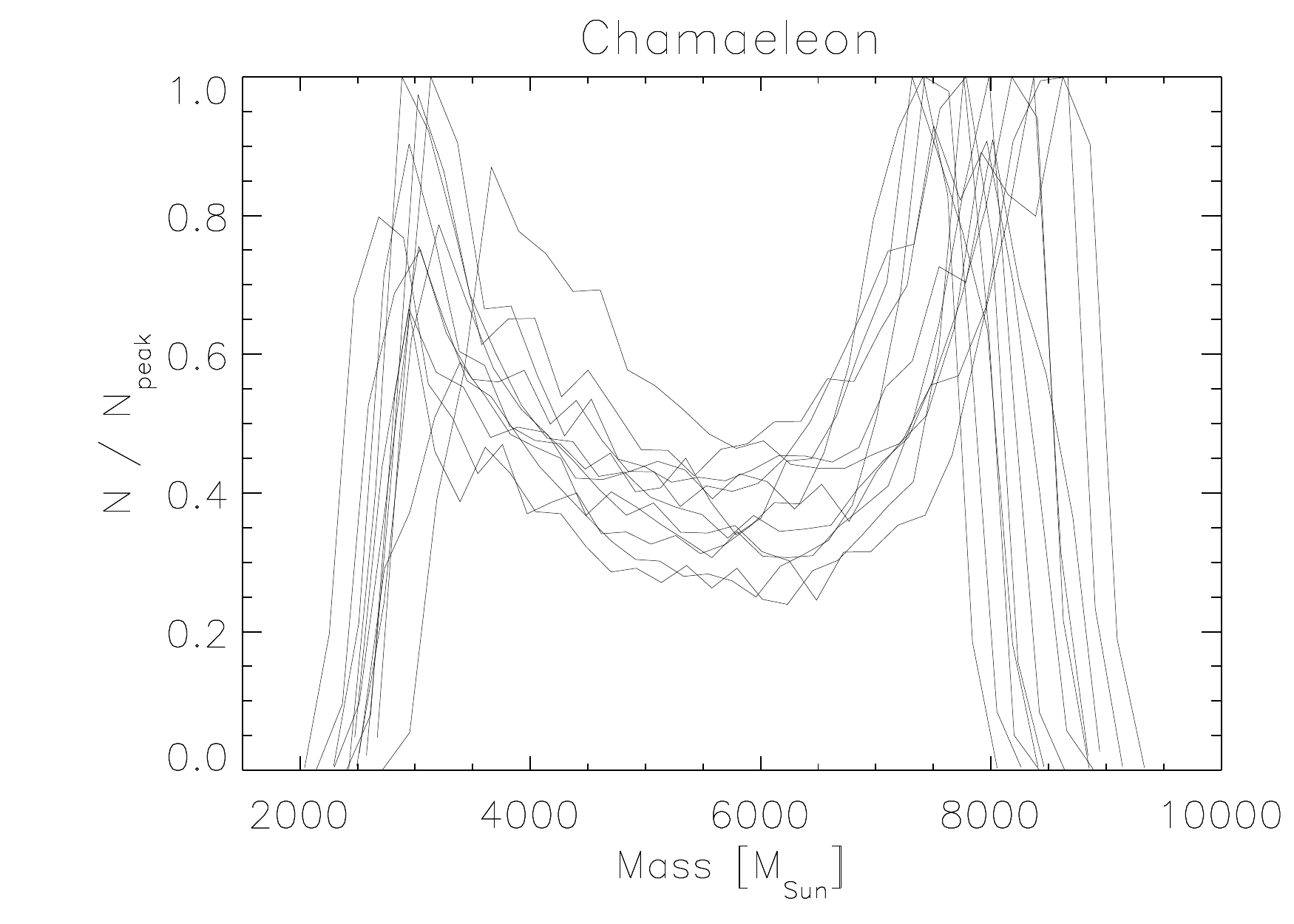}\includegraphics[width=0.33\textwidth]{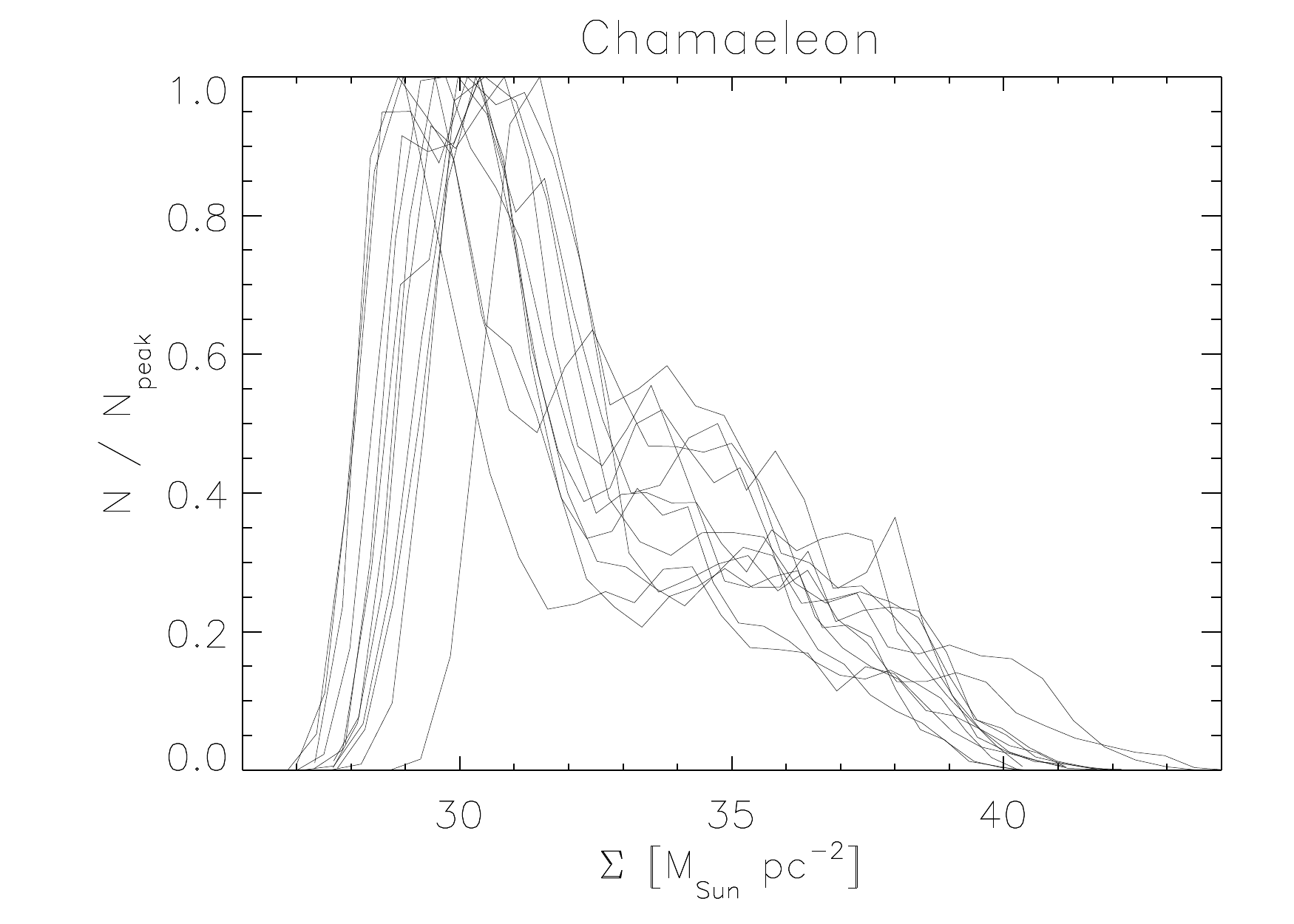}
\caption{The probability distributions of area (left panel), mass (centre panel) and gas surface density (right panel) derived from the 12 posterior samples of the \citet{Leike2020} data for Chamaeleon. For visibility, the distributions are shown as lines instead of histogram columns.}
\label{fig:appendix_allsamples}
\end{figure*}

Figure \ref{fig:appendix_allsamples} shows the areas, masses, and surface density probability distributions for the 12 posterior samples of Chamaeleon. In all cases, the shapes of the distributions are qualitatively similar. For Chamaeleon, the distribution is characterised by two peaks; this behaviour is recovered by all samples. The ranges of the parameters vary on the 10\% level, which is clearly lower than the variance due to the viewing angles (typically 100\% and often more, cf., Table \ref{tab:parameters}). The POS and face-on areas and masses vary between 5\%-8\%, and the gas surface densities about 2\%. In conclusion, the test indicates that our results are not significantly affected by the inherent uncertainty of the \citet{Leike2020} data.

\section{Effect of the extinction threshold}
\label{sec:appendix_threshold}

\begin{figure*}
\centering
\includegraphics[width=0.32\textwidth]{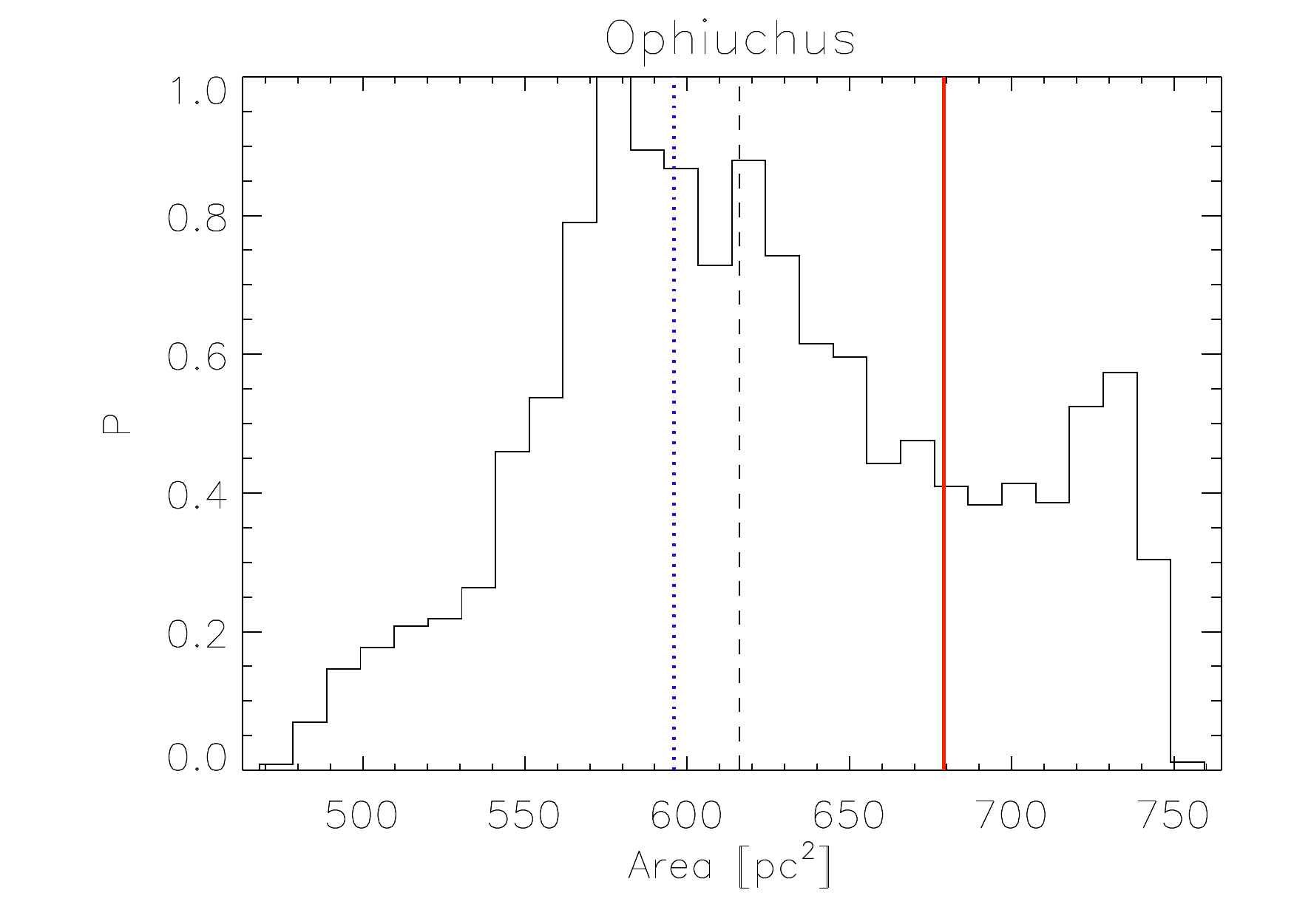}\includegraphics[width=0.32\textwidth]{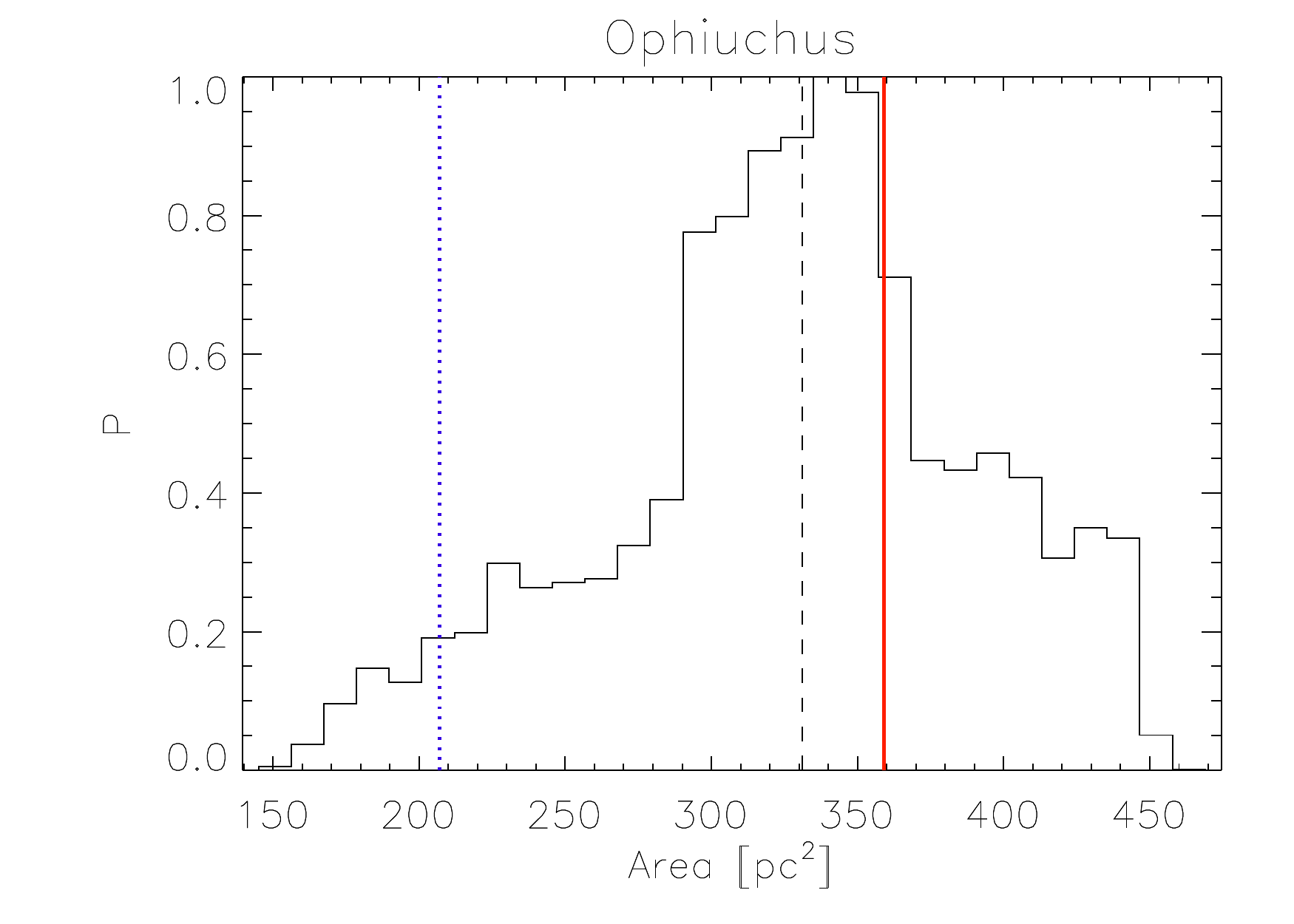}\includegraphics[width=0.32\textwidth]{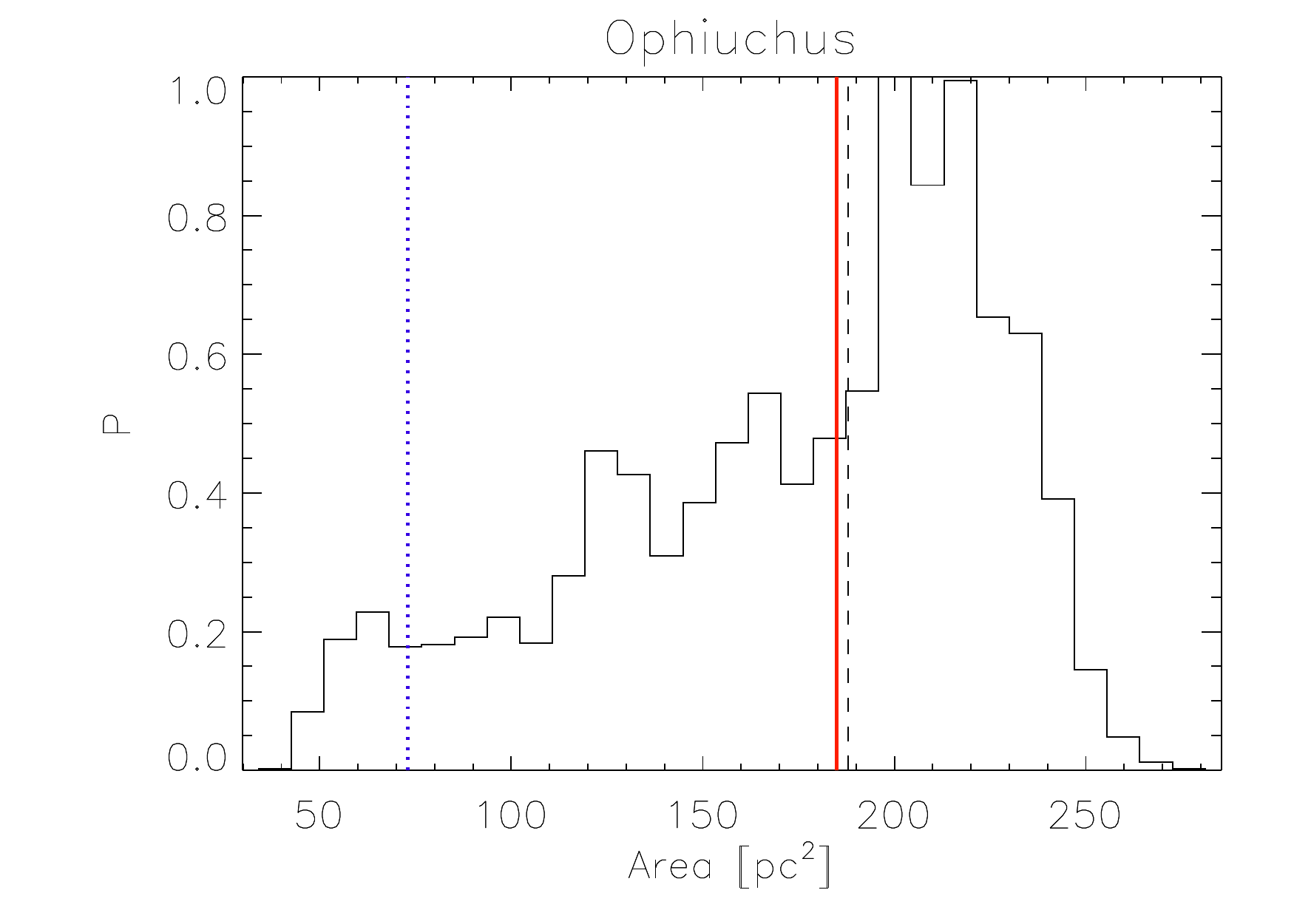}
\includegraphics[width=0.32\textwidth]{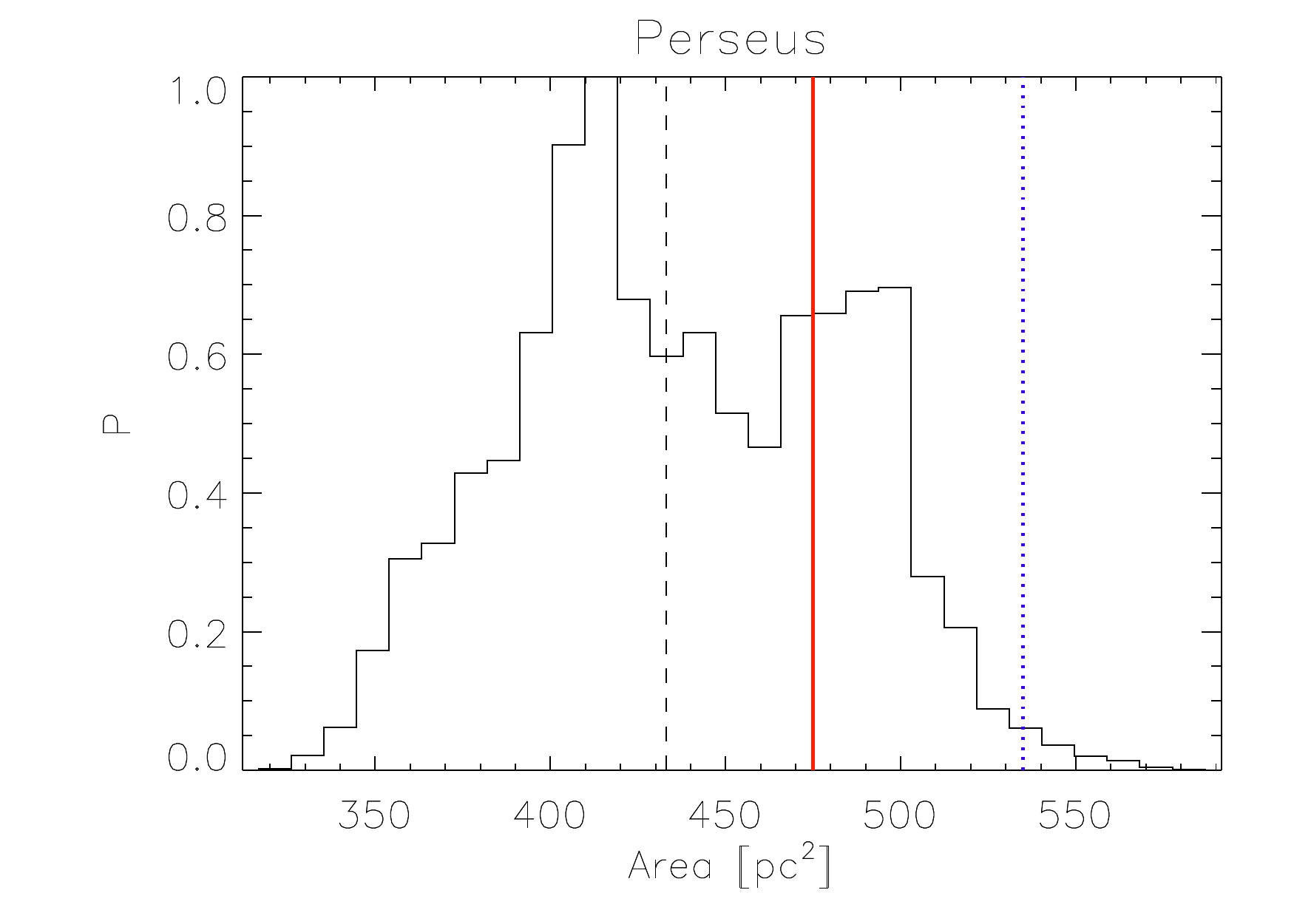}\includegraphics[width=0.32\textwidth]{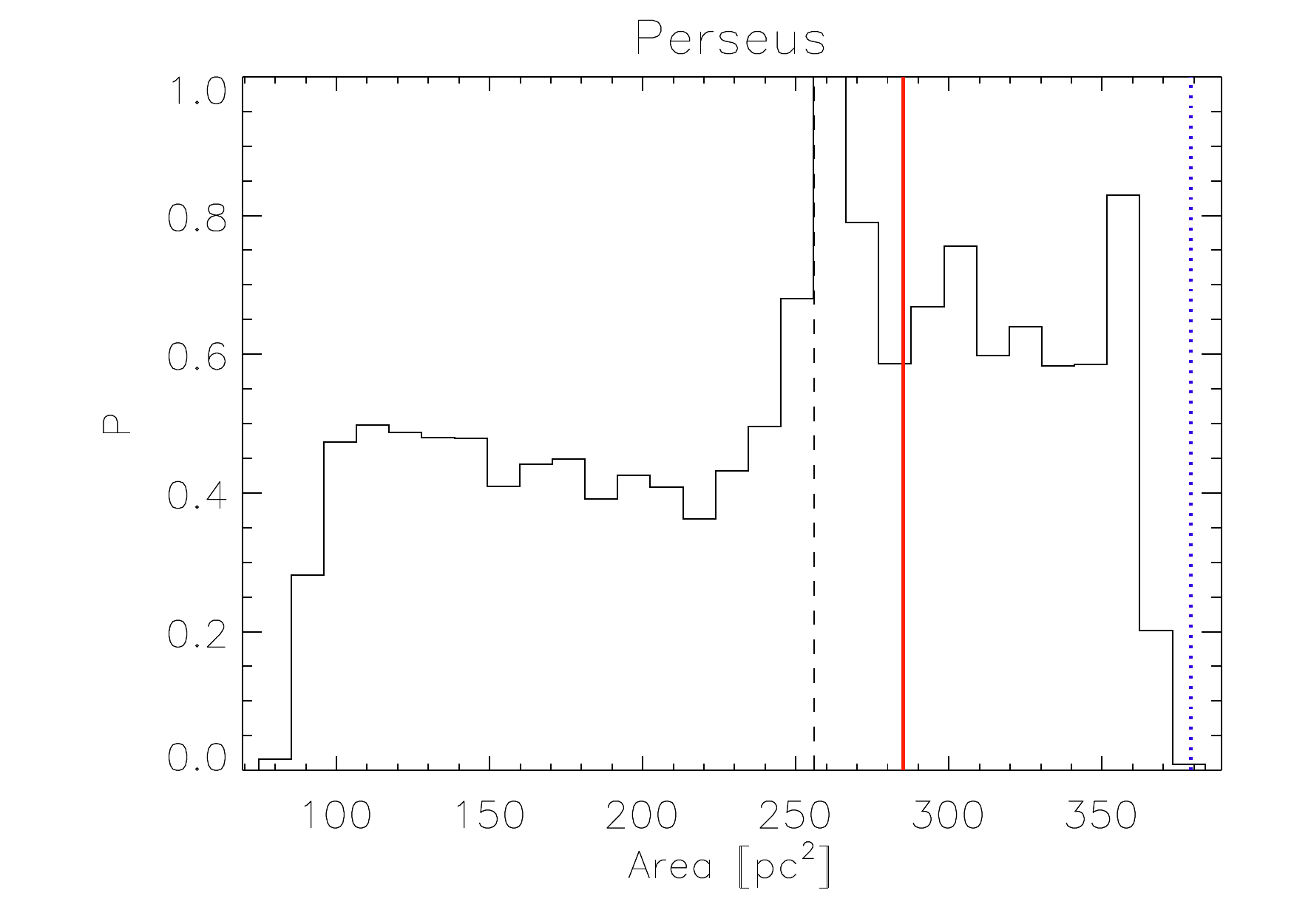}\includegraphics[width=0.32\textwidth]{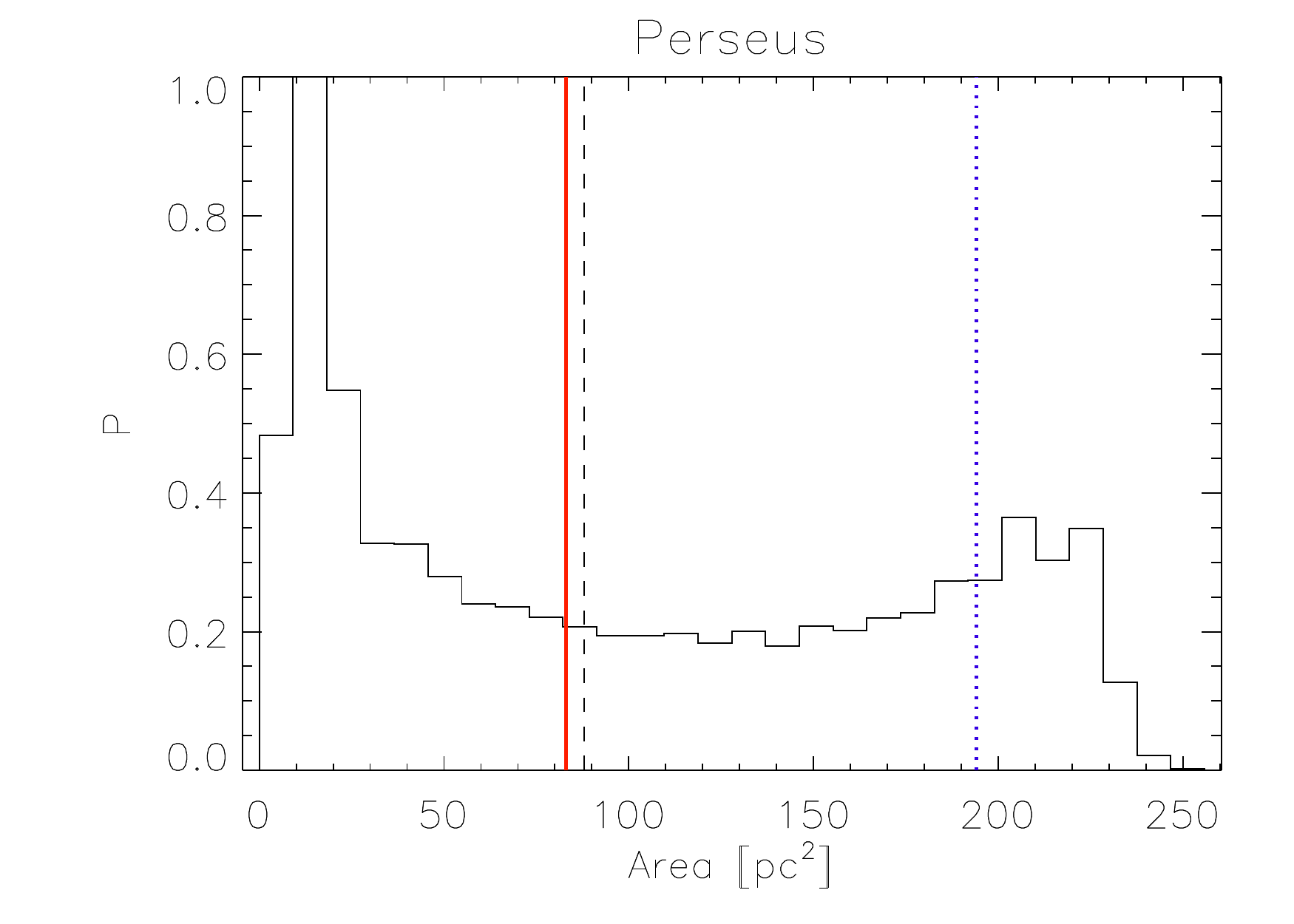}
\caption{Frequency distributions of cloud areas for Ophiuchus and Perseus, derived using different threshold levels: $A_\mathrm{G} = 0.5$ mag (left panels), 0.75 mag (center panels), and 1 mag (right panels). The dashed vertical line indicates the median value. The red line shows the area from the POS perspective (i.e., the observed area) and the blue dotted line from the face-on perspective (perpendicular to the Galactic disk). }
\label{fig:thcomparison}
\end{figure*}

We explore the effect of the extinction threshold used to define the clouds by using three different thresholds, $A_\mathrm{G} = \{0.5, 0.75, 1\}$ mag. Figure \ref{fig:viewingangles} demonstrates the effect of the threshold in the POS and face-on viewing angles for one cloud (Chamaeleon). It shows how the different thresholds lead to different structures being identified as clouds, and thus, included in the calculation of total mass and area of the cloud. Figure \ref{fig:thcomparison} describes how the threshold affects the probability distributions of areas for two example clouds (Ophiuchus and Perseus). It show how the shape of the distribution can change, even quite dramatically as in the case of Perseus, with the threshold. Clearly, the choice of the threshold has a non-trivial effect on the mass and area statistics. This indicates potential problems if cloud properties derived using different thresholds are compared with each others, e.g., when comparing cloud catalogues from different literature studies.  

\begin{figure*}
\centering
\includegraphics[width=0.32\textwidth]{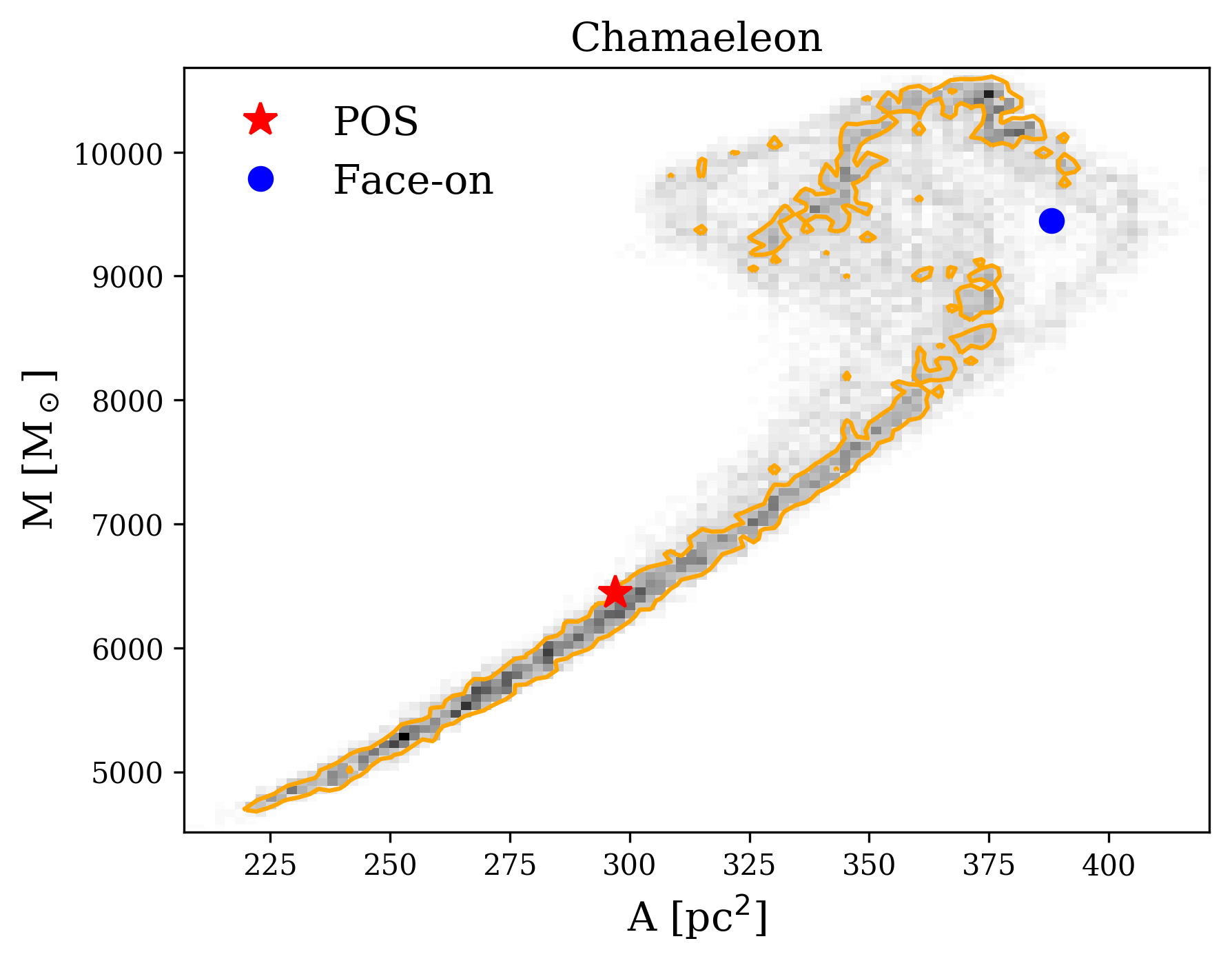}
\includegraphics[width=0.32\textwidth]{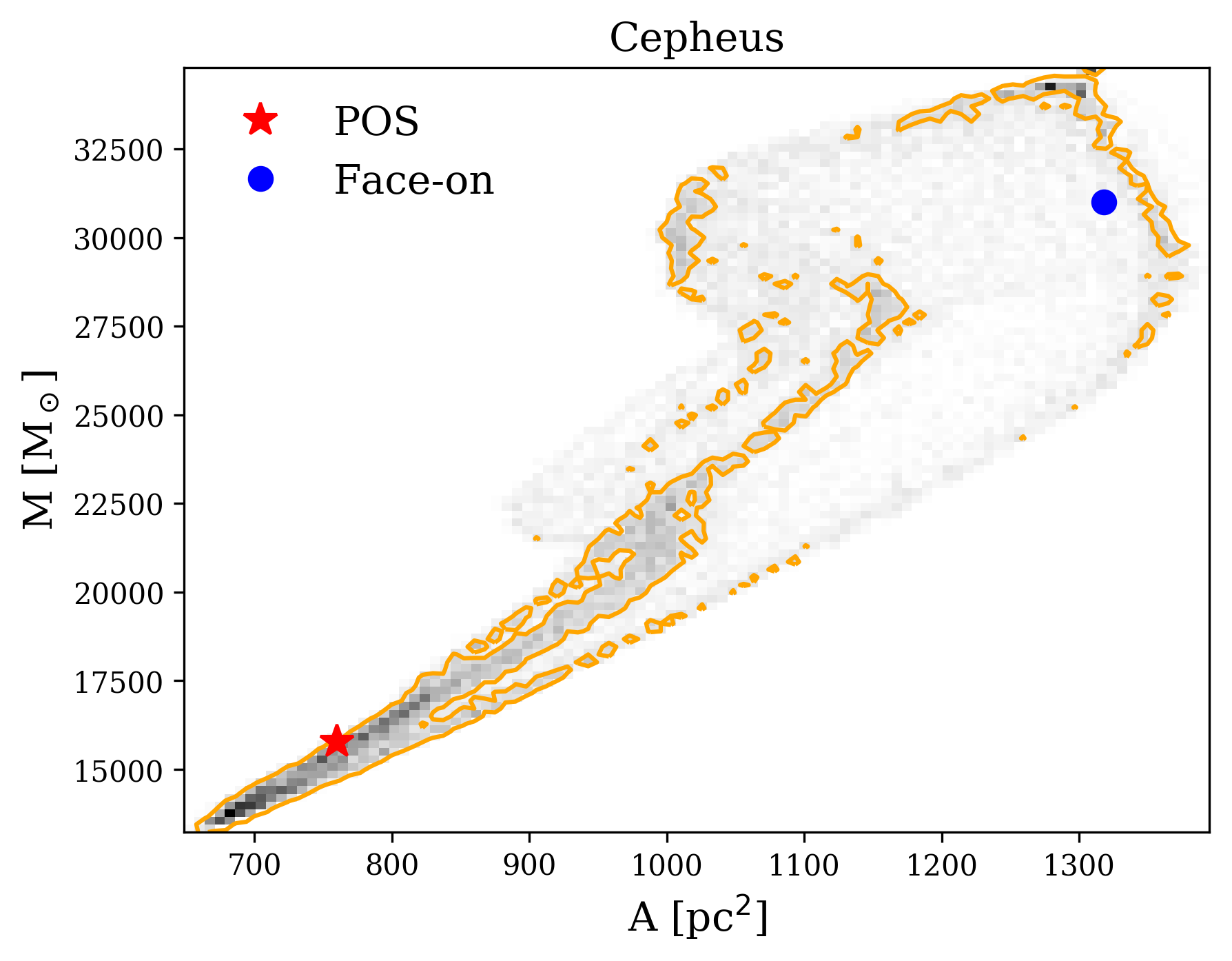}
\includegraphics[width=0.32\textwidth]{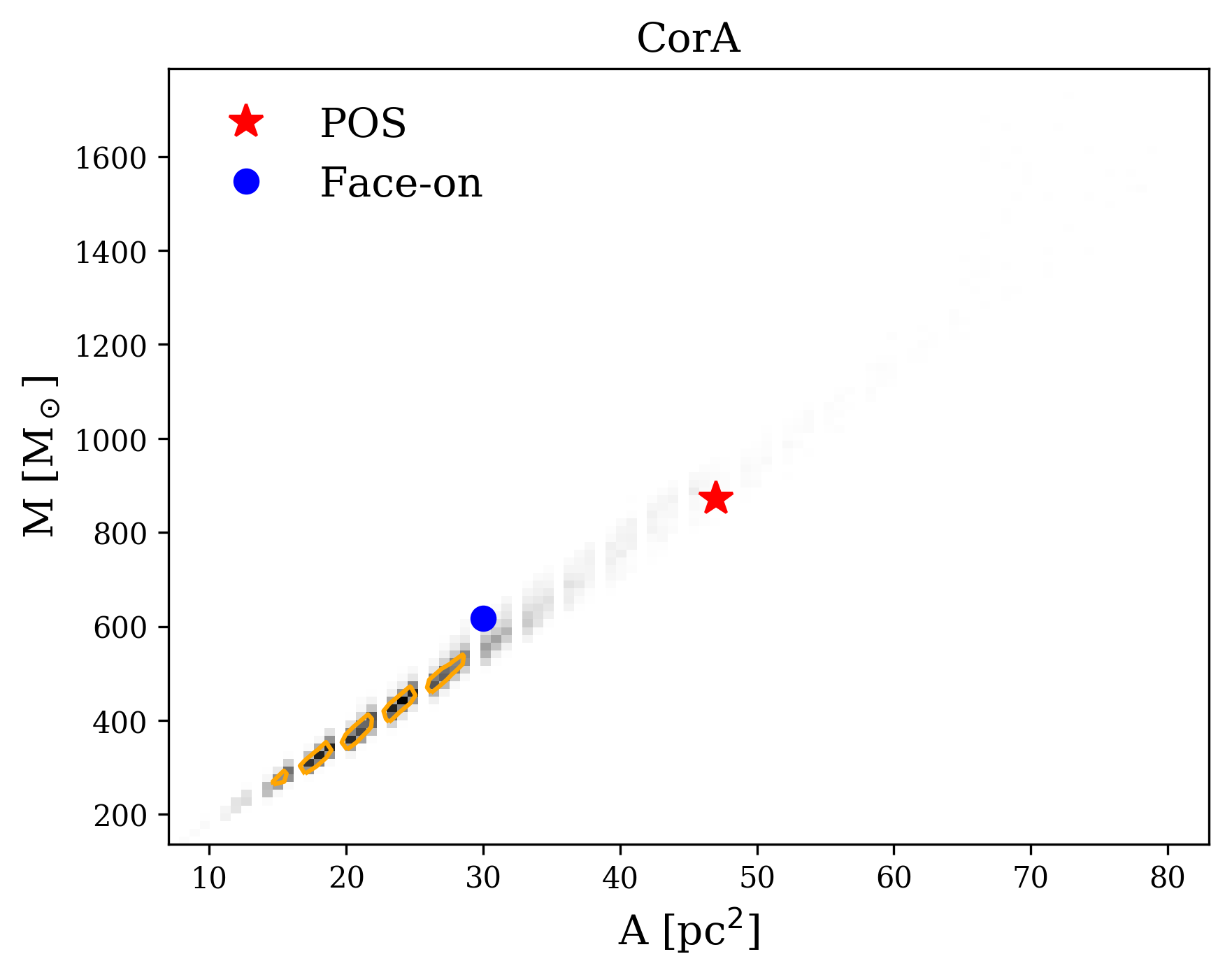}
\includegraphics[width=0.32\textwidth]{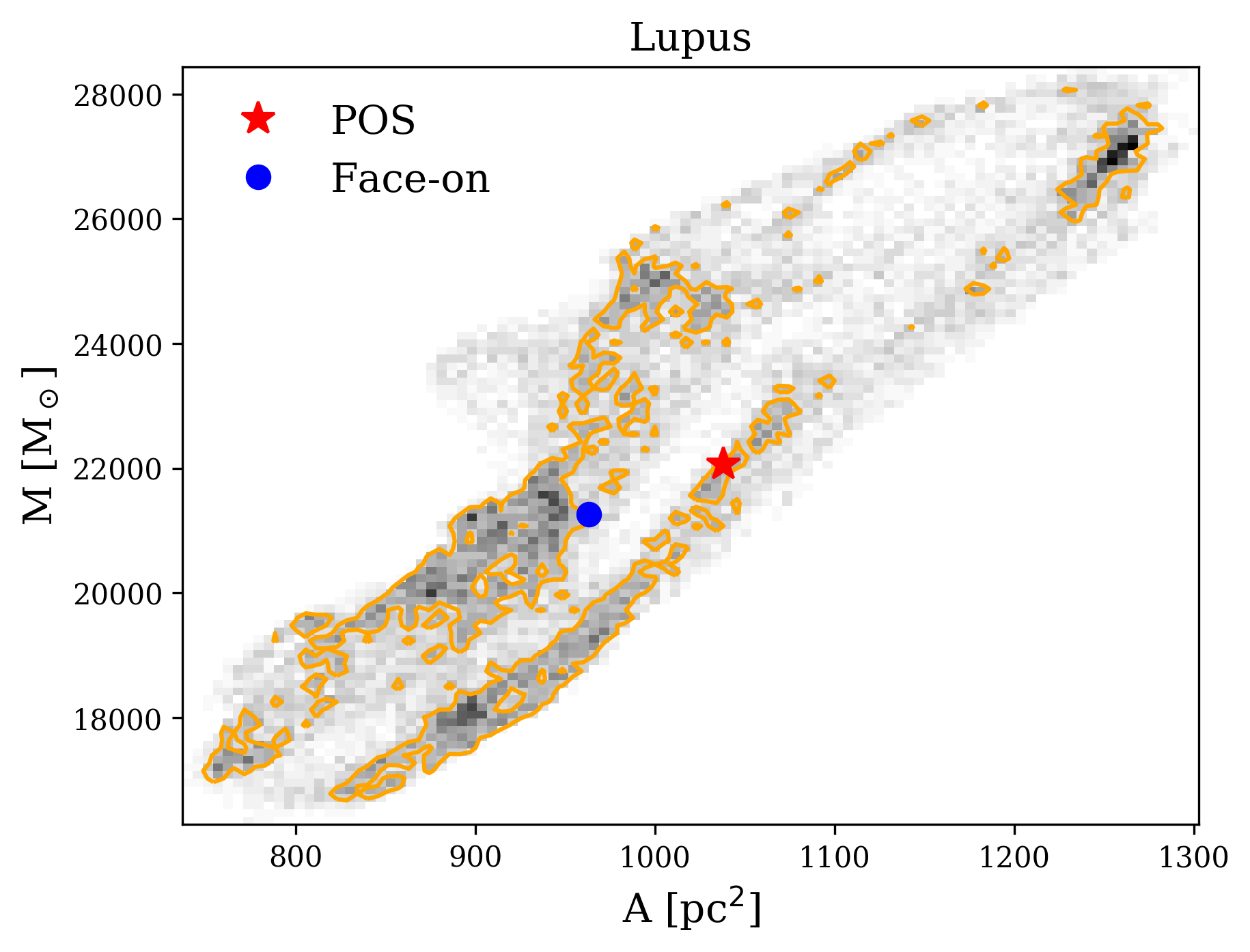}
\includegraphics[width=0.32\textwidth]{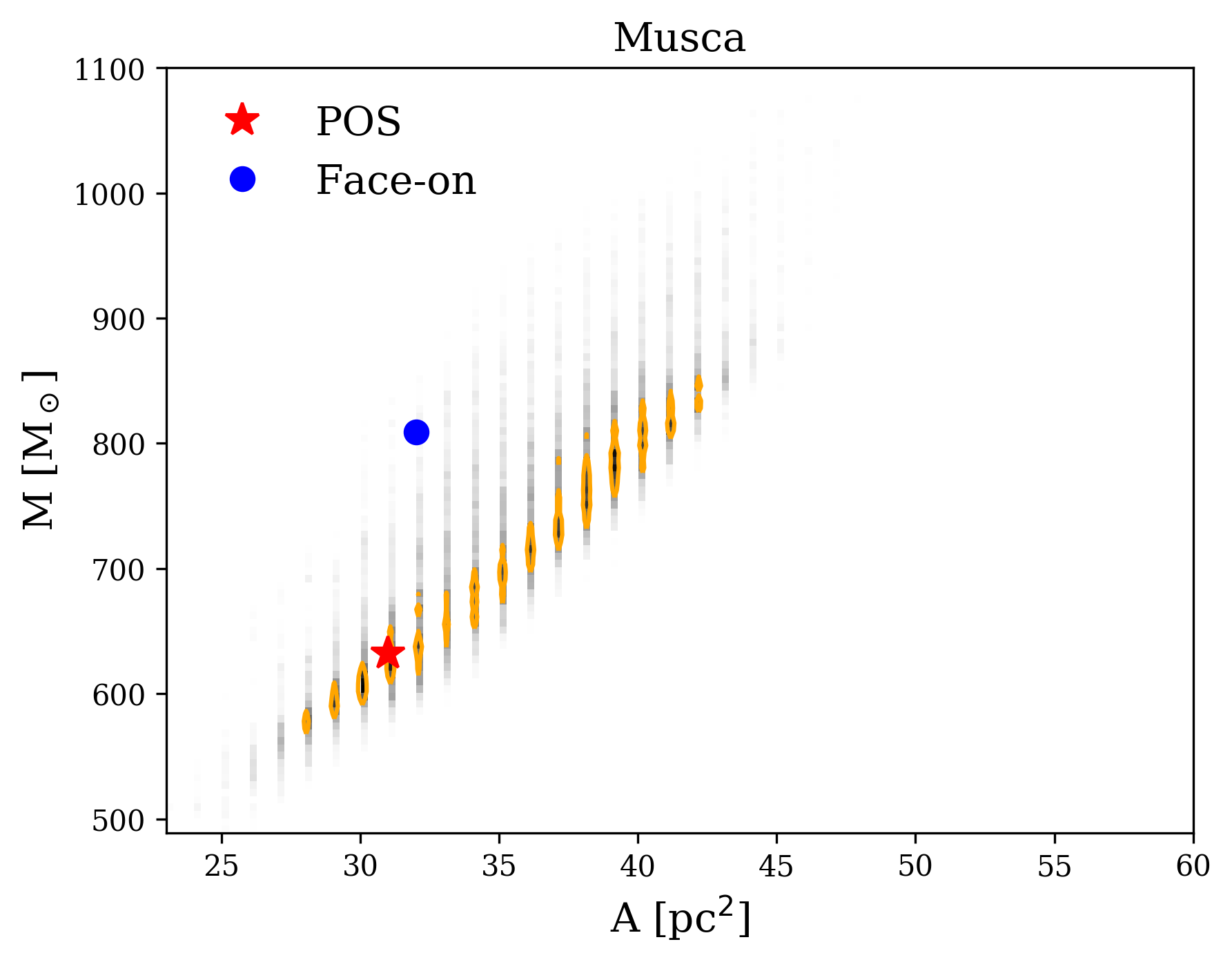}
\includegraphics[width=0.32\textwidth]{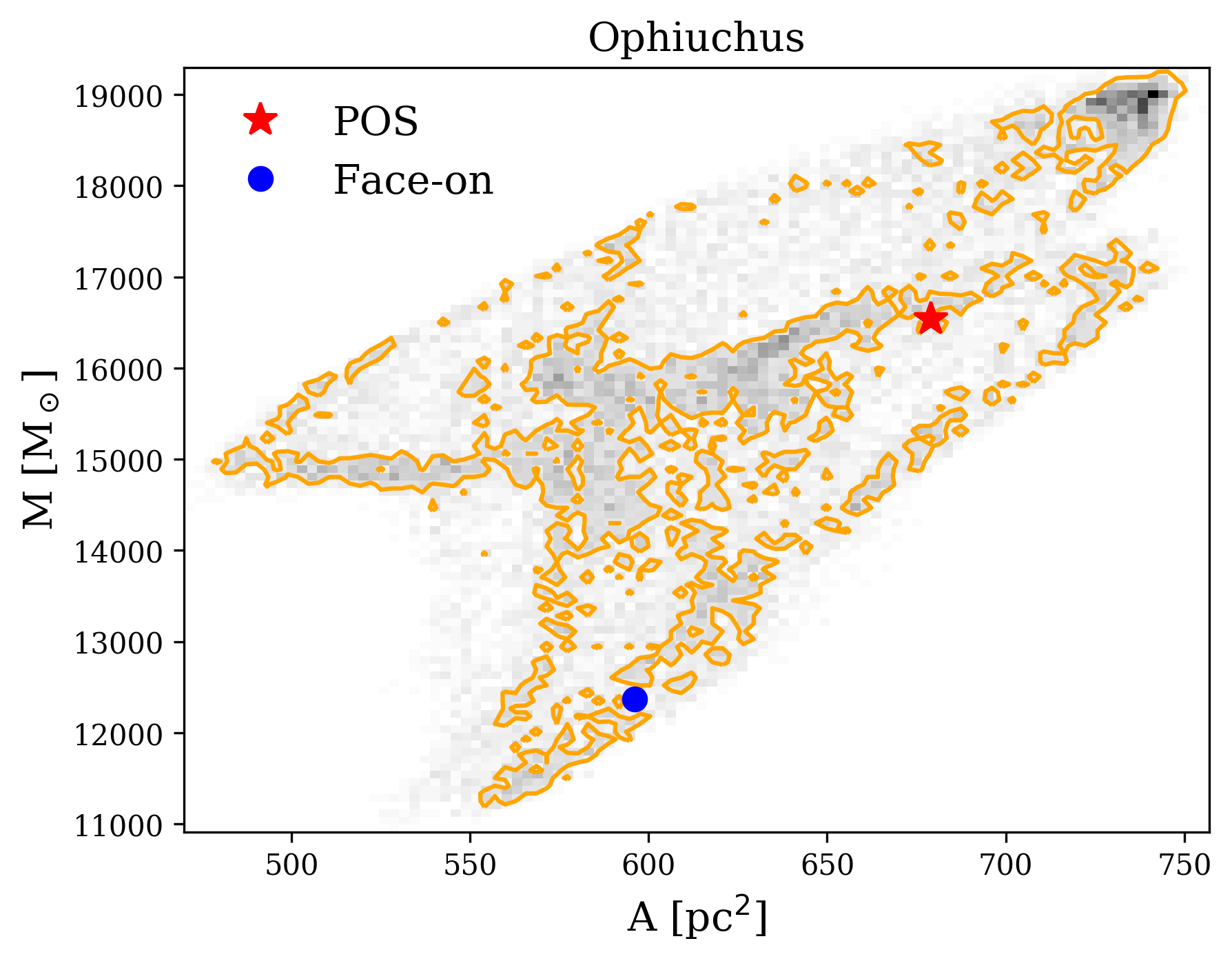}
\includegraphics[width=0.32\textwidth]{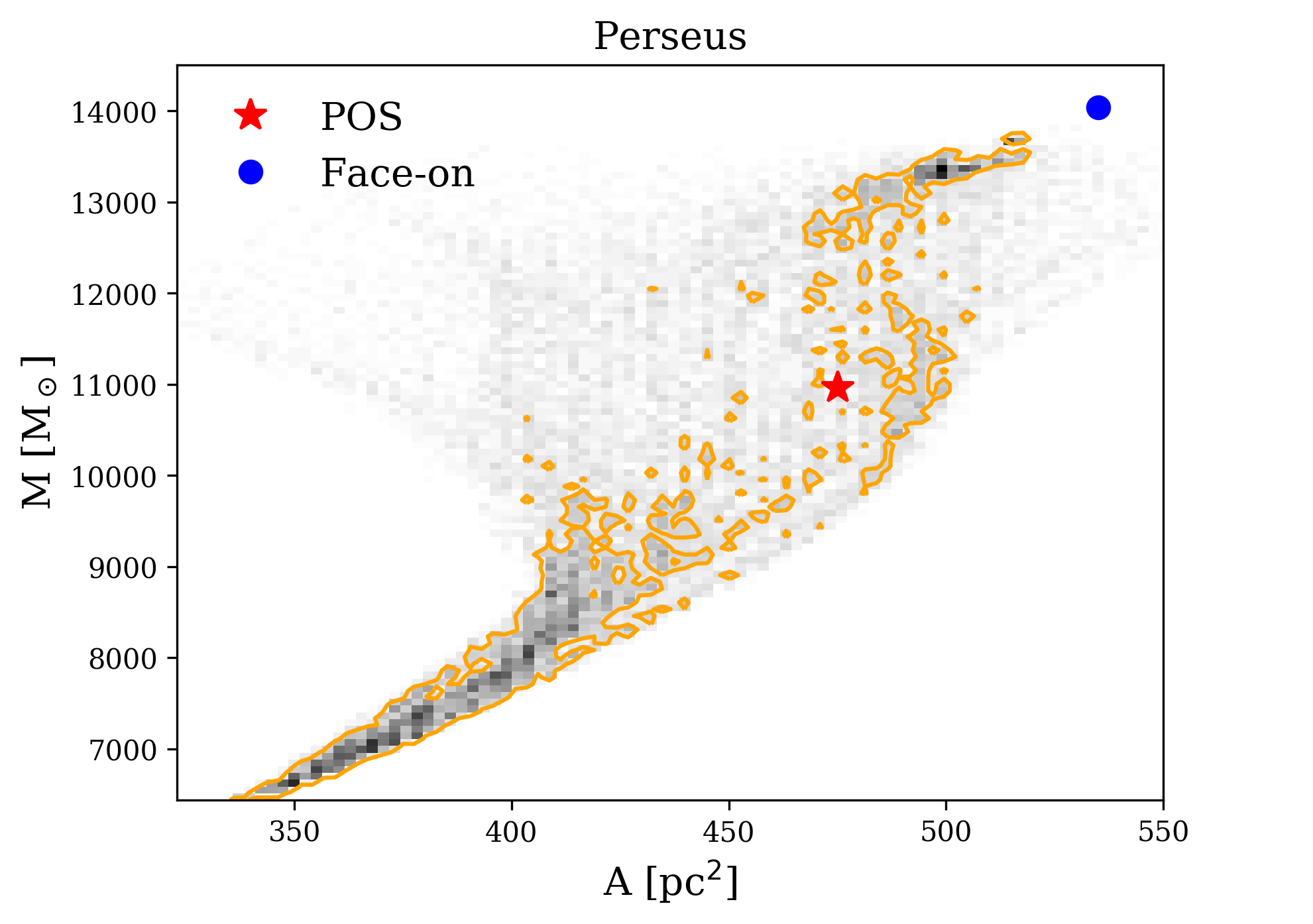}
\includegraphics[width=0.32\textwidth]{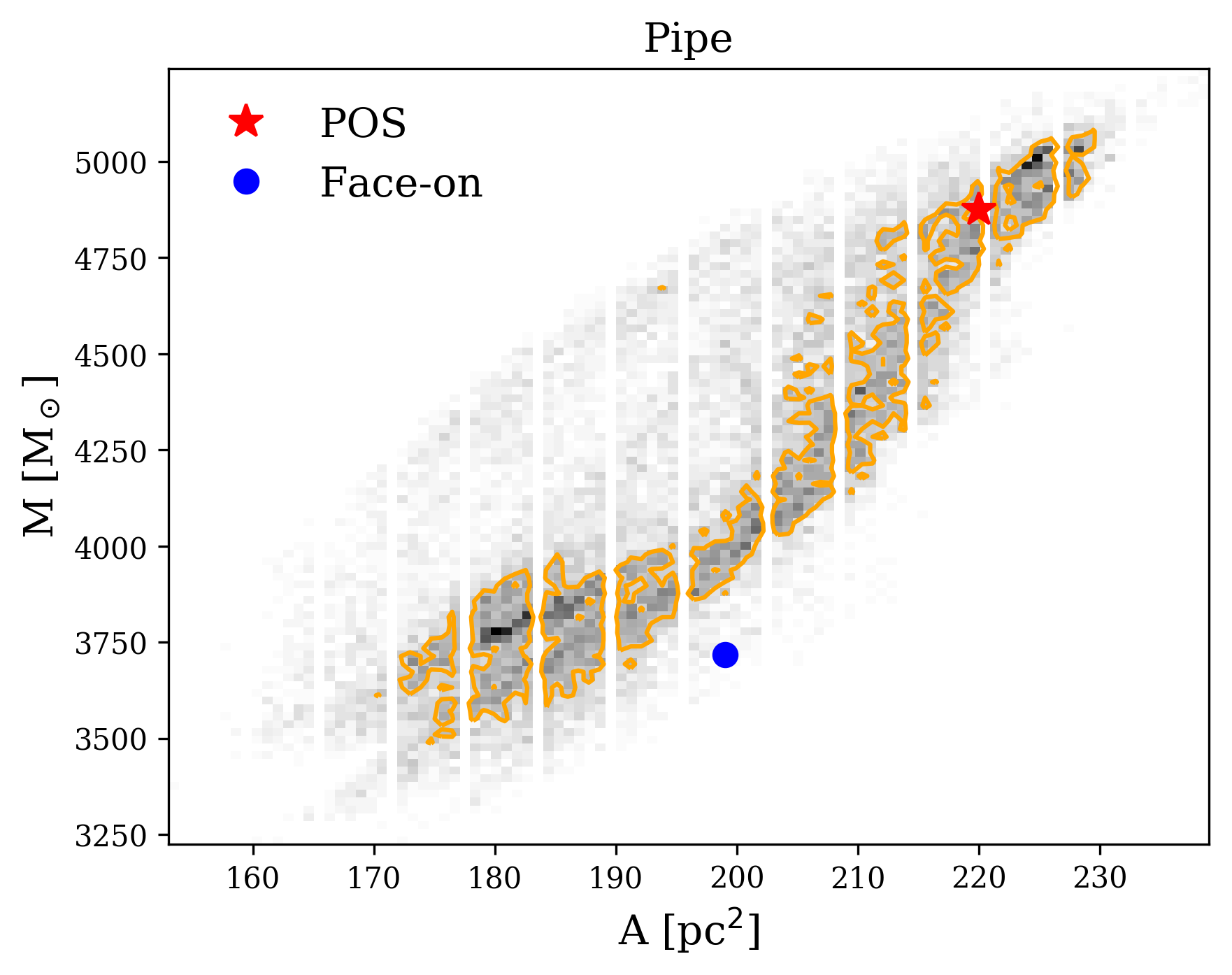}
\includegraphics[width=0.32\textwidth]{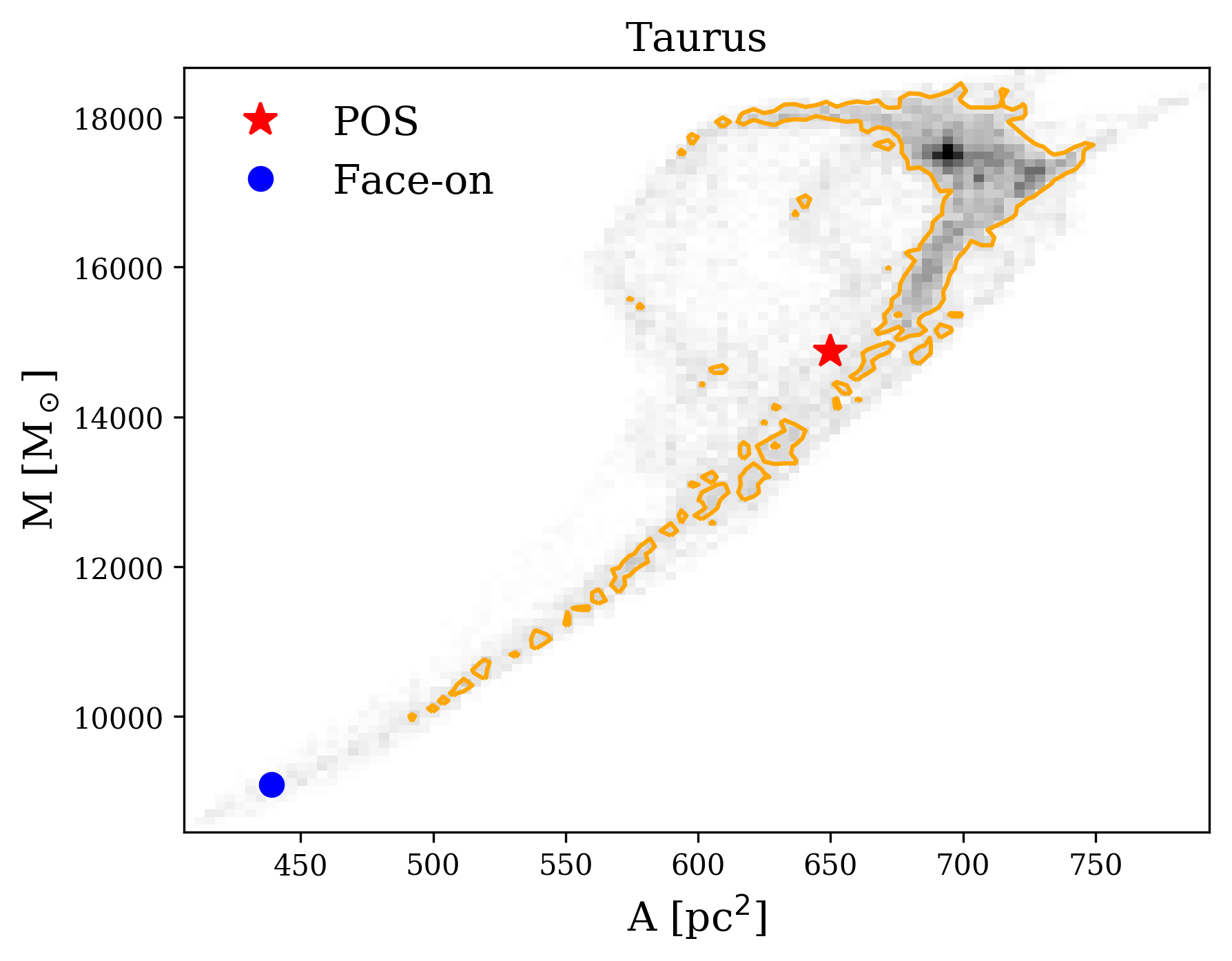}
\includegraphics[width=\columnwidth]{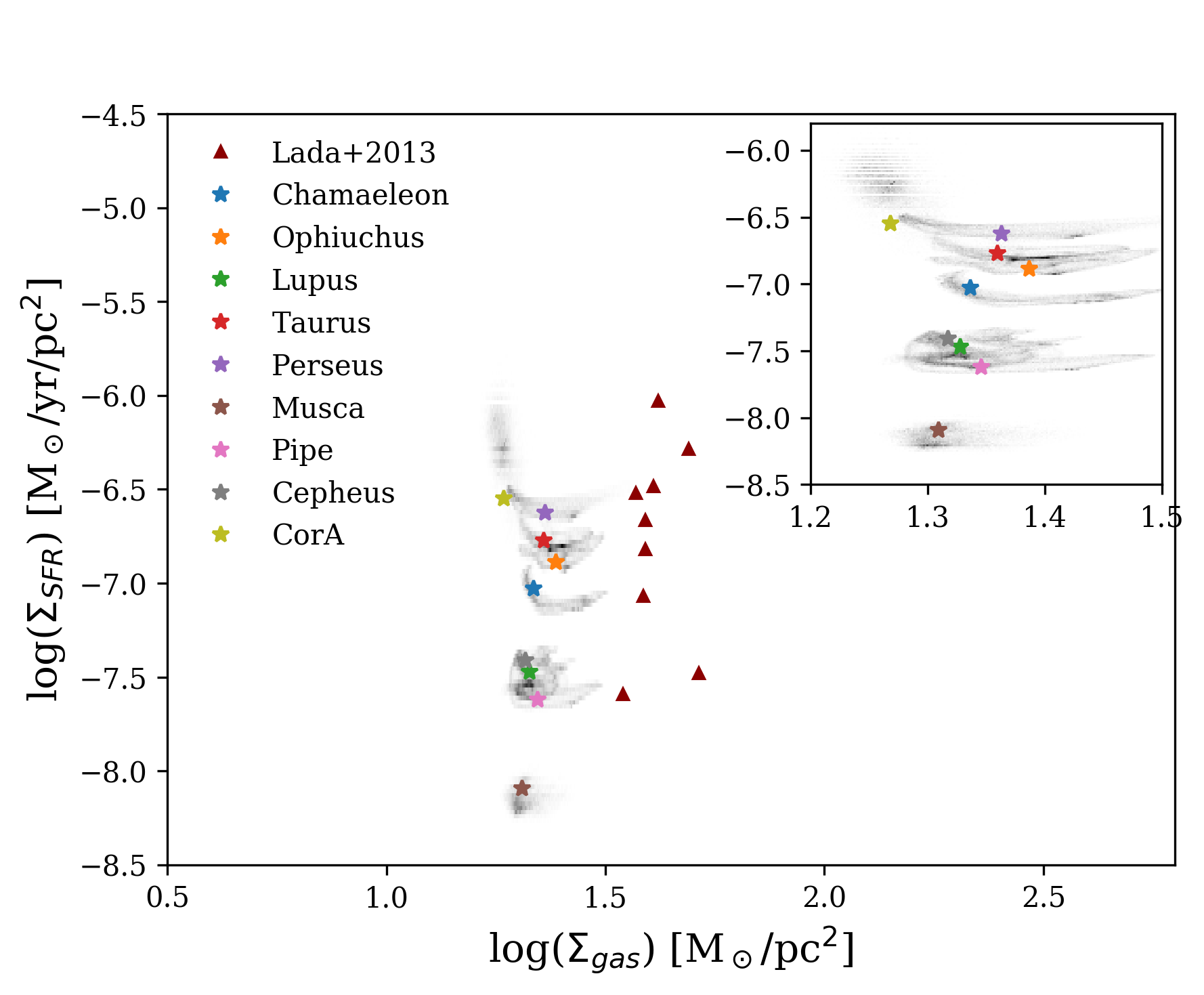}
\caption{\emph{Top $3\times3$ panels: }Joint probability distributions of the projected mass and area for the nine clouds of our sample. The red star shows the value from the POS angle and the blue circle from the face-on angle. The orange contour delineates the area within which the cumulative probability is 50\% (computed in the order of descending probability from the peak). \emph{Bottom panel: }The corresponding probability distributions in the KS-relation. All data were derived using the extinction threshold $A_\mathrm{G} = 0.5$ mag.}
\label{fig:2dhist_t0_50}
\end{figure*}

\begin{figure*}
\centering
\includegraphics[width=0.32\textwidth]{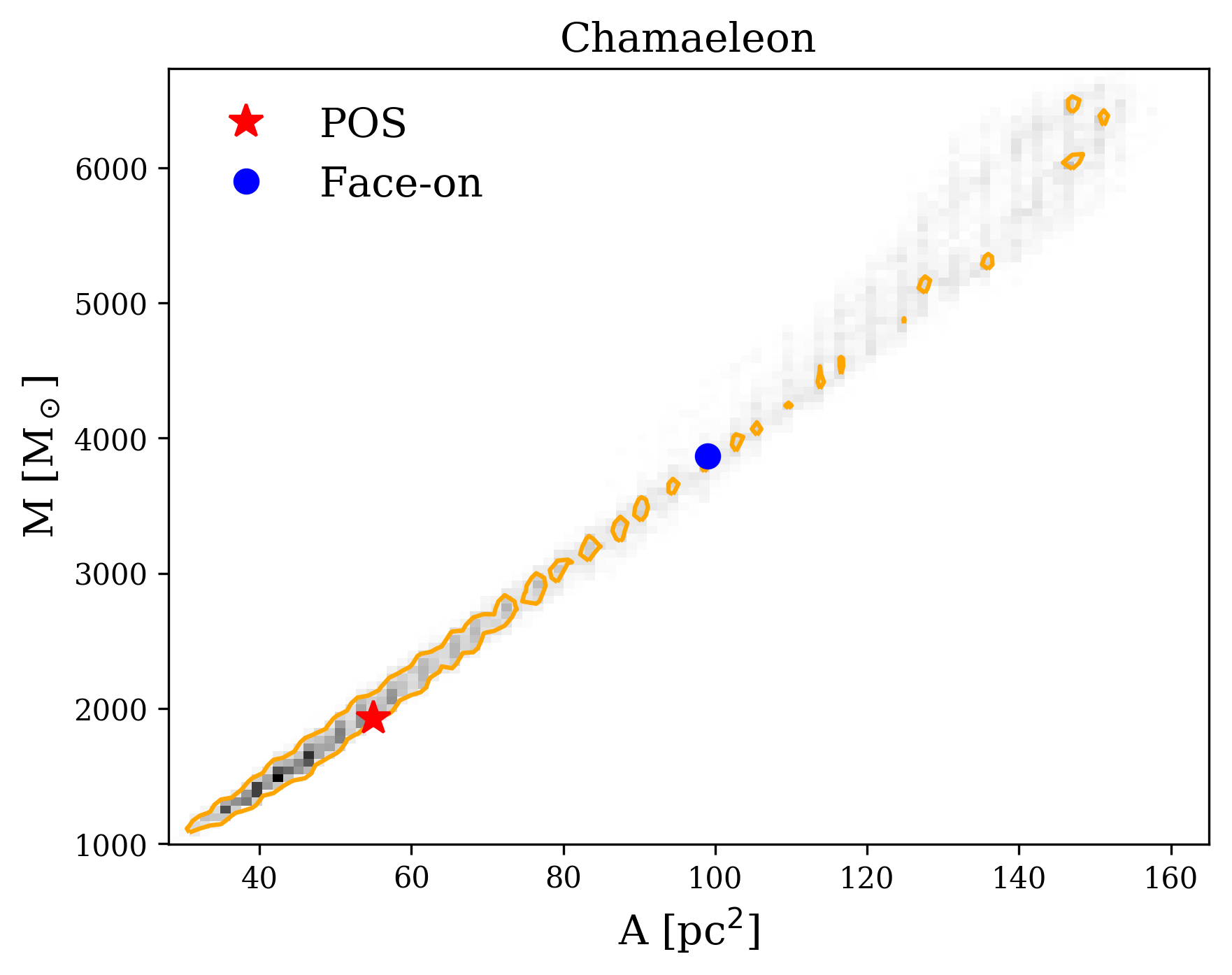}
\includegraphics[width=0.32\textwidth]{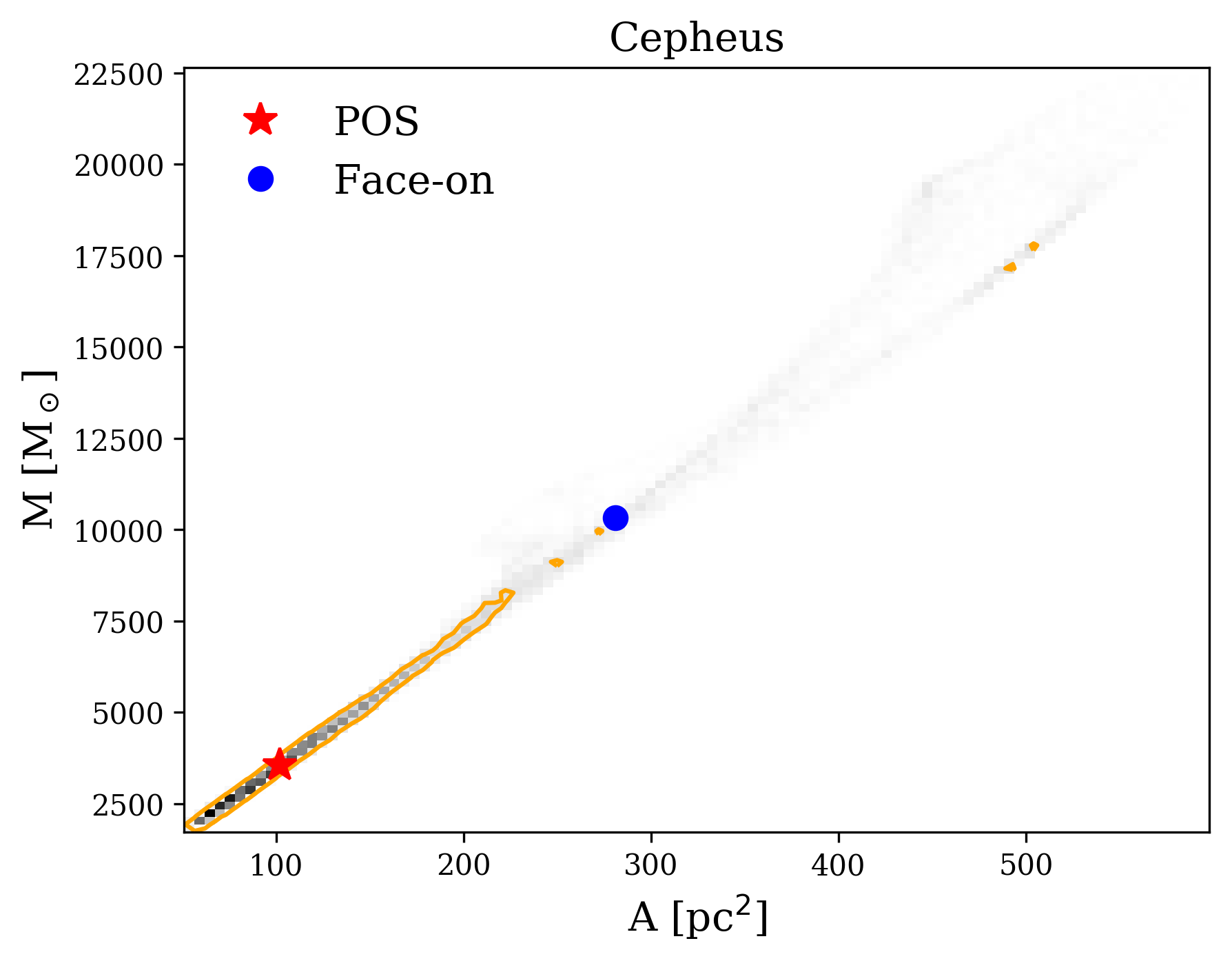}
\includegraphics[width=0.32\textwidth]{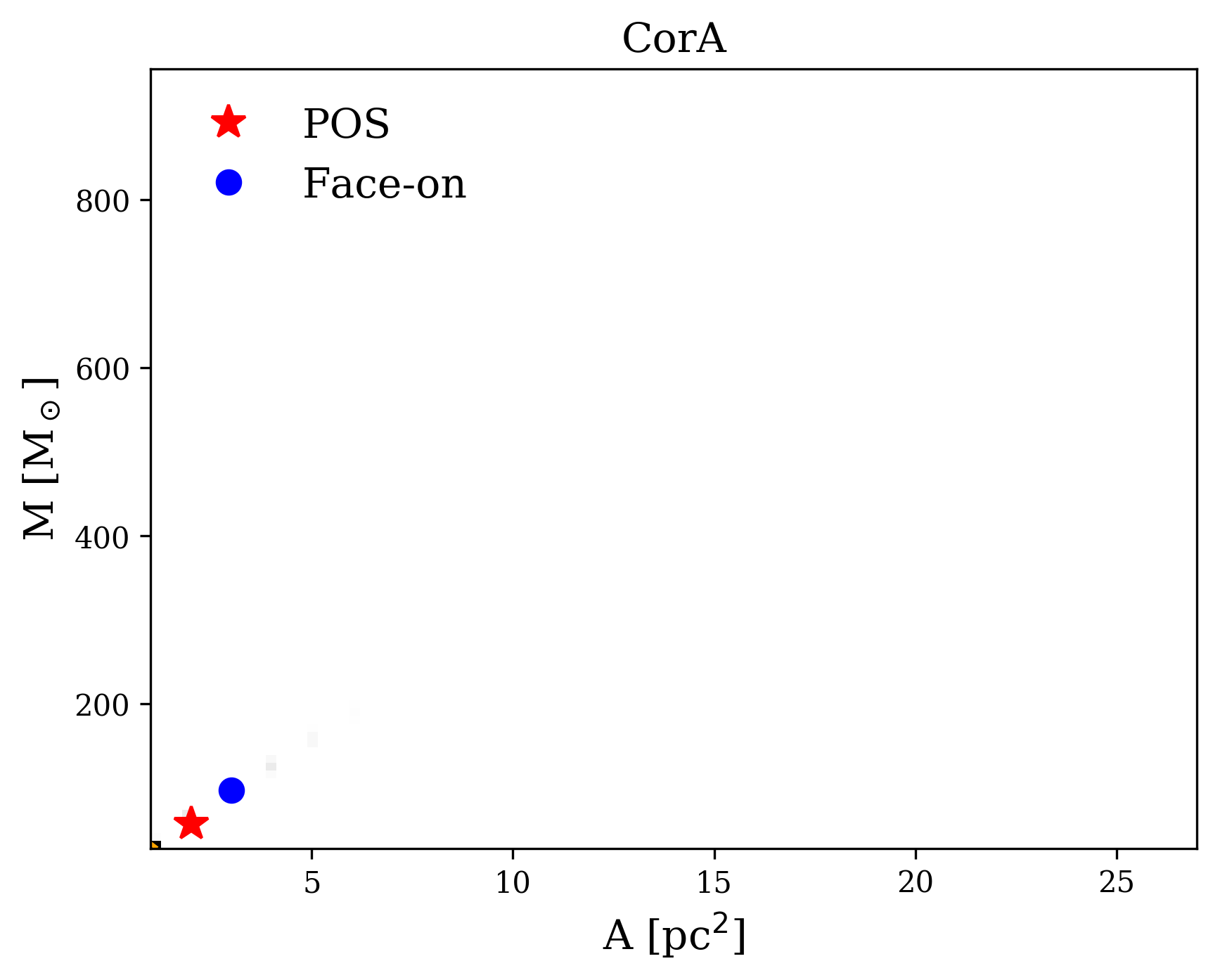}
\includegraphics[width=0.32\textwidth]{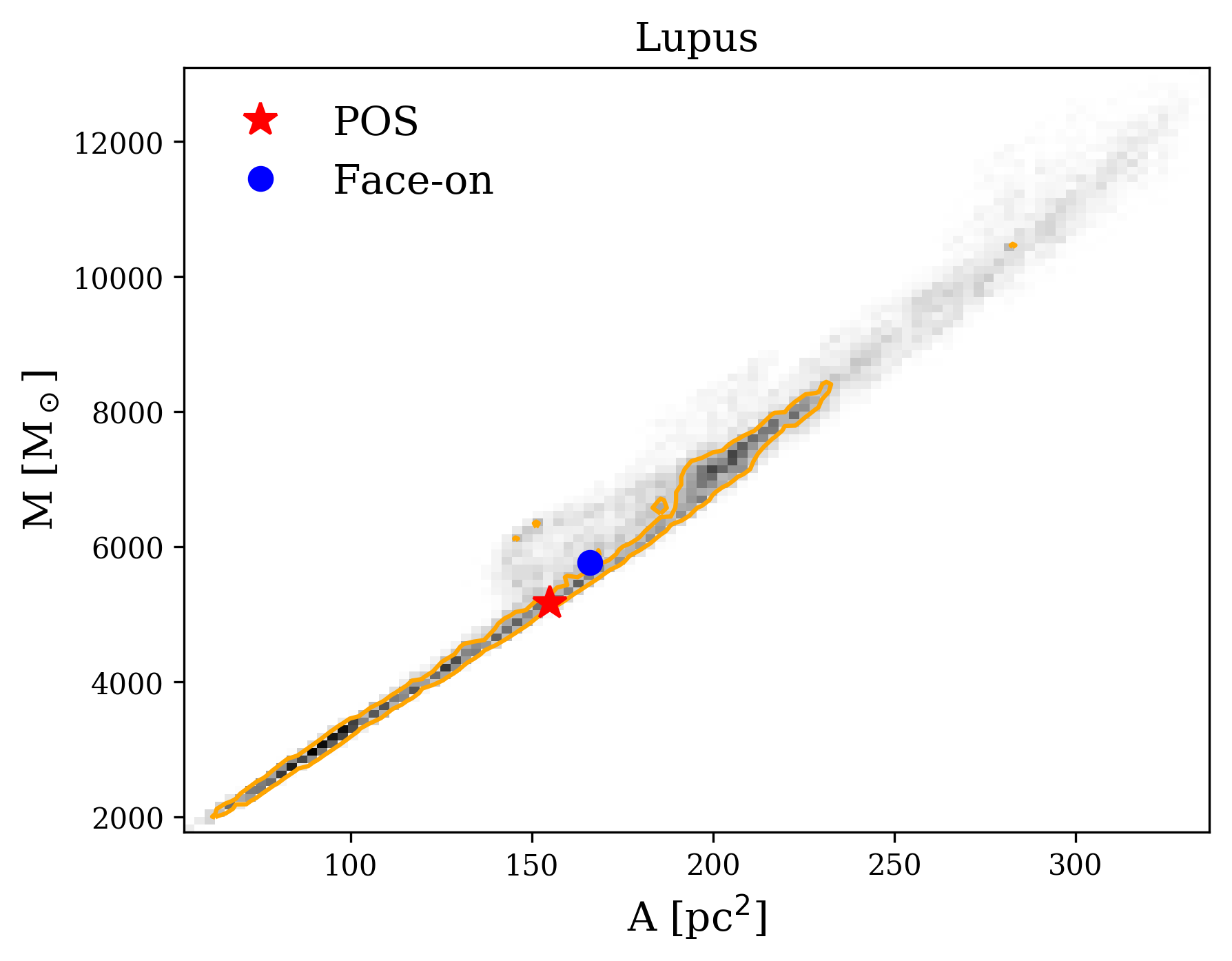}
\includegraphics[width=0.32\textwidth]{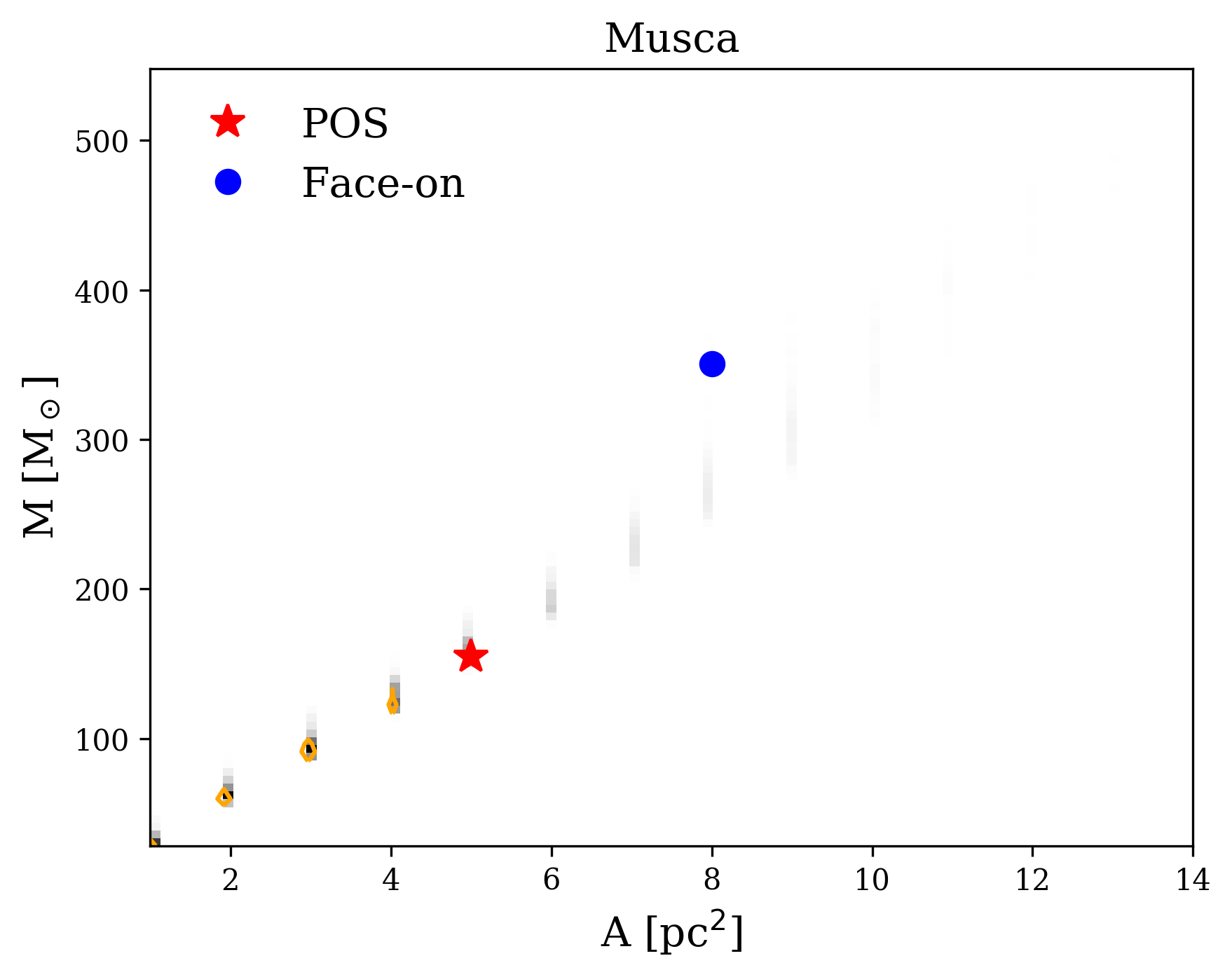}
\includegraphics[width=0.32\textwidth]{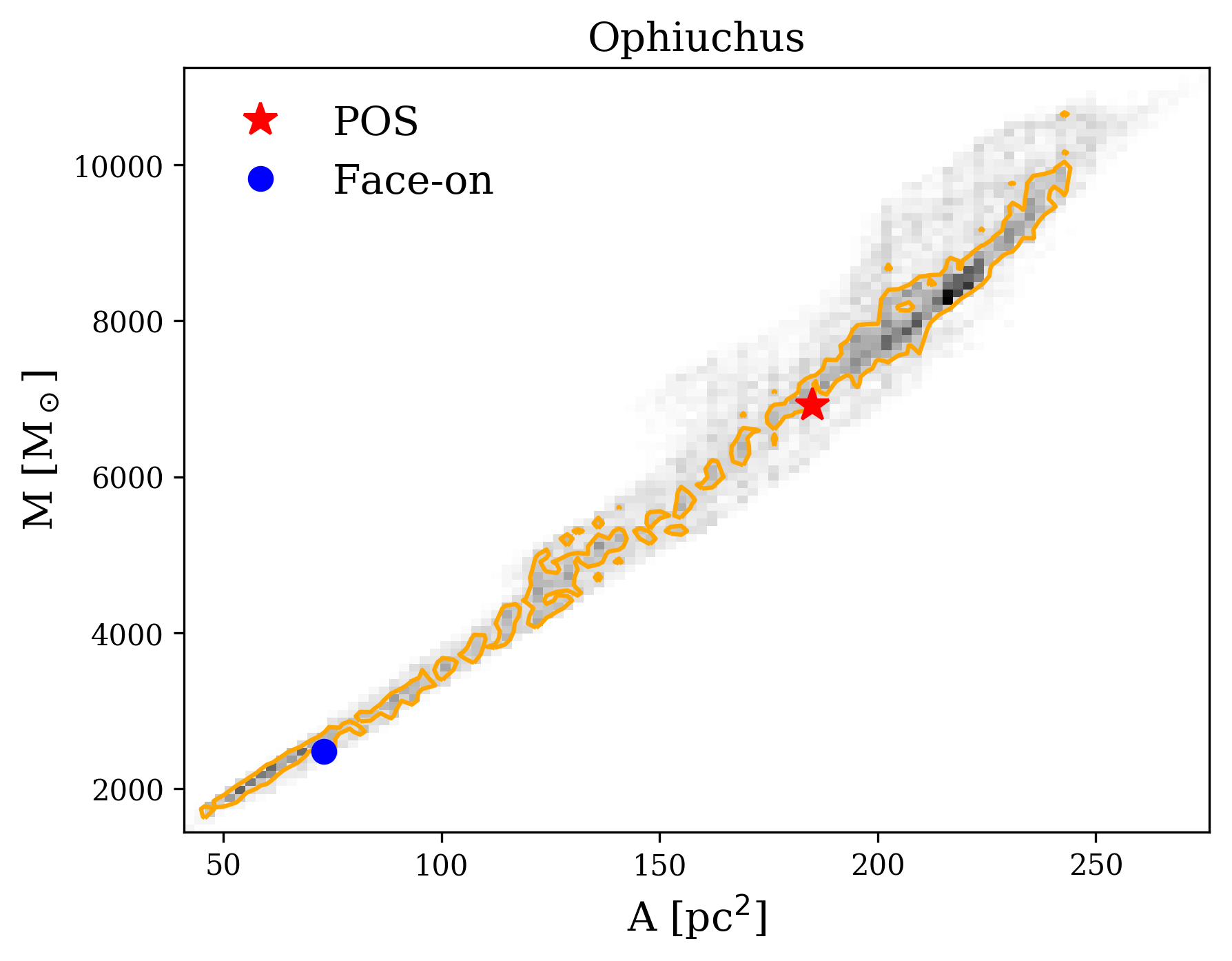}
\includegraphics[width=0.32\textwidth]{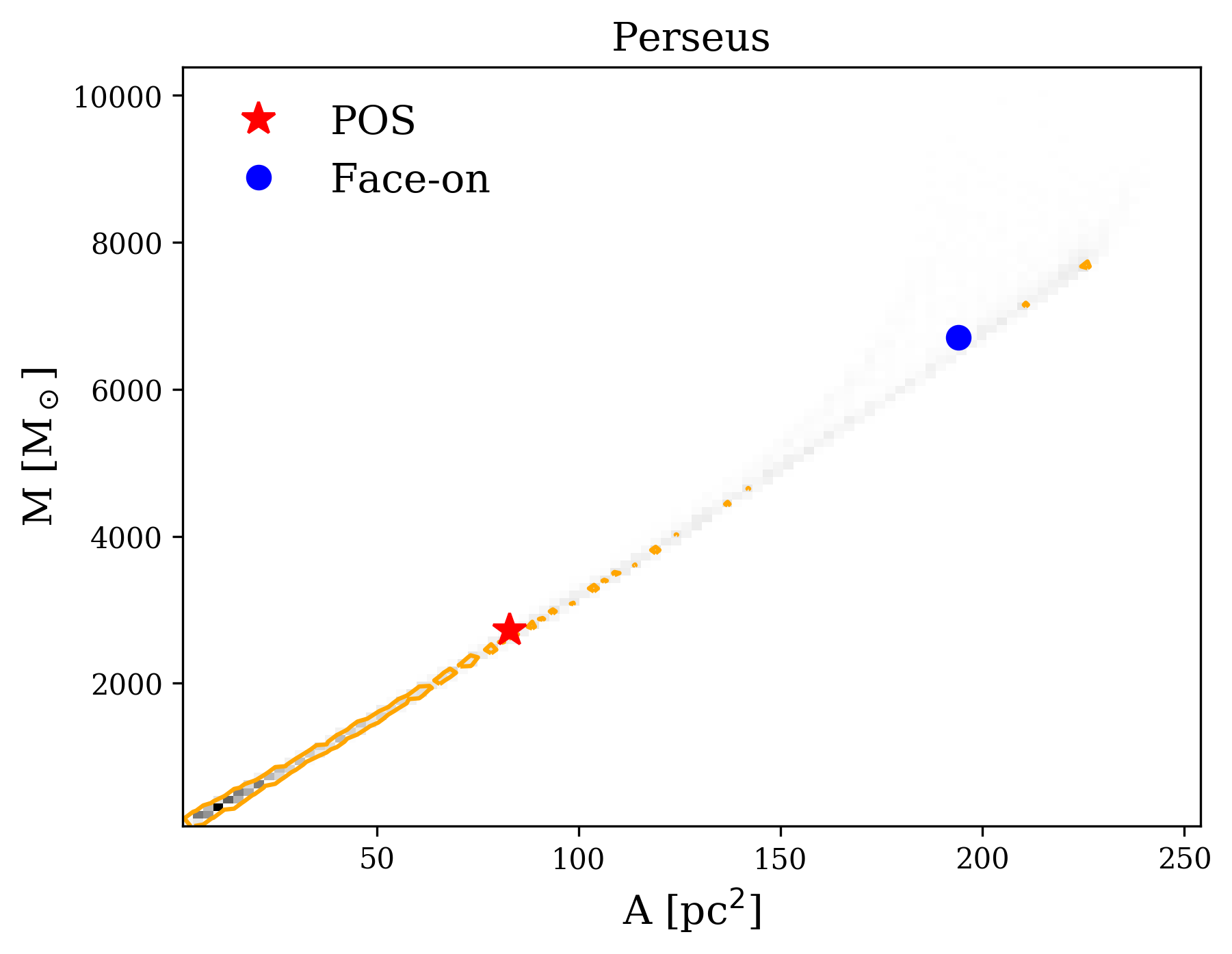}
\includegraphics[width=0.32\textwidth]{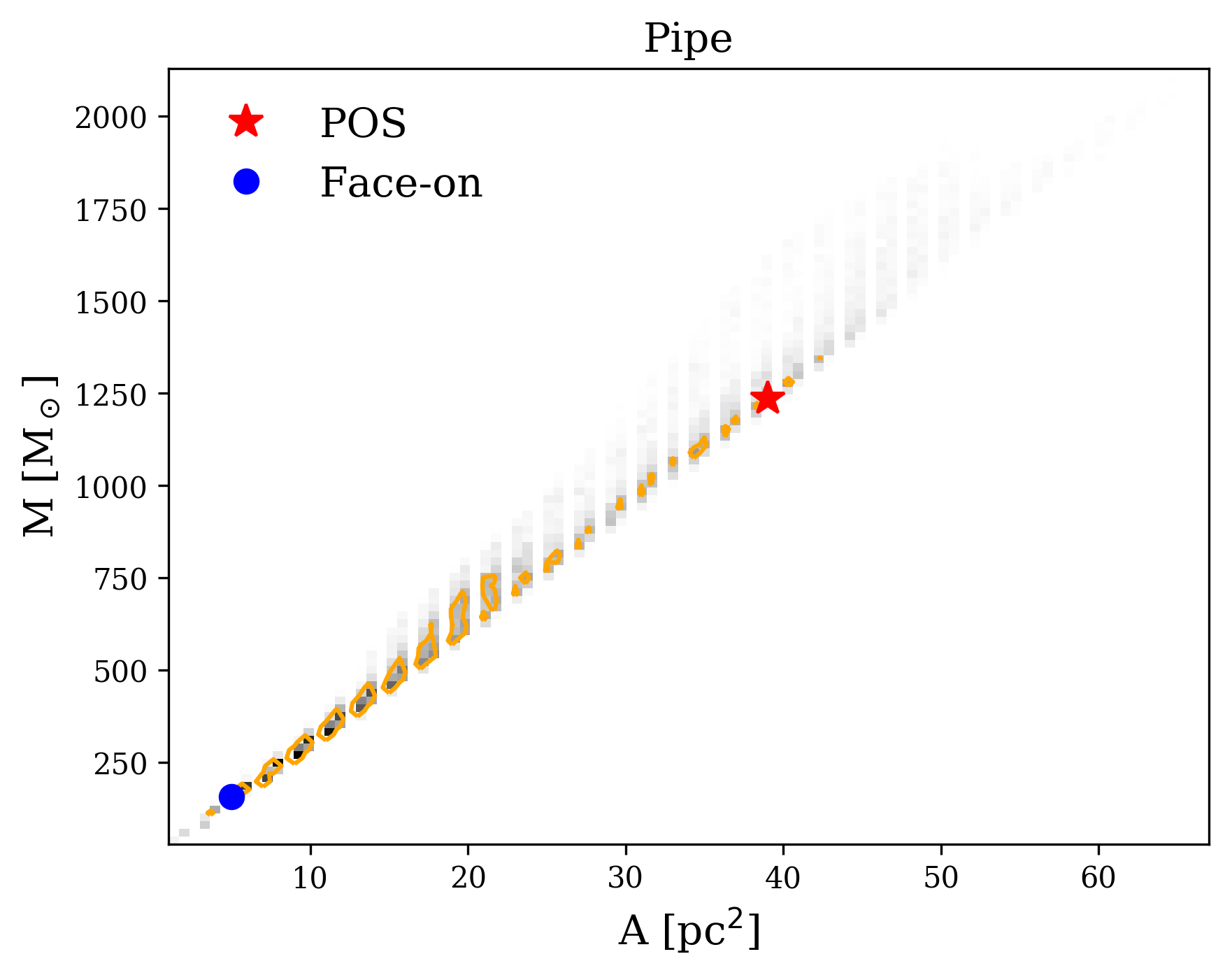}
\includegraphics[width=0.32\textwidth]{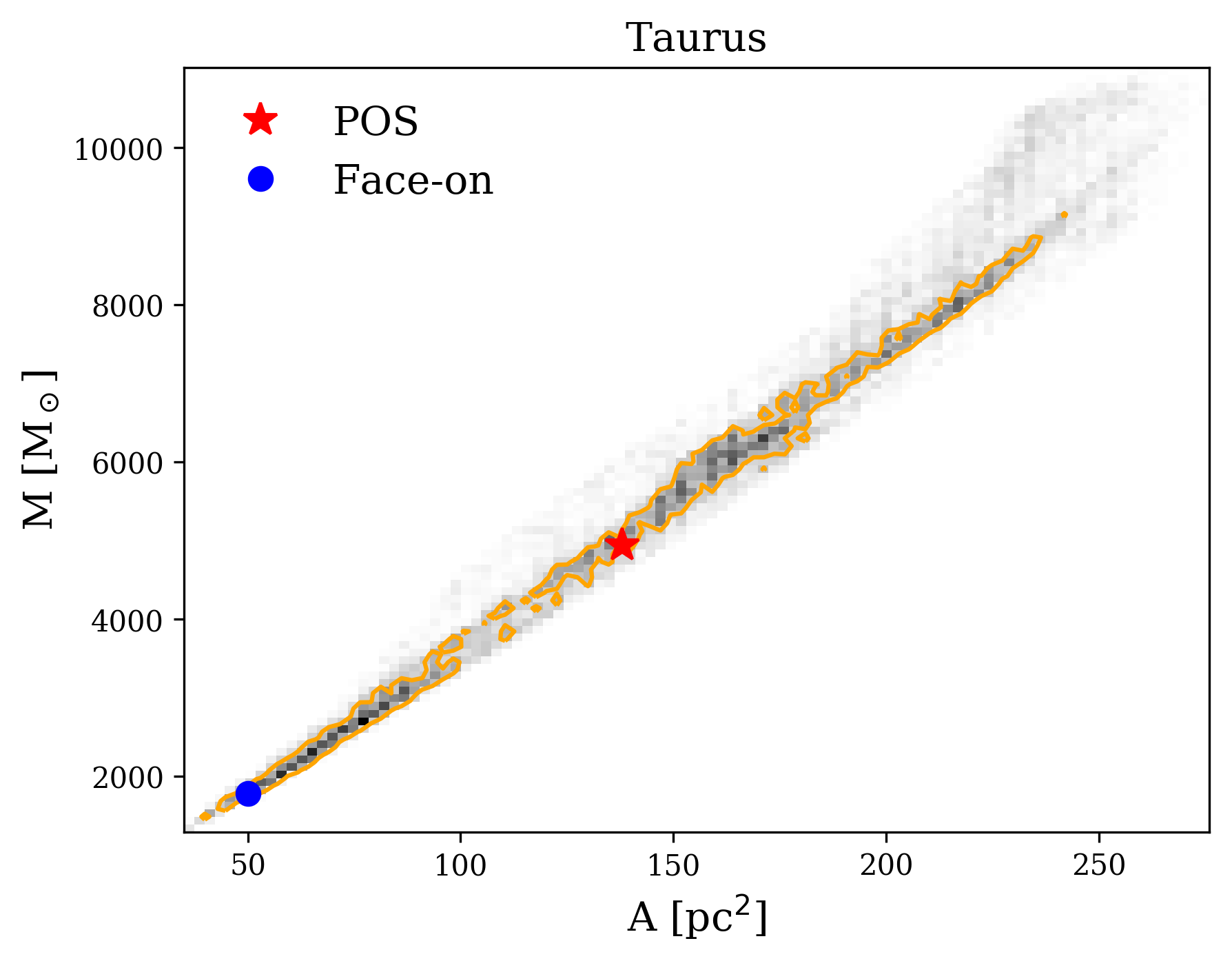}
\includegraphics[width=\columnwidth]{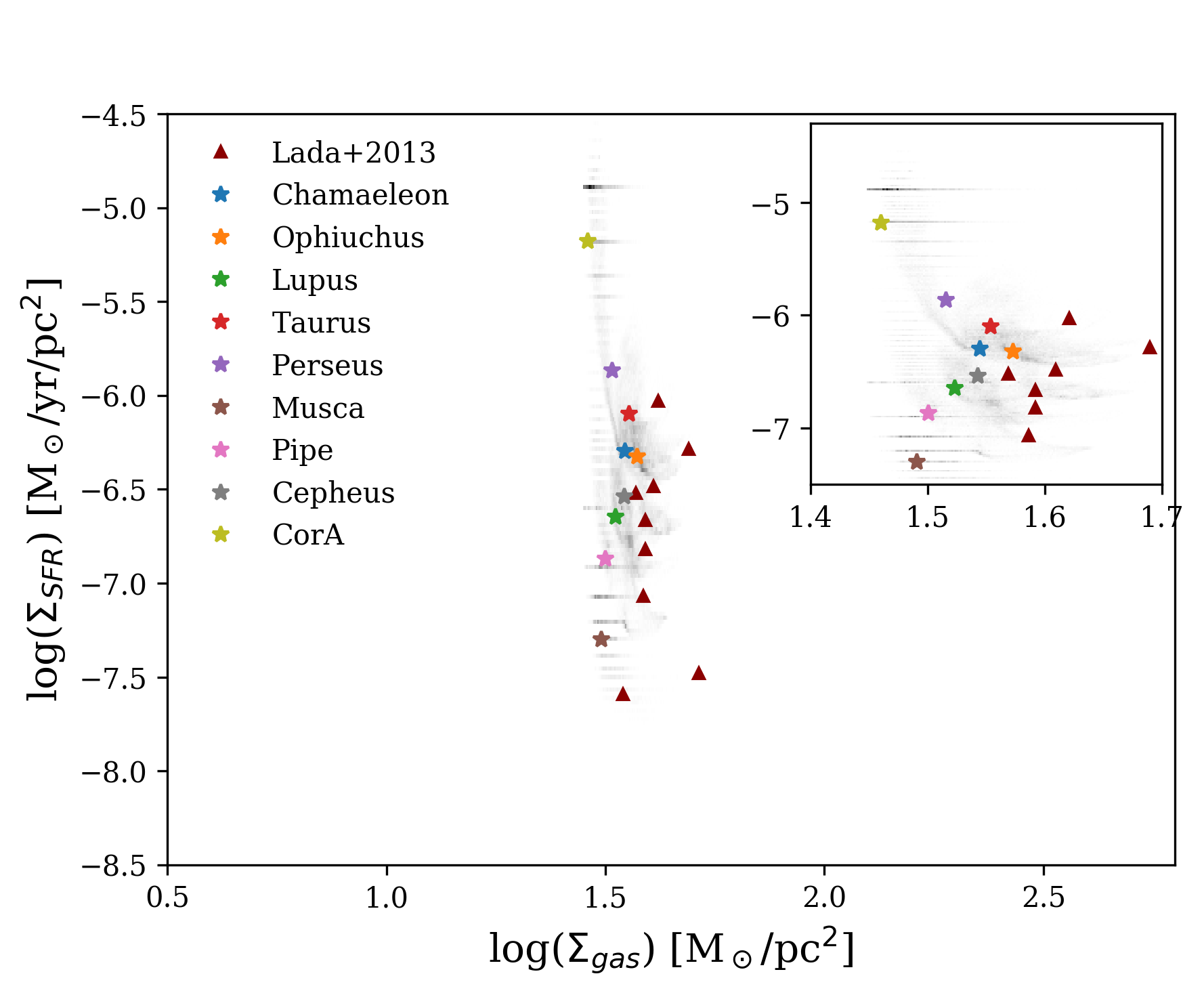}
\caption{\emph{Top $3\times3$ panels: }Joint probability distributions of the projected mass and area for the nine clouds of our sample. The red star shows the value from the POS angle and the blue circle from the face-on angle. The orange contour delineates the area within which the cumulative probability is 50\% (computed in the order of descending probability from the peak). \emph{Bottom panel: }The corresponding probability distributions in the KS-relation. All data were derived using the extinction threshold $A_\mathrm{G} = 1.0$ mag.}
\label{fig:2dhist_t1}
\end{figure*}

We describe the effect of the extinction threshold for the joint probability distributions of masses and areas and on the KS-relation in Figs. \ref{fig:2dhist_t0_50} and \ref{fig:2dhist_t1}. These figures show that as the threshold decreases the joint distributions become increasingly more complex and flat. However, the effect of this on the KS-relation is relatively small, owing to the co-variance of the mass and area probability distributions.

\section{Comparison of POS and face-on perspectives}
\label{sec:appendix_POS-vs-faceon}

\begin{figure}
\centering
\includegraphics[width=\columnwidth]{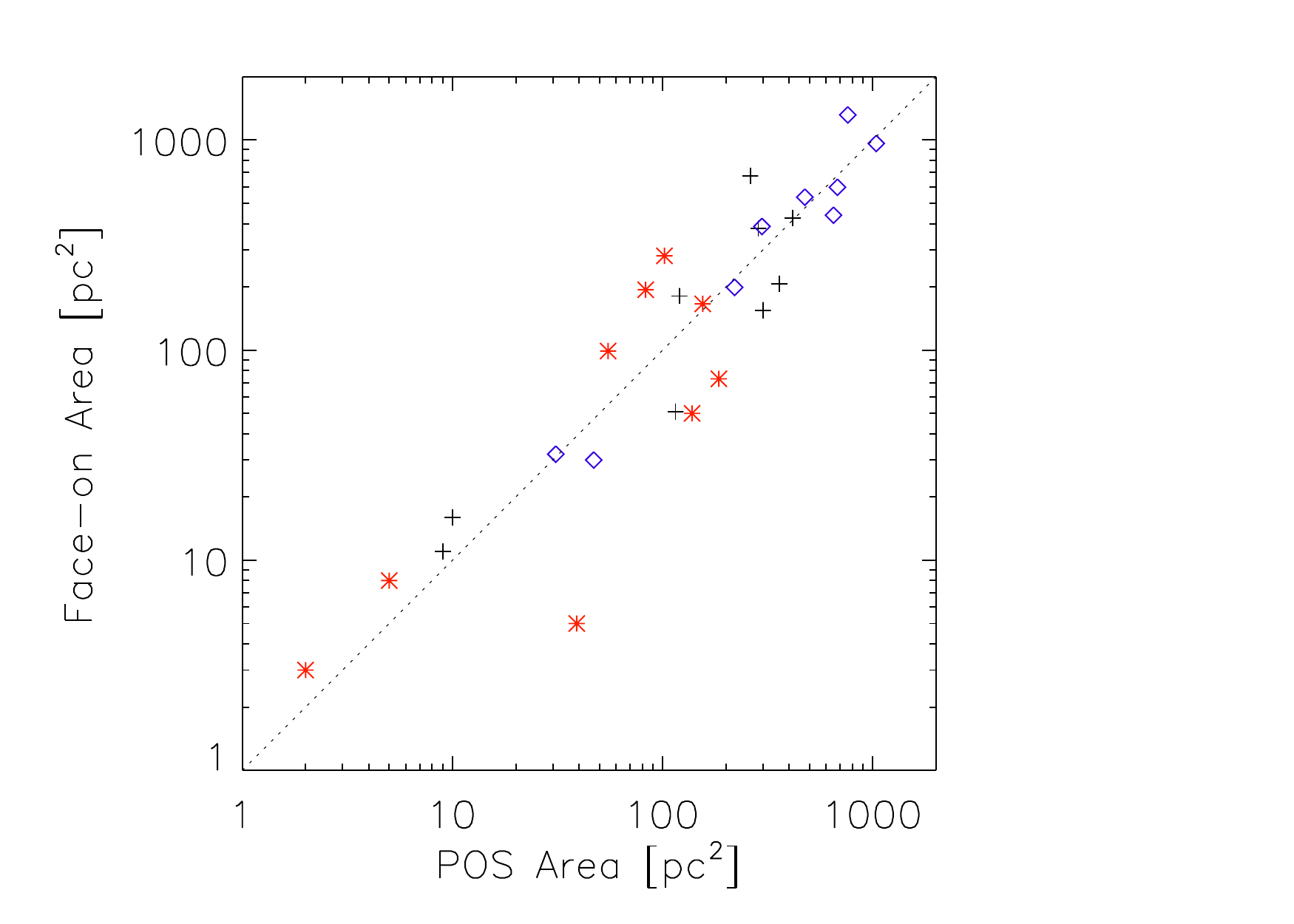}
\includegraphics[width=\columnwidth]{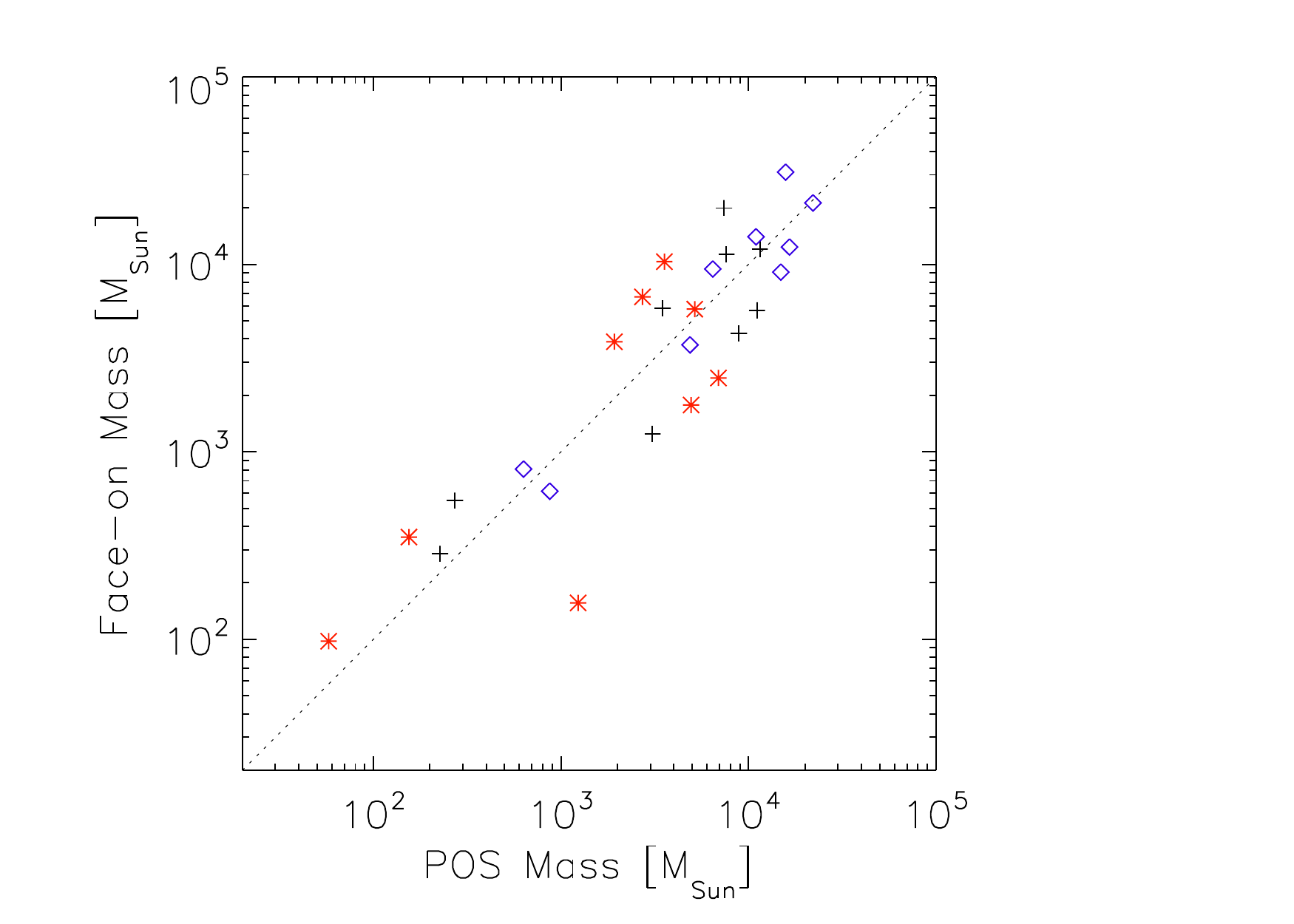}
\includegraphics[width=\columnwidth]{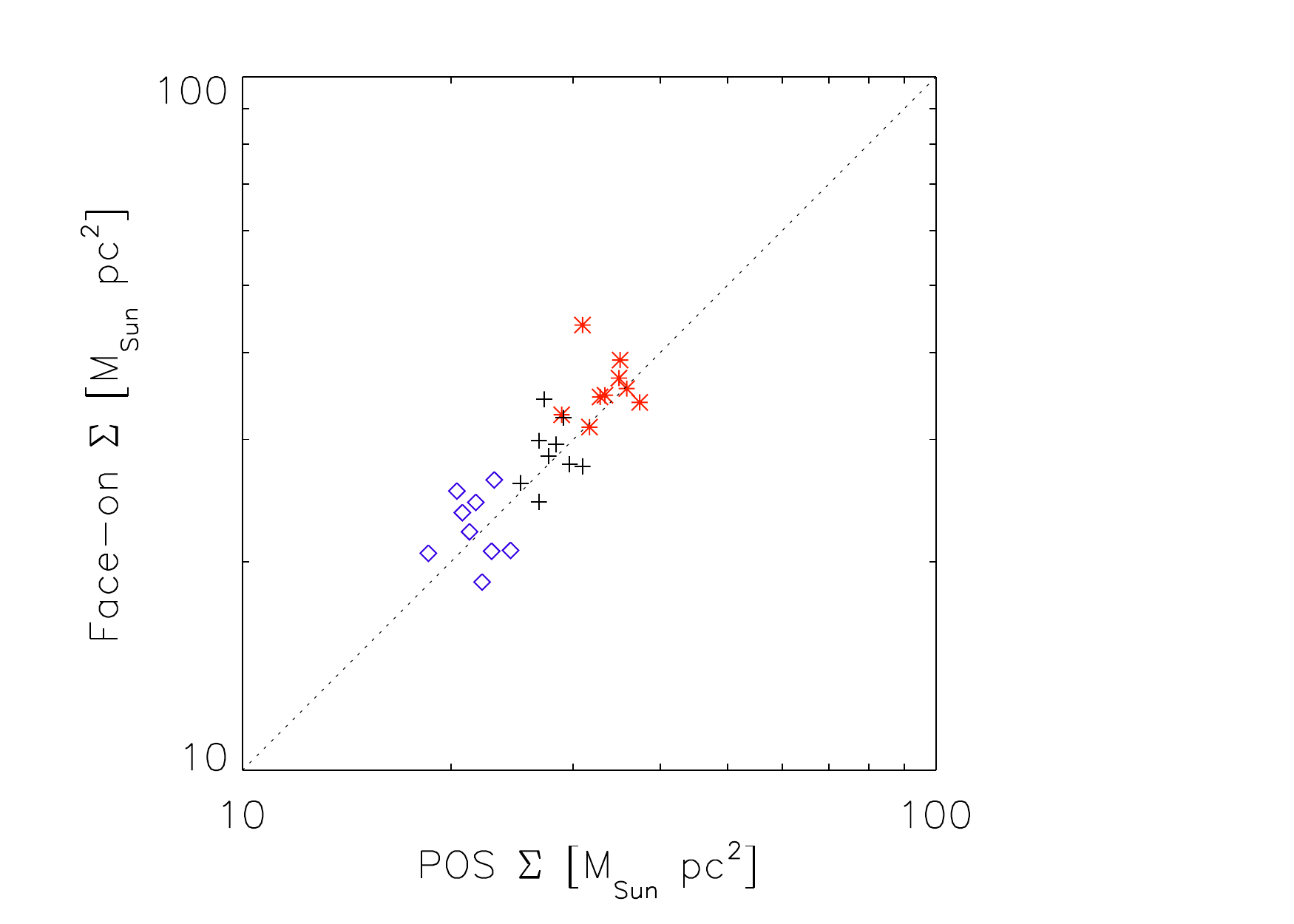}
\caption{The relationship between cloud areas (top), masses (centre), and surface densities (bottom) from the POS and face-on angles for the nine clouds in our sample. The are derived using the thresholds of $A_\mathrm{G}$=0.5 mag (blue diamonds), 0.75 mag (black pluses), and 1 mag (red asterisks). The dotted line shows the one-to-one relation. }
\label{fig:pos_vs_faceon}
\end{figure}

We present here a comparison of the areas, masses, and surface densities of the nine sample clouds from the two specific viewing angles: the POS and face-on angles. Figure \ref{fig:pos_vs_faceon} presents the comparison for all three extinction threshold used in the study, i.e., $A_\mathrm{G}=\{0.5, 0.75, 1\}$ mag. The comparison shows that even though the individual clouds can show clearly different properties when viewed from the POS and face-on angles, the sample of nine clouds on average aligns around the one-to-one relation. This suggests that the average masses, areas, and surface densities of the cloud samples are not significantly different when viewed from POS or face-on perspectives. This is an relevant for studies that aim at comparing, e.g., cloud properties from extra-galactic and galactic studies. However, the effect of the used threshold is clearly visible in all measures; this again points out the importance of homogenising the cloud definition processes when comparing different cloud catalogues. 


\end{document}